\documentclass[
  aps,
  prb,
  reprint,
  hyperref,
  colorlinks=true,
  linkcolor=blue,
  citecolor=blue,
  urlcolor=blue
]{revtex4-2}
\usepackage{newunicodechar}
\newunicodechar{₄}{$_4$}

\usepackage{xurl}

\usepackage{algpseudocode}
\usepackage{graphicx}
\usepackage{float}
\usepackage{wasysym}
\usepackage{amssymb}
\usepackage{pgfplots}
\usepackage{bbold}
\usepackage{tcolorbox}
\let\iint\relax
\let\iiint\relax
\usepackage{amsmath}
\usepackage{braket}
\newcommand{\MyEnsure}{\item[\textbf{Ensure:}]}
\usepackage{threeparttable}
\usepackage{longtable}
\usepackage{listings}
\usepackage{quantikz}
\usepackage{placeins}
\lstdefinestyle{pythonstyle}{
  language=Python,
  basicstyle=\ttfamily\footnotesize,
  keywordstyle=\color{blue},
  commentstyle=\color{teal},
  stringstyle=\color{magenta},
  backgroundcolor=\color{pink!10}, 
  frame=single,
  rulecolor=\color{pink},          
  numbers=left,
  numberstyle=\tiny,
  stepnumber=1,
  numbersep=5pt,
  showspaces=false,
  showstringspaces=false,
  tabsize=4,
  breaklines=true,
  captionpos=b
}
\newcommand{\be}{\begin{equation}}
\newcommand{\ee}{\end{equation}}
\usepackage[unicode=true,pdfusetitle,
 bookmarks=true,bookmarksnumbered=false,bookmarksopen=false,
breaklinks=false,pdfborder={0 0 1},backref=false,colorlinks=false]{hyperref}

\usepackage[makeroom]{cancel}
\usepackage[table,xcdraw]{xcolor}
\usepackage{array}
\usepackage{booktabs}
\usepackage{tikz}
\usetikzlibrary{positioning}
\usepackage{placeins}

\begin{document}

\renewcommand{\thetable}{\arabic{table}}
\renewcommand{\thefigure}{\arabic{figure}}

\title{Quantum Matrix-Element Estimators for Spin-Coupled Generalized Valence Bond (SCGVB) Wavefunctions}

\author{Bruna G. M. Araújo}
\email{bruna.gabriellyma@gmail.com}
\altaffiliation{Current address: Davidson School of Chemical Engineering, Purdue University, Forney Hall of Chemical Engineering, 480 Stadium Mall Drive, West Lafayette, IN 47907-2100, USA; and Instituto de Física, Universidade Federal da Bahia, Campus Universitário de Ondina, Salvador, 40210-340, BA, Brazil.}

 \affiliation{Department of Chemistry, Materials and Chemical Engineering ”Giulio Natta”,
 Politecnico di Milano, Via Bassini 6, 20133 Milano, Italy}

\begin{abstract}
\textbf{Abstract.}\ 
Valence-bond wavefunctions such as spin-coupled generalized valence bond (SCGVB) provide compact and chemically interpretable descriptions of strong correlation, but their nonorthogonal determinant structure makes quantum evaluation of overlaps and Hamiltonian matrix elements difficult. Here we introduce an ancilla-free, measurement-driven framework that reformulates these quantities as vacuum expectation values of Pauli-string operators, enabling evaluation with shallow circuits, local Clifford rotations, and computational-basis measurements, without controlled operations.  We validate the method for H$_4$ along a dissociation pathway and for a 70-determinant C$_2$ active-valence benchmark. The reconstructed matrices agree with the classical Löwdin-based reference, while the full nonorthogonal Chirgwin--Coulson analysis preserves the expected chemical dissociation trend. These results establish the proposed estimators as a low-depth, circuit-compatible formulation for evaluating SCGVB matrix elements, while making no claim of quantum computational advantage.

\end{abstract}

\maketitle

\section{Introduction}

Generalized valence bond (GVB) theory has evolved into a family of chemically transparent multiconfigurational approaches for describing static correlation in terms of localized bonding patterns. Among its most powerful extensions is the Spin-Coupled Generalized Valence Bond (SCGVB) method~\cite{thorsteinsson1995,hiberty2002}, which enforces exact total-spin eigenfunctions through the explicit coupling of the spin parts of electron pairs. This construction provides a compact and chemically intuitive framework for describing singlet--triplet mixing, multibond correlation, and bond reorganization processes, while retaining a localized picture of chemical bonding~\cite{ShaikHiberty2007}. Related developments, including the Breathing-Orbital Valence Bond (BOVB) method~\cite{shaik2012} and multiconfigurational SCGVB variants~\cite{hiberty2002}, further increase the flexibility of the valence-bond description by allowing orbital relaxation across different valence-bond structures.

Beyond its conceptual appeal, GVB-based theories are also computationally competitive with modern multireference approaches. For simple yet paradigmatic systems such as H$_2$, H$_4$, and related molecules, GVB treatments can reproduce near-exact dissociation curves with only a small number of determinants~\cite{goddard1973,szabo1996}, whereas full configuration interaction generally requires an exponentially growing expansion. Within this framework, the resulting generalized eigenvalue problem [Eq.~(\ref{eq:gen_eig})] naturally accommodates orthogonalization procedures such as L\"owdin symmetric orthogonalization~\cite{lowdin1950,helgaker2000} and enables quantitative wavefunction analyses, including Chirgwin--Coulson weighting schemes~\cite{chirgwin1950,shaik2012}.

Despite these advantages, the evaluation of overlap and off-diagonal Hamiltonian matrix elements between nonorthogonal determinants remains a major obstacle, especially in the context of quantum implementations of valence-bond wavefunctions. This difficulty reflects a long-standing challenge already present in classical multiconfigurational electronic-structure theory, where nonorthogonality considerably complicates the algebra of many-electron operators and matrix elements~\cite{kutzelnigg1997}. Recent progress in fermion-to-qubit mappings for nonorthogonal orbital bases has opened new possibilities for quantum simulation in such settings~\cite{marruzzo2025nojw}. However, existing quantum subroutines for matrix-element estimation typically rely on ancilla qubits, Hadamard tests, or deep controlled-unitary circuits, all of which are poorly suited to noisy intermediate-scale quantum (NISQ) hardware~\cite{Baek2023NO}.

In this work, we introduce two ancilla-free quantum estimators tailored to SCGVB determinant spaces. The central idea is to reformulate the required overlap and Hamiltonian matrix elements as vacuum expectation values of Pauli-string operators, so that nonorthogonality, spin coupling, and determinant algebra are treated classically, while the quantum register is used only for shallow measurement tasks. The resulting estimators rely exclusively on local Clifford operations and computational-basis measurements, thereby avoiding ancilla qubits, controlled unitaries, and Hadamard-test-based protocols. In this sense, the present approach should be viewed not as a full quantum realization of the SCGVB ansatz, but rather as a measurement-based quantum backend for nonorthogonal electronic-structure methods such as SCGVB, formulated within a nonorthogonal Jordan--Wigner framework~\cite{marruzzo2025nojw}.

From the perspective of computational chemistry, this separation between many-electron algebra and quantum-state manipulation constitutes the main conceptual contribution of the present work. Rather than attempting to prepare the full nonorthogonal wavefunction on quantum hardware, the method directly targets the central bottleneck of SCGVB calculations, namely the reliable evaluation of overlap and Hamiltonian matrix elements between nonorthogonal determinants. Although the H$_4$ cluster considered here is chemically minimal, within the SCGVB framework, it already generates a nonorthogonal determinant basis and a nontrivial generalized eigenvalue problem, making it a meaningful proof-of-principle benchmark for algorithms designed to handle nonorthogonal determinant algebra. The resulting framework is therefore complementary to classical SCGVB implementations and may be naturally integrated into broader hybrid workflows involving orbital optimization, tensor-network embeddings, or selected configuration-interaction expansions.

The purpose of the present framework is not to replace classical Löwdin-rule evaluation of nonorthogonal determinant matrix elements. Rather, it constitutes a first step toward the implementation of SCGVB methodologies on quantum hardware. The present study isolates the matrix-element measurement layer and formulates it using shallow, ancilla-free circuits that avoid the controlled operations required by conventional Hadamard-test constructions. In the current proof-of-principle implementation, the nonorthogonal determinant algebra and Pauli expansions are generated classically, while the quantum register realizes the corresponding measurement primitives. Because the present measurement circuits contain only local Clifford operations acting on computational-basis states, they are efficiently classically simulable and do not establish a computational need for quantum hardware. A conceptually different extension would combine the present matrix-element formulation with explicit coherent preparation of nonorthogonal states and interference-based transition measurements; that extension belongs to a different resource regime and is discussed only as an outlook below.

The present work belongs to a broader research program devoted to the compactification of electronic wavefunctions in strongly correlated systems. In this sense, it is aligned with the direction initiated in our previous study ~\cite{araujo2025compactifyingelectronicwavefunctionsi}, although the two works are methodologically independent and address distinct algorithmic settings.

The remainder of this paper is organized as follows. Section~\ref{background} introduces the theoretical framework of the SCGVB wavefunction and its representation on a nonorthogonal determinant basis. Section~\ref{sec:SCGVB-NOQA} presents the ancilla-free quantum estimators for the evaluation of determinant overlaps and Hamiltonian matrix elements. Section~\ref{details} describes the computational implementation, including the Pauli-string reconstruction workflow and implementation-level reductions used for larger determinant spaces. Finally, Section~\ref{Results} presents the numerical validation of the proposed estimators for H$_4$ along a dissociation pathway and for a 70-determinant C$_2$ active-valence benchmark, analyzing the accuracy of the reconstructed overlap and Hamiltonian matrices against L\"owdin rules as reference.

\section{Theoretical Framework}
\label{background}
\subsection{Spin-Coupled Generalized Valence Bond (SCGVB) Wavefunction}
\label{SCGVB_wf}

The Generalized Valence Bond (GVB) approach provides a compact and chemically intuitive framework bridging traditional valence bond (VB) and molecular orbital (MO) descriptions of electronic structure.  
Originally developed by Goddard and co-workers~\cite{goddard1973}, GVB extends the Heitler--London and Pauling VB pictures by allowing the orbitals associated with each electron pair to be nonorthogonal and variationally optimized.  
This flexibility enables an efficient description of static and dynamic correlation while maintaining a localized, bond-centered interpretation~\cite{hiberty2002,shaik2012}.

In contrast to delocalized Hartree--Fock molecular orbitals, GVB employs localized orbital pairs that describe both covalent and ionic contributions.  For a singlet-coupled electron pair, the GVB pair function is
\begin{equation}
\Phi_{\mathrm{pair}} = N \left[ \phi_a(1)\phi_b(2) + \phi_b(1)\phi_a(2) \right] \chi_{s=0},
\label{eq:gvb_pair}
\end{equation}
where $\phi_a$ and $\phi_b$ are generally nonorthogonal spatial orbitals, $\chi_{s=0}$ is the singlet spin function, and $N$ is a normalization constant depending on the orbital overlap $S_{ab}=\langle\phi_a|\phi_b\rangle$.  
Variation of the pair orbitals in the presence of all other electrons leads to coupled equations analogous to those of Hartree--Fock theory~\cite{goddard1973,szabo1996,helgaker2000}.  
Because the orbitals are nonorthogonal, the resulting variational problem takes the form of a generalized eigenvalue equation,
\begin{equation}
\mathbf{H C} = \mathbf{S C E},
\label{eq:gen_eig}
\end{equation}
where $\mathbf{H}$ is the Hamiltonian matrix, $\mathbf{S}$ the overlap matrix, and $\mathbf{C}$ the coefficient matrix describing the linear combination of nonorthogonal configurations $\{\Theta_\mu\}$.  This nonorthogonality is crucial for properly describing near-degeneracies, bond breaking, and diradical character in molecular systems~\cite{hiberty2002,shaik2012}.

\begin{figure*}[t]
\centering
\begin{tikzpicture}[
    xscale=1,
    yscale=1.15,
    node/.style={
        circle, draw=black, very thick,
        inner sep=1pt, minimum size=8mm,
        font=\scriptsize\bfseries,
        fill opacity=0.95
    },
    edge/.style={line width=0.6pt}
]

\definecolor{softpink}{RGB}{255,205,230}
\definecolor{softgreen}{RGB}{195,255,200}

\newcommand{\spinnode}[5]{%
  \pgfmathsetmacro{\ratio}{min(max(#4/7,0),1)}%
  \pgfmathsetmacro{\percent}{int(100*\ratio)}%
  \node[node, fill=softpink!\percent!softgreen] (#1) at (#2,#3) {#5};}

\draw[->,thick] (0,0) -- (0,7.6) node[above] {$S$};
\foreach \sy/\lab in {0/0,1/1,2/2,3/3,4/4,5/5,6/6,7/7} {
  \node[left=6pt, font=\scriptsize] at (0,\sy) {$\lab$};
}
\foreach \sy/\lab in {
  0.5/\tfrac12,
  1.5/\tfrac32,
  2.5/\tfrac52,
  3.5/\tfrac72,
  4.5/\tfrac92,
  5.5/\tfrac{11}{2},
  6.5/\tfrac{13}{2}}{
  \node[left=6pt, font=\scriptsize] at (0,\sy) {$\lab$};
}

\draw[->,thick] (0,0) -- (15,0) node[right] {$N$};

\foreach \x in {0,...,14} {
  \node[font=\scriptsize] at (\x,-0.5) {$\x$};
}


\spinnode{N1S0p5}{1}{0.5}{0.5}{1}

\spinnode{N2S0}{2}{0}{0}{1}
\spinnode{N2S1p0}{2}{1.0}{1}{1}

\spinnode{N3S0p5}{3}{0.5}{0.5}{2}
\spinnode{N3S1p5}{3}{1.5}{1.5}{1}

\spinnode{N4S0}{4}{0}{0}{2}
\spinnode{N4S1p0}{4}{1.0}{1}{3}
\spinnode{N4S2p0}{4}{2.0}{2}{1}

\spinnode{N5S0p5}{5}{0.5}{0.5}{5}
\spinnode{N5S1p5}{5}{1.5}{1.5}{4}
\spinnode{N5S2p5}{5}{2.5}{2.5}{1}

\spinnode{N6S0}{6}{0}{0}{5}
\spinnode{N6S1p0}{6}{1.0}{1}{9}
\spinnode{N6S2p0}{6}{2.0}{2}{5}
\spinnode{N6S3p0}{6}{3.0}{3}{1}

\spinnode{N7S0p5}{7}{0.5}{0.5}{14}
\spinnode{N7S1p5}{7}{1.5}{1.5}{14}
\spinnode{N7S2p5}{7}{2.5}{2.5}{6}
\spinnode{N7S3p5}{7}{3.5}{3.5}{1}

\spinnode{N8S0}{8}{0}{0}{14}
\spinnode{N8S1p0}{8}{1.0}{1}{28}
\spinnode{N8S2p0}{8}{2.0}{2}{20}
\spinnode{N8S3p0}{8}{3.0}{3}{7}
\spinnode{N8S4p0}{8}{4.0}{4}{1}

\spinnode{N9S0p5}{9}{0.5}{0.5}{42}
\spinnode{N9S1p5}{9}{1.5}{1.5}{48}
\spinnode{N9S2p5}{9}{2.5}{2.5}{27}
\spinnode{N9S3p5}{9}{3.5}{3.5}{8}
\spinnode{N9S4p5}{9}{4.5}{4.5}{1}

\spinnode{N10S0}{10}{0}{0}{42}
\spinnode{N10S1p0}{10}{1.0}{1}{90}
\spinnode{N10S2p0}{10}{2.0}{2}{75}
\spinnode{N10S3p0}{10}{3.0}{3}{35}
\spinnode{N10S4p0}{10}{4.0}{4}{9}
\spinnode{N10S5p0}{10}{5.0}{5}{1}

\spinnode{N11S0p5}{11}{0.5}{0.5}{132}
\spinnode{N11S1p5}{11}{1.5}{1.5}{165}
\spinnode{N11S2p5}{11}{2.5}{2.5}{110}
\spinnode{N11S3p5}{11}{3.5}{3.5}{44}
\spinnode{N11S4p5}{11}{4.5}{4.5}{10}
\spinnode{N11S5p5}{11}{5.5}{5.5}{1}

\spinnode{N12S0p0}{12}{0.0}{0}{132}
\spinnode{N12S1p0}{12}{1.0}{1}{297}
\spinnode{N12S2p0}{12}{2.0}{2}{275}
\spinnode{N12S3p0}{12}{3.0}{3}{154}
\spinnode{N12S4p0}{12}{4.0}{4}{54}
\spinnode{N12S5p0}{12}{5.0}{5}{11}
\spinnode{N12S6p0}{12}{6.0}{6}{1}

\spinnode{N13S0p5}{13}{0.5}{0.5}{429}
\spinnode{N13S1p5}{13}{1.5}{1.5}{572}
\spinnode{N13S2p5}{13}{2.5}{2.5}{429}
\spinnode{N13S3p5}{13}{3.5}{3.5}{208}
\spinnode{N13S4p5}{13}{4.5}{4.5}{65}
\spinnode{N13S5p5}{13}{5.5}{5.5}{12}
\spinnode{N13S6p5}{13}{6.5}{6.5}{1}

\spinnode{N14S0p0}{14}{0.0}{0}{429}
\spinnode{N14S1p0}{14}{1.0}{1}{1001}
\spinnode{N14S2p0}{14}{2.0}{2}{1001}
\spinnode{N14S3p0}{14}{3.0}{3}{637}
\spinnode{N14S4p0}{14}{4.0}{4}{273}
\spinnode{N14S5p0}{14}{5.0}{5}{77}
\spinnode{N14S6p0}{14}{6.0}{6}{13}
\spinnode{N14S7p0}{14}{7.0}{7}{1}


\draw[edge] (0,0) -- (N1S0p5);

\draw[edge] (N1S0p5) -- (N2S0);
\draw[edge] (N1S0p5) -- (N2S1p0);

\draw[edge] (N2S0)   -- (N3S0p5);
\draw[edge] (N2S1p0) -- (N3S0p5);
\draw[edge] (N2S1p0) -- (N3S1p5);

\draw[edge] (N3S0p5) -- (N4S0);
\draw[edge] (N3S0p5) -- (N4S1p0);
\draw[edge] (N3S1p5) -- (N4S1p0);
\draw[edge] (N3S1p5) -- (N4S2p0);

\draw[edge] (N4S0)   -- (N5S0p5);
\draw[edge] (N4S1p0) -- (N5S0p5);
\draw[edge] (N4S1p0) -- (N5S1p5);
\draw[edge] (N4S2p0) -- (N5S1p5);
\draw[edge] (N4S2p0) -- (N5S2p5);

\draw[edge] (N5S0p5) -- (N6S0);
\draw[edge] (N5S0p5) -- (N6S1p0);
\draw[edge] (N5S1p5) -- (N6S1p0);
\draw[edge] (N5S1p5) -- (N6S2p0);
\draw[edge] (N5S2p5) -- (N6S2p0);
\draw[edge] (N5S2p5) -- (N6S3p0);

\draw[edge] (N6S0)   -- (N7S0p5);
\draw[edge] (N6S1p0) -- (N7S0p5);
\draw[edge] (N6S1p0) -- (N7S1p5);
\draw[edge] (N6S2p0) -- (N7S1p5);
\draw[edge] (N6S2p0) -- (N7S2p5);
\draw[edge] (N6S3p0) -- (N7S2p5);
\draw[edge] (N6S3p0) -- (N7S3p5);

\draw[edge] (N7S0p5) -- (N8S0);
\draw[edge] (N7S0p5) -- (N8S1p0);
\draw[edge] (N7S1p5) -- (N8S1p0);
\draw[edge] (N7S1p5) -- (N8S2p0);
\draw[edge] (N7S2p5) -- (N8S2p0);
\draw[edge] (N7S2p5) -- (N8S3p0);
\draw[edge] (N7S3p5) -- (N8S3p0);
\draw[edge] (N7S3p5) -- (N8S4p0);

\draw[edge] (N8S0)   -- (N9S0p5);
\draw[edge] (N8S1p0) -- (N9S0p5);
\draw[edge] (N8S1p0) -- (N9S1p5);
\draw[edge] (N8S2p0) -- (N9S1p5);
\draw[edge] (N8S2p0) -- (N9S2p5);
\draw[edge] (N8S3p0) -- (N9S2p5);
\draw[edge] (N8S3p0) -- (N9S3p5);
\draw[edge] (N8S4p0) -- (N9S3p5);
\draw[edge] (N8S4p0) -- (N9S4p5);

\draw[edge] (N9S0p5) -- (N10S0);
\draw[edge] (N9S0p5) -- (N10S1p0);
\draw[edge] (N9S1p5) -- (N10S1p0);
\draw[edge] (N9S1p5) -- (N10S2p0);
\draw[edge] (N9S2p5) -- (N10S2p0);
\draw[edge] (N9S2p5) -- (N10S3p0);
\draw[edge] (N9S3p5) -- (N10S3p0);
\draw[edge] (N9S3p5) -- (N10S4p0);
\draw[edge] (N9S4p5) -- (N10S4p0);
\draw[edge] (N9S4p5) -- (N10S5p0);

\draw[edge] (N10S0)   -- (N11S0p5);
\draw[edge] (N10S1p0) -- (N11S0p5);
\draw[edge] (N10S1p0) -- (N11S1p5);
\draw[edge] (N10S2p0) -- (N11S1p5);
\draw[edge] (N10S2p0) -- (N11S2p5);
\draw[edge] (N10S3p0) -- (N11S2p5);
\draw[edge] (N10S3p0) -- (N11S3p5);
\draw[edge] (N10S4p0) -- (N11S3p5);
\draw[edge] (N10S4p0) -- (N11S4p5);
\draw[edge] (N10S5p0) -- (N11S4p5);
\draw[edge] (N10S5p0) -- (N11S5p5);

\draw[edge] (N11S0p5) -- (N12S0p0);
\draw[edge] (N11S0p5) -- (N12S1p0);
\draw[edge] (N11S1p5) -- (N12S1p0);
\draw[edge] (N11S1p5) -- (N12S2p0);
\draw[edge] (N11S2p5) -- (N12S2p0);
\draw[edge] (N11S2p5) -- (N12S3p0);
\draw[edge] (N11S3p5) -- (N12S3p0);
\draw[edge] (N11S3p5) -- (N12S4p0);
\draw[edge] (N11S4p5) -- (N12S4p0);
\draw[edge] (N11S4p5) -- (N12S5p0);
\draw[edge] (N11S5p5) -- (N12S5p0);
\draw[edge] (N11S5p5) -- (N12S6p0);

\draw[edge] (N12S0p0) -- (N13S0p5);
\draw[edge] (N12S1p0) -- (N13S0p5);
\draw[edge] (N12S1p0) -- (N13S1p5);
\draw[edge] (N12S2p0) -- (N13S1p5);
\draw[edge] (N12S2p0) -- (N13S2p5);
\draw[edge] (N12S3p0) -- (N13S2p5);
\draw[edge] (N12S3p0) -- (N13S3p5);
\draw[edge] (N12S4p0) -- (N13S3p5);
\draw[edge] (N12S4p0) -- (N13S4p5);
\draw[edge] (N12S5p0) -- (N13S4p5);
\draw[edge] (N12S5p0) -- (N13S5p5);
\draw[edge] (N12S6p0) -- (N13S5p5);
\draw[edge] (N12S6p0) -- (N13S6p5);

\draw[edge] (N13S0p5) -- (N14S0p0);
\draw[edge] (N13S0p5) -- (N14S1p0);
\draw[edge] (N13S1p5) -- (N14S1p0);
\draw[edge] (N13S1p5) -- (N14S2p0);
\draw[edge] (N13S2p5) -- (N14S2p0);
\draw[edge] (N13S2p5) -- (N14S3p0);
\draw[edge] (N13S3p5) -- (N14S3p0);
\draw[edge] (N13S3p5) -- (N14S4p0);
\draw[edge] (N13S4p5) -- (N14S4p0);
\draw[edge] (N13S4p5) -- (N14S5p0);
\draw[edge] (N13S5p5) -- (N14S5p0);
\draw[edge] (N13S5p5) -- (N14S6p0);
\draw[edge] (N13S6p5) -- (N14S6p0);
\draw[edge] (N13S6p5) -- (N14S7p0);

\end{tikzpicture}

\caption{
Branching diagram of spin couplings for $N=1,\dots,14$ electrons.  
Each node at coordinates $(N,S)$ is labeled by $f_S^N = \dfrac{(2S+1)N!}{(\frac{N}{2}+S+1)!\,(\frac{N}{2}-S)!}$, which gives the number of linearly independent spin eigenfunctions $\Theta_{S,M;k}^N$ for that electron number and total spin~\cite{McWeeny1992,Pauncz1979,cooper_rumer}.  
Edges indicate the allowed couplings when a spin-$\tfrac12$ electron is added, connecting $(N,S)$ to $(N+1,S\pm\tfrac12)$.  
Thus, the integer at each node equals the range of the index $k$ in Eq.~(\ref{eq1}), and therefore the number of spin-coupled structures entering the SCGVB expansion for that $(N,S)$.  
The color gradient (from light pink to light green) indicates increasing total spin $S$.  
Adapted from R.~McWeeny~\cite{McWeeny1992} and R.~Pauncz~\cite{Pauncz1979}.
}
\label{fig:branching-color}.
\end{figure*}
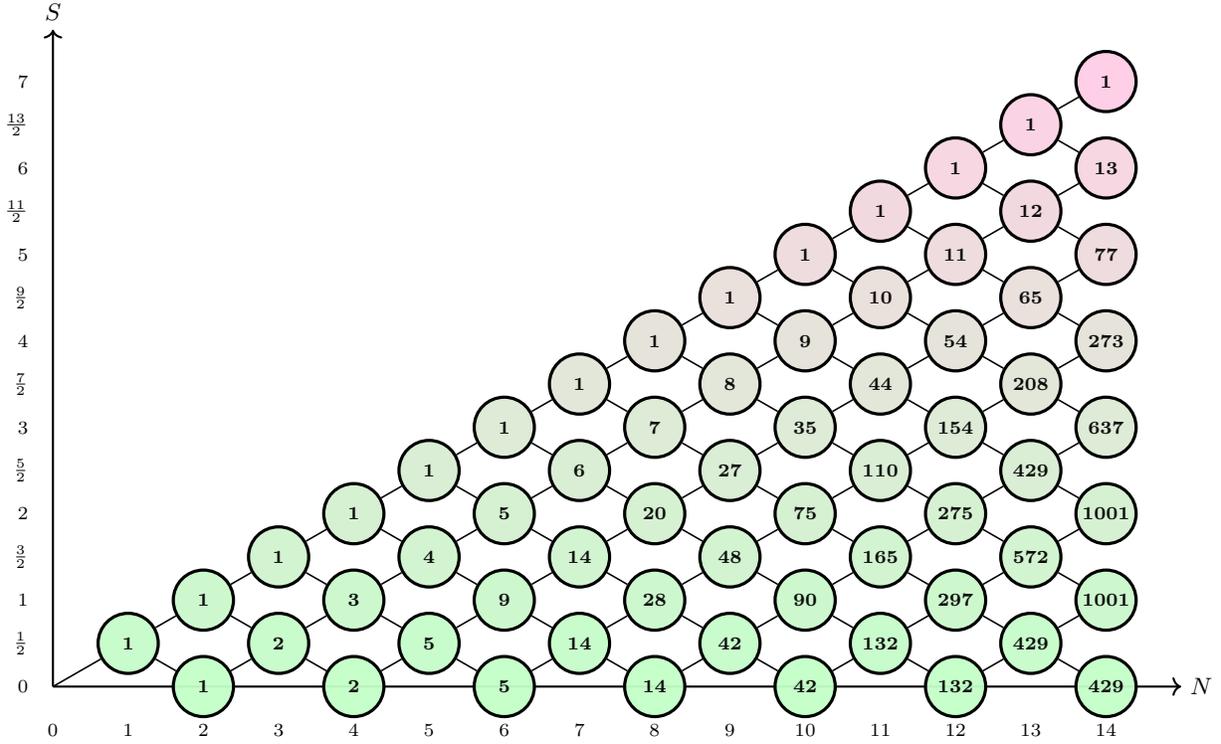

In the spin-coupled generalized valence bond (SCGVB) approach, the many-electron wavefunction for an $N$-electron system in a spin state characterized by the quantum numbers $(S,M)$ is written as an explicitly spin-adapted linear combination of spin-coupled (SC) structures constructed from generally nonorthogonal spatial orbitals~\cite{goddard1973,cooper_rumer,cooper_introVB,dunning2021_scgvb,dunning2023_scgvb}.  
The SCGVB wavefunction can be expressed as
\begin{eqnarray}
\Psi_{\mathrm{SCGVB}} 
& = & \mathcal{N}\; \sum_{k=1}^{f_S^N}\; c_{S,k} \; \psi_{S,M;k}^N \nonumber \\
& = & \mathcal{N}\; \sum_{k=1}^{f_S^N}\; c_{S,k} \; 
\hat{A}\Bigl(\phi_1 \,\phi_2\, \dots \, \phi_N\, \Theta_{S,M;k}^N\Bigr),
\label{eq1}
\end{eqnarray}
where $f_S^N$ is the number of linearly independent $N$-electron spin eigenfunctions $\Theta_{S,M;k}^N$ for a given $S$ (see below and Fig.~\ref{fig:branching-color}), $c_{S,k}$ are spin-coupling coefficients, and $\mathcal{N}$ is a normalization constant ensuring $\langle \Psi_{\mathrm{SCGVB}} | \Psi_{\mathrm{SCGVB}} \rangle = 1$.  Each spin-coupled structure $\psi_{S,M;k}^N$ is obtained by applying the antisymmetrizer $\hat{A}$ to a product of $N$ spatial orbitals and one spin eigenfunction. The antisymmetrizer is defined as~\cite{szabo1996,helgaker2000}
\begin{equation}
  \hat{A}=\frac{1}{\sqrt{N!}}\;\sum_{\hat{\mathcal{P}} \in S_N} \; (-1)^p \; \hat{\mathcal{P}},
  \label{eq2}
\end{equation}
where the sum runs over all permutations $\hat{\mathcal{P}}$ of the $N$ electrons belonging to the symmetric group $S_N$, and $(-1)^p$ is the parity of the permutation.  
The set $\{\phi_i\}$ in Eq.~(\ref{eq1}) denotes the spin-coupled spatial orbitals, which are in general nonorthogonal,
$\langle \phi_i | \phi_j \rangle \neq \delta_{ij}$,
and are expanded in a one-electron atomic basis $\{\chi_\mu\}$,
\begin{equation}
  \phi_i(\mathbf{r})=\sum_{\mu=1}^{N_{bf}}\,C_{\mu i}\;\chi_{\mu}(\mathbf{r}),
  \label{eq3}
\end{equation}
where $N_{bf}$ is the number of basis functions and $C_{\mu i}$ are the orbital expansion coefficients.  
In practical SCGVB calculations, both the spin-coupling coefficients $\{c_{S,k}\}$ and the orbital coefficients $\{C_{\mu i}\}$ (or, equivalently, the spin-coupled orbitals $\{\phi_i\}$) are determined variationally by energy minimization~\cite{thorsteinsson1995,dunning2021_scgvb,dunning2023_scgvb}.

It is important to emphasize that the first line of Eq.~(\ref{eq1}) represents a formally exact expansion of an $N$-electron eigenfunction in a complete set of spin eigenfunctions $\{\Theta_{S,M;k}^N\}$.  The approximation in SCGVB enters through the second line of Eq.~(\ref{eq1}), where each spin-coupled structure $\psi_{S,M;k}^N$ is represented in terms of a finite set of spin-coupled orbitals $\{\phi_i\}$ and a chosen spin basis $\{\Theta_{S,M;k}^N\}$.  
In the most general formulation, an additional set of doubly occupied core orbitals may be included.  
For the hydrogen clusters H$_n$ considered in this work, however, no core orbitals are present, and this complication can be omitted~\cite{hiberty2002,shaik2012}.

The branching diagram shown in Fig.~\ref{fig:branching-color} provides a convenient visual summary of the spin space entering Eq.~(\ref{eq1}).  
Each node at coordinates $(N,S)$ is labelled by
\begin{equation}
  f_S^N = \frac{(2S+1)N!}{\bigl(\tfrac{N}{2}+S+1\bigr)!\,\bigl(\tfrac{N}{2}-S\bigr)!},
\end{equation}
which is the number of linearly independent spin eigenfunctions $\Theta^N_{S,M;k}$ for that electron number and total spin~\cite{McWeeny1992,Pauncz1979,cooper_rumer}.  
This number is identical to the range of the index $k$ in Eq.~(\ref{eq1}).  
For example, the $N=4$, $S=1$ node is labelled by $f_1^4=3$, indicating that the triplet SCGVB wavefunction for four electrons involves three linearly independent spin-coupled structures.  
Similarly, the $N=6$, $S=0$ node is labelled $f_0^6=5$, so the six-electron singlet SCGVB expansion contains five spin-coupled structures.  
Edges in Fig.~\ref{fig:branching-color} connect $(N,S)$ to $(N+1,S\pm\tfrac12)$ and represent the allowed couplings when a new spin-$\tfrac12$ electron is added.  
Thus, the diagram not only encodes the values of $f_S^N$ but also illustrates how the various spin manifolds are generated by successive spin couplings.

\paragraph{Spin eigenfunctions and primitive spin functions.}

Each spin eigenfunction $\Theta_{S,M;k}^N$ can be written as a fixed linear combination of spin primitive functions $\{\theta_i^N\}$, i.e., simple products of one-electron spin functions $\alpha$ and $\beta$ that are eigenfunctions of $\hat{S}_z$ but not of $\hat{S}^2$,
\begin{equation}
  \Theta_{S,M;k}^N=\sum_{i=1}^{N_d} \, d_{i,k}^S \; \theta_{i}^N ,
 \label{eq4}
\end{equation}
where $N_d$ is the number of determinants (or primitive spin functions) with the appropriate spin projection $M$,
\begin{equation}
  N_d={N \choose N_{\alpha}}={N \choose N_{\beta}},
\end{equation}
and $N_{\alpha}$ and $N_{\beta}$ are the numbers of $\alpha$ and $\beta$ spins, respectively.  The coefficients $\{d_{i,k}^S\}$ depend on the chosen spin basis (such as Kotani, Serber, or Rumer bases) and can be obtained from standard angular-momentum coupling or group-theoretical arguments~\cite{cooper_introVB,Pauncz1979,cooper_rumer}.  
In particular, the Rumer basis provides an irreducible representation of the symmetric group $S_N$, with coefficients $d_{i,k}^S$ restricted to $0$ and $\pm 1$, which is especially convenient in practical SCGVB implementations. Substituting Eq.~(\ref{eq4}) into Eq.~(\ref{eq1}), the SCGVB wavefunction can be recast as
\begin{equation}
\Psi_{\mathrm{SCGVB}}=\mathcal{N}\;\sum_{i=1}^{N_d}\;b_{S,i}\;\hat{A}\Bigl(\phi_1\,\phi_2\, \dots \, \phi_N\, \theta_{i}^N\Bigr),
  \label{eq6}
\end{equation}
where the coefficients $b_{S,i}$ are given by
\begin{equation}
  b_{S,i}=\sum_{k=1}^{f_S^N}\, c_{S,k} \; d_{i,k}^S.
\end{equation}
Since each primitive spin function $\theta_i^N$ is a product of $N$ one-electron spin functions, the product
$\phi_1\,\phi_2\, \dots \, \phi_N\, \theta_{i}^N$ is simply a product of $N$ spin orbitals,
\begin{equation}
  \phi_1\,\phi_2\, \dots \, \phi_N\, \theta_{i}^N =  \phi_1^i\,\phi_2^i\, \dots \, \phi_N^i,
\end{equation}
with $\phi_j^i(\mathbf{r},\sigma)=\phi_j(\mathbf{r})\,\omega_j^i(\sigma)$, where $\omega_j^i$ is either $\alpha$ or $\beta$ in the $i$th primitive function.  
Application of the antisymmetrizer to a product of spin orbitals generates a Slater determinant, so Eq.~(\ref{eq6}) becomes
\begin{equation}
 \Psi_{\mathrm{SCGVB}}=\mathcal{N}\;\sum_{i=1}^{N_d}\;\frac{b_{S,i}}{\sqrt{N!}} \; 
  \big|\phi_1^i\,\phi_2^i\, \dots \, \phi_N^i\big|
  \equiv \mathcal{N}\;\sum_{i=1}^{N_d}\;b_{S,i}\;\Omega_i,
  \label{eq9}
\end{equation}
where $\{\Omega_i\}$ denotes a set of non-normalized Slater determinants built from nonorthogonal spin orbitals.  
Equation~(\ref{eq9}) shows that the SCGVB wavefunction can be viewed as a particular linear combination of nonorthogonal determinants with spin-coupling information embedded in the coefficients $b_{S,i}$~\cite{thorsteinsson1995,dunning2021_scgvb,dunning2023_scgvb}.

\paragraph{Normalization and matrix elements.}

Because the determinants $\{\Omega_i\}$ are constructed from nonorthogonal spin orbitals, they are themselves nonorthogonal and not normalized.  
Using L\"owdin’s rules for overlaps and matrix elements between determinants of nonorthogonal spin orbitals~\cite{lowdin1950,szabo1996,helgaker2000}, the normalization constant can be written as
\begin{equation}
  \mathcal{N}=\mathcal{D}^{-1/2},
\end{equation}
with
\begin{equation}
  \mathcal{D}=\sum_{i,j=1}^{N_d}\,b_{S,i}\;b_{S,j}\; \det\bigl[\mathbf{O}_{ij}\bigr],
\end{equation}
where $\mathbf{O}_{ij}$ is the matrix of overlap integrals between the spin orbitals entering determinants
$\Omega_i$ and $\Omega_j$.  
Accordingly, Eq.~(\ref{eq9}) may also be written as
\begin{align}
  \Psi_{\mathrm{SCGVB}}
  &= \mathcal{D}^{-1/2}\;\sum_{i=1}^{N_d}\;\frac{b_{S,i}}{\sqrt{N!}} \;
  \big|\phi_1^i\,\phi_2^i\, \dots \, \phi_N^i\big|, \label{eq12a} \\
  &= \mathcal{D}^{-1/2}\;\sum_{i=1}^{N_d}\;b_{S,i}\;\Omega_i. \label{eq12b}
\end{align}

The SCGVB wavefunction is determined variationally by minimizing the energy functional
\begin{equation}
  E\bigl[\Psi_{\mathrm{SCGVB}} \bigr]=
  \frac{\bigl\langle \Psi_{\mathrm{SCGVB}} \, \big| \, \hat{H} \, \big| \,\Psi_{\mathrm{SCGVB}} \bigr\rangle}
       {\bigl\langle \Psi_{\mathrm{SCGVB}} \, \big|\, \Psi_{\mathrm{SCGVB}} \, \bigr\rangle},
  \label{eq13}
\end{equation}
is minimized with respect to both orbital and spin-coupling coefficients.  
Evaluation of the required overlaps and Hamiltonian matrix elements constitutes the dominant computational bottleneck of SCGVB and related VB methods~\cite{dunning2021_scgvb,valence2019}, motivating alternative strategies such as quantum computing.

\subsubsection*{L\"owdin Symmetric Orthogonalization of the VB Basis}

To simplify the solution of Eq.~(\ref{eq:gen_eig}), it is convenient to transform the nonorthogonal basis $\{\Theta_\mu\}$ into an orthonormal one, $\{\Theta'_\mu\}$, by means of a symmetric orthogonalization transformation~\cite{lowdin1950, lowdin1955, lowdin1962}.  
This procedure, introduced by L\"owdin in 1950, guarantees that the orthogonalization preserves the symmetry and near-local character of the original VB structures, while minimizing the distortion of the orbital subspace~\cite{lowdin1950, lowdin1955, lowdin1962,szabo1996,helgaker2000}. The transformation is defined as
\begin{equation}
\Theta'_\mu = \sum_{\nu} \Theta_\nu X_{\nu\mu},
\label{eq:theta_prime}
\end{equation}
or, in matrix notation,
\begin{equation}
\mathbf{\Theta'} = \mathbf{\Theta X},
\label{eq:theta_matrix}
\end{equation}
where the transformation matrix $\mathbf{X}$ satisfies the orthonormality condition
\begin{equation}
\mathbf{X}^\dagger \mathbf{S} \mathbf{X} = \mathbf{I}.
\label{eq:X_cond}
\end{equation}
The unique positive-definite Hermitian solution that defines the symmetric L\"owdin transformation is
\begin{equation}
\mathbf{X} = \mathbf{S}^{-1/2}.
\label{eq:S_half}
\end{equation}
Because $\mathbf{S}^{-1/2}$ is Hermitian, this transformation ensures that the orthogonalized basis retains the symmetry properties of the original VB structures~\cite{lowdin1950,helgaker2000}.  
Such orthogonalization is routinely employed in many-body electronic structure theory, including configuration interaction, coupled cluster, and tensor network formulations~\cite{szabo1996,helgaker2000, perezobiol2022}.

\subsubsection*{Transformation of the Hamiltonian and Coefficients}

Upon orthogonalization, the total wavefunction can be equivalently written as
\begin{equation}
\Psi_i = \mathbf{\Theta'} \mathbf{C}' = \mathbf{\Theta X C}',
\label{eq:psi_prime}
\end{equation}
which implies that the coefficients in the nonorthogonal and orthogonal representations are related by
\begin{equation}
\mathbf{C} = \mathbf{X C}' = \mathbf{S}^{-1/2} \mathbf{C}'.
\label{eq:C_relation}
\end{equation}
Substituting this into Eq.~(\ref{eq:gen_eig}) yields
\begin{equation}
\mathbf{H X C}' = \mathbf{S X C}' \mathbf{E}.
\label{eq:HX_eq}
\end{equation}
Multiplying both sides by $\mathbf{X}^\dagger$ and using Eq.~(\ref{eq:X_cond}), we obtain the orthogonalized secular equation
\begin{equation}
\mathbf{H'} \mathbf{C}' = \mathbf{C}' \mathbf{E},
\label{eq:orth_eig}
\end{equation}
where the transformed Hamiltonian is defined as
\begin{equation}
\mathbf{H'} = \mathbf{X}^\dagger \mathbf{H X} = \mathbf{S}^{-1/2} \mathbf{H S}^{-1/2}.
\label{eq:H_prime}
\end{equation}
Diagonalization of $\mathbf{H'}$ yields the eigenvalues $\mathbf{E}$ and the orthogonalized coefficients $\mathbf{C}'$, from which the original coefficients $\mathbf{C}$ can be recovered via Eq.~(\ref{eq:C_relation}).  
This transformation formalism forms the mathematical foundation of spin-coupled VB theory and its modern extensions to multireference and tensor-based frameworks~\cite{thorsteinsson1995,hiberty2002,shaik2012,helgaker2000}.

Once the orthogonalized coefficients $\mathbf{C}'$ and the overlap matrix $\mathbf{S}$ are available, the quantitative analysis of individual VB structure contributions to the total wavefunction can be performed using the \textit{Chirgwin--Coulson weights}~\cite{chirgwin1950,thorsteinsson1995,shaik2012,penotti2019,wiki_cc}.  These weights provide a rigorous way to measure the relative importance of each spin-coupled structure in the final SCGVB state, offering both interpretability and a direct connection to chemical bonding concepts.  In the following section, we present a detailed derivation and discussion of these weights, along with their numerical evaluation and interpretational limitations.

\subsubsection*{Chirgwin--Coulson Weights}
Chirgwin–Coulson weights are introduced here because they provide a chemically interpretable diagnostic that will be used in the analysis of the quantum results later.

Once the generalized eigenvalue problem (Eq.~\ref{eq:gen_eig}) is solved, the coefficients $\mathbf{C}$ obtained in the nonorthogonal valence-bond (VB) basis $\{\Theta_\mu\}$ can be used to analyze the relative importance of individual VB structures.  
A standard quantitative measure for this purpose is provided by the \textit{Chirgwin--Coulson (CC) weights}, originally introduced by Chirgwin and Coulson~\cite{chirgwin1950}.  
For a nonorthogonal expansion
\begin{equation}
|\Psi\rangle = \sum_i C_i\,|\Phi_i\rangle,
\end{equation}
with overlaps $S_{ij}=\langle\Phi_i|\Phi_j\rangle$, the CC weight associated with structure $i$ is defined as
\begin{equation}
W_i = \sum_j C_i\,C_j\,S_{ij}.
\label{eq:cc_definition}
\end{equation}
In an orthonormal basis ($S_{ij}=\delta_{ij}$), this expression reduces to the familiar form $W_i=|C_i|^2$.

In the general nonorthogonal case, the overlap terms introduce interference contributions, so that $W_i$ depends both on coefficient magnitudes and on mutual overlaps between structures.  
As a result, CC weights are not strictly probabilistic quantities, but rather relative indicators of structural importance in the total wavefunction~\cite{thorsteinsson1995,shaik2012,penotti2019,wiki_cc}.  
For the real coefficients used in this work, normalization with respect to the overlap metric gives $\mathbf{C}^{\mathrm{T}}\mathbf{S}\mathbf{C}=1$, and therefore the CC weights satisfy $\sum_i W_i=1$.  
Nonorthogonality nevertheless means that individual CC weights are not constrained to the interval $[0,1]$ and may become negative or exceed unity for strongly overlapping structures~\cite{thorsteinsson1995,penotti2019}.

In chemically well-behaved cases, CC weights typically remain positive, allowing a clear distinction between dominant and secondary VB structures~\cite{shaik2012}.  
When individual weights become negative or exceed unity, it is customary either to report this behavior explicitly or to complement the analysis with alternative weighting schemes.  
Two commonly used modified definitions are the L\"owdin weights,
\begin{equation}
W_i^{(\mathrm{L})}
= \sum_{j,k} S_{ij}^{1/2}\,C_j\,S_{ik}^{1/2}\,C_k,
\label{eq:lowdin}
\end{equation}
obtained after symmetric orthogonalization of the VB basis, and the inverse weights,
\begin{equation}
W_i^{(\mathrm{inv})}
= \frac{C_i^2/(S^{-1})_{ii}}{\sum_j C_j^2/(S^{-1})_{jj}},
\label{eq:inverse}
\end{equation}
which suppress the contribution of strongly overlapping structures and enforce $0\le W_i^{(\mathrm{inv})}\le1$~\cite{lowdin1950,shaik2012}.  
Although these definitions differ numerically, they generally preserve the qualitative ordering of structure importance and serve as consistency checks for CC analyses.

In this work, CC weights are computed directly from Eq.~(\ref{eq:cc_definition}) using the coefficients and overlap matrix obtained from Eq.~(\ref{eq:gen_eig}).  
{For the two-structure H$_4$ case, the spin-coupled structures used below are not individually normalized, so their diagonal structural overlaps are not unity.  The explicit form used in the numerical analysis is therefore
\begin{align}
W_1 &= C_1^2 S_{11} + C_1 C_2 S_{12},\\
W_2 &= C_2^2 S_{22} + C_1 C_2 S_{21},
\label{eq:cc_two_structure_full}
\end{align}
which is exactly the two-structure specialization of Eq.~(\ref{eq:cc_definition}).  The commonly used expressions $W_1=C_1^2+C_1C_2S_{12}$ and $W_2=C_2^2+C_1C_2S_{21}$ apply only when the two basis structures are individually normalized ($S_{11}=S_{22}=1$).}
The practical procedure consists of constructing the overlap matrix, solving the generalized eigenvalue problem, evaluating the CC weights, and verifying their physical consistency.  
Dominant structures typically exhibit $W_i \gtrsim 0.5$, while smaller values indicate secondary contributions.  
When CC weights become negative or exceed unity, this signals strong overlap effects, and the alternative definitions in Eqs.~(\ref{eq:lowdin})--(\ref{eq:inverse}) should be preferred for quantitative interpretation.

Throughout this work we restrict attention to ground states within a fixed total-spin sector.  
Extensions to excited states would require substantially larger spin-coupling spaces and are beyond the scope of the present near-term quantum algorithm.

\subsection{Nonorthogonal Jordan--Wigner mapping}
\label{subsec:NOJW}

To represent the SCGVB Hamiltonian and determinant operators on a qubit register,
we adopt the nonorthogonal Jordan-Wigner (NOJW) fermion--qubit mapping
introduced in Ref.~\cite{marruzzo2025nojw}.
The SCGVB ansatz introduced above is expressed in a generally nonorthogonal
spin-orbital basis and therefore leads naturally to a nonorthogonal
second-quantized Hamiltonian.

\subsubsection*{Nonorthogonal second quantization and auxiliary bases}

Let $\{\varphi_p(x)\}_{p=1}^M$ be a set of nonorthogonal spin orbitals with overlap
matrix
\begin{equation}
  S_{pq} = \int dx\, \varphi_p^*(x)\,\varphi_q(x),
  \qquad \mathbf{S} = [S_{pq}],
  \label{eq:NO_overlap}
\end{equation}
where $x$ collects spatial and spin coordinates.
Creation operators $\{\hat{a}_p^\dagger\}$ and their adjoints
$\{\hat{a}_p\}$ obey the generalized anticommutation relations
\begin{align}
  \{\hat{a}_p,\hat{a}_q\}_+ &= 0, \\
  \{\hat{a}_p^\dagger,\hat{a}_q^\dagger\}_+ &= 0, \\
  \{\hat{a}_p,\hat{a}_q^\dagger\}_+ &= S_{pq},
  \label{eq:NO_anticom}
\end{align}
which reduce to the canonical CAR when $\mathbf{S} = \mathbf{I}$.
The electronic Hamiltonian in this basis reads
\begin{equation}
  \hat{H}
  = \sum_{p,q=1}^M h_{pq}\,\hat{a}_p^\dagger \hat{a}_q
    + \frac{1}{2}\sum_{p,q,r,s=1}^M g_{pqrs}\,
      \hat{a}_p^\dagger \hat{a}_q^\dagger \hat{a}_s \hat{a}_r,
  \label{eq:H_NO_raw}
\end{equation}
with one- and two-electron integrals evaluated in the nonorthogonal orbitals,
so that in general $h_{pq} \neq h_{qp}$ and $g_{pqrs} \neq g_{rspq}$.

For later use, we introduce two auxiliary orbital sets.
First, a symmetrically orthogonalized basis is obtained via L\"owdin
orthogonalization,
\begin{equation}
  \tilde{\varphi}_p(x)
  = \sum_{q=1}^M \varphi_q(x)\,[S^{-1/2}]_{qp},
  \qquad \langle \tilde{\varphi}_p | \tilde{\varphi}_q \rangle = \delta_{pq},
  \label{eq:LO_orth}
\end{equation}
with associated operators
$\{\hat{\tilde{a}}_p^\dagger,\hat{\tilde{a}}_p\}$ satisfying the canonical CAR.
The two operator sets are related by
\begin{align}
  \hat{\tilde{a}}_p^\dagger
  &= \sum_{q=1}^M \hat{a}_q^\dagger\,[S^{-1/2}]_{qp},
  \label{eq:a_tilde_from_a} \\
  \hat{a}_p^\dagger
  &= \sum_{q=1}^M \hat{\tilde{a}}_q^\dagger\,[S^{1/2}]_{qp},
  \label{eq:a_from_a_tilde}
\end{align}
and similarly for their adjoints.
Thus $\hat{a}_p$ is the Hermitian adjoint of $\hat{a}_p^\dagger$ but not a
canonical annihilation operator when $\mathbf{S} \neq \mathbf{I}$.

Second, we define the biorthogonal partner orbitals
\begin{equation}
  \varphi^p(x)
  = \sum_{q=1}^M \varphi_q(x)\,[S^{-1}]_{qp},
  \qquad \langle \varphi^p | \varphi_q \rangle = \delta^p_q,
\end{equation}
with corresponding annihilation operators $\{\hat{b}_p\}$ satisfying
\begin{equation}
  \{\hat{b}_p,\hat{a}_q^\dagger\}_+ = \delta_{pq},
\end{equation}
and
\begin{equation}
  \hat{b}_p = \sum_{q=1}^M [S^{-1}]_{pq}\,\hat{a}_q.
\end{equation}
In the biorthogonal operator set $\{\hat{a}_p^\dagger,\hat{b}_p\}$ the
Hamiltonian can be rewritten as
\begin{equation}
  \hat{H}
  = \sum_{p,q=1}^M \tilde{h}_{pq}\,\hat{a}_p^\dagger \hat{b}_q
    + \frac{1}{2}\sum_{p,q,r,s=1}^M \tilde{g}_{pqrs}\,
      \hat{a}_p^\dagger \hat{a}_q^\dagger \hat{b}_s \hat{b}_r,
  \label{eq:H_biorth}
\end{equation}
with effective integrals $\tilde{h}_{pq}$ and $\tilde{g}_{pqrs}$ obtained by
contraction with $\mathbf{S}^{-1}$ (see, e.g.,
Refs.~\cite{lowdin1950,McWeeny1992}).
This representation will be particularly convenient for encoding $\hat{H}$
on qubits.

\subsubsection*{Jordan--Wigner encoding for nonorthogonal orbitals}

We now outline the fermion--qubit mapping suitable for the operator sets
$\{\hat{a}_p^\dagger,\hat{a}_p\}$ and $\{\hat{b}_p\}$, following the
nonorthogonal Jordan--Wigner construction of Ref.~\cite{marruzzo2025nojw}.
Each spin orbital is associated with a qubit, and we use the usual ON basis
$\{|k_1,\dots,k_M\rangle\}$, where $k_p\in\{0,1\}$ denotes the occupation of
the orbital $p$.

On an orthonormal basis, the standard Jordan-Wigner encoding
\cite{szabo1996,helgaker2000} maps the creation operator to
\begin{equation}
  \hat{\tilde{a}}_p^\dagger
  \;\longleftrightarrow\;
  Z_1 \otimes \cdots \otimes Z_{p-1} \otimes
  \frac{X_p - i Y_p}{2}
  \otimes I_{p+1} \otimes \cdots \otimes I_M,
  \label{eq:JW_orth_create}
\end{equation}
with the annihilation operator obtained by Hermitian conjugation.
Here $X_p$, $Y_p$, and $Z_p$ are Pauli matrices acting on qubit $p$, and $I_p$
is the identity.

Because $\{\hat{\tilde{a}}_p^\dagger,\hat{\tilde{a}}_p\}$ obey the canonical
CAR, Eq.~\eqref{eq:JW_orth_create} defines a valid fermion--qubit mapping.
Using Eqs.~\eqref{eq:a_tilde_from_a}–\eqref{eq:a_from_a_tilde}, any
nonorthogonal operator can be written as a linear combination of orthonormal
operators and thus as a linear combination of JW strings.
In particular, by absorbing the orbital transformation into the definition of
the $\varphi_p$ and of the integrals, the nonorthogonal creation operators
inherit the same JW string structure as in the orthogonal case,
\begin{equation}
  \hat{a}_p^\dagger
  \;\longleftrightarrow\;
  Z_1 \otimes \cdots \otimes Z_{p-1} \otimes
  \frac{X_p - i Y_p}{2}
  \otimes I_{p+1} \otimes \cdots \otimes I_M.
  \label{eq:NOJW_create}
\end{equation}

The adjoint operators $\hat{a}_p$ require a different treatment.
Using Eq.~\eqref{eq:NO_anticom}, one can show that applying $\hat{a}_p$ to an
ON state $|k_1,\dots,k_M\rangle$ yields a linear combination of states in which
one of the occupied orbitals $q$ is emptied, with coefficients proportional to
$S_{pq}$.
This motivates the following JW-like encoding:
\begin{equation}
  \hat{a}_p
  \;\longleftrightarrow\;
  \sum_{q=1}^M S_{pq}\,
  \Bigl[
    Z_1 \otimes \cdots \otimes Z_{q-1} \otimes
    \frac{X_q + i Y_q}{2}
    \otimes I_{q+1} \otimes \cdots \otimes I_M
  \Bigr],
  \label{eq:NOJW_adj}
\end{equation}
that is, a weighted sum of standard JW annihilation strings, with weights given
by the overlap matrix elements $S_{pq}$.
One can verify explicitly that the images of $\hat{a}_p^\dagger$ and
$\hat{a}_q$ under Eqs.~\eqref{eq:NOJW_create}–\eqref{eq:NOJW_adj} satisfy the
generalized anticommutation relations~\eqref{eq:NO_anticom} when acting on the
qubit basis.

The mapping in Eq.~\eqref{eq:NOJW_adj} provides a direct encoding of
wavefunctions built from nonorthogonal determinants, such as the SCGVB
determinants introduced earlier, onto a quantum register.
However, because each $\hat{a}_p$ is represented as a sum of Pauli strings, the
number of distinct strings appearing in the qubit Hamiltonian of
Eq.~\eqref{eq:H_NO_raw} scales in the worst case as $\mathcal{O}(M^3)$ for the
one-electron term and $\mathcal{O}(M^6)$ for the two-electron term, which is
significantly less favorable than the $\mathcal{O}(M^2)$ and
$\mathcal{O}(M^4)$ scalings of the conventional JW mapping in an orthonormal
basis.

While the nonorthogonal Jordan--Wigner mapping defines how operators are
represented on qubits, it does not address the dominant NISQ bottleneck: the
evaluation of overlaps and off-diagonal Hamiltonian matrix elements between
nonorthogonal determinants.
In practice, this task is commonly addressed via Hadamard-test--based schemes,
which require ancilla qubits, controlled operations, and deep circuits.
In the following, we introduce quantum estimators that resolve this bottleneck
using only shallow, ancilla-free, and control-free circuits.

\section{SCGVB-Inspired Quantum Algorithms}
\label{sec:SCGVB-NOQA}

The nonorthogonal generalized valence bond (SCGVB) ansatz requires the
evaluation of two classes of quantities:
(i)~overlaps between nonorthogonal Slater determinants
$\langle \psi_j|\psi_i\rangle$, and
(ii)~Hamiltonian matrix elements
$H_{ij}=\langle \psi_j|\hat{H}|\psi_i\rangle$.
On a quantum device, overlaps and expectation values are commonly evaluated
using ancilla-based Hadamard tests or modified Hadamard-test variants relying
on controlled-unitary operations~\cite{Kitaev1997,NielsenChuang,Cleve1998,Baek2023NO},
which introduce additional qubit and coherence overhead.

Although powerful, these approaches require controlled unitaries,
entangling gates, and circuits whose depth scales with the Pauli decomposition
of the operators, making them poorly suited for NISQ hardware.
Here we introduce a pair of low-depth quantum measurement estimators designed to minimize ancilla and controlled-operation overhead that compute
both classes of quantities without using variational parameters,
controlled operations, or ancilla qubits.

We define two primitives:
\begin{itemize}
    \item \textbf{SCGVB Determinant--Overlap Estimator (DOE):}
    a direct probability-estimation scheme for evaluating
    $\langle \psi_j|\psi_i\rangle$
    from the probability of the all-zero outcome after applying the Pauli
    strings associated with the SCGVB determinants.
    \item \textbf{SCGVB Pauli-Grouped Hamiltonian Estimator (PGHE):}
    a measurement-based expectation-value estimator for
    $\langle \psi_j|\hat{H}|\psi_i\rangle$,
    based on qubit-wise commuting (QWC) grouping and local basis rotations.
\end{itemize}

In contrast to ancilla-based approaches, the SCGVB--NOQA framework introduced
here replaces all controlled operations by purely local, single-qubit
Clifford circuits.
Both estimators are ancilla-free, non-variational, and contain no
entangling gates.
All measurements reduce to sampling in rotated computational bases, and all
Pauli groups are evaluated via shallow, constant-depth circuits.
The resulting construction is a fully local, ancilla-free measurement formulation tailored to SCGVB determinant spaces; no claim of computational advantage over direct classical evaluation is implied by the circuit structure alone.

\paragraph*{Resource comparison with ancilla-based estimators.}
Before detailing the estimators, we briefly contrast their resource
requirements with those of ancilla-based approaches.
Ancilla-based methods for evaluating overlaps or transition amplitudes,
such as the Hadamard test, require an additional control qubit on top of the
system register, resulting in $n+1$ qubits for an $n$-qubit system.
Modified Hadamard-test schemes used to estimate general matrix elements of the
form $\langle \psi | \hat{O} | \phi \rangle$ typically require two system
registers together with an ancilla qubit, leading to $2n+1$ qubits.

In contrast, the estimators introduced in this work act directly on a single
system register and require only $n$ qubits.
Moreover, the proposed circuits avoid controlled unitaries entirely and
consist solely of local single-qubit Clifford operations followed by
computational-basis measurements, resulting in constant-depth circuit
implementations.

This section now presents the complete formulation of the two estimators,
including their circuit constructions, measurement rules, and asymptotic
scaling.
\subsection{SCGVB Determinant–Overlap Estimator (DOE)}
\label{alg:overlap_1}
We evaluate the overlap matrix elements of the form
\begin{equation}
    \langle \psi_j | \psi_i \rangle,
    \qquad
    |\psi_i\rangle = \hat{f}_i \,|{\rm vac}\rangle, \quad
    \langle \psi_j| = \langle {\rm vac}| \,\hat{w}_j,
\end{equation}
where $|{\rm vac}\rangle = |0\rangle^{\otimes n}$ denotes the computational-basis vacuum.  
Each determinant operator $\hat{f}_i$ or $\hat{w}_j$ corresponds to a product of fermionic creation and annihilation operators, respectively.
After applying the nonorthogonal Jordan--Wigner mapping into $\hat{w}_j$  and the standard JW into $\hat{f}_i$,
each operator becomes a linear combination of $n$-qubit Pauli strings:
\begin{align}
    \hat{f}_i = \sum_{a} c^{(f)}_{i,a} Q_{i,a},
    \qquad
    \hat{w}_j = \sum_{b} c^{(w)}_{j,b} P_{j,b},
\end{align}
where $Q_{i,a}$ and $P_{j,b}$ are tensor products of $\{X,Y,Z,I\}$ operators.
These Pauli terms are stored externally and loaded directly by the quantum routine.

\paragraph{Circuit construction.}
Expanding both operators yields
\begin{equation}
    \langle \psi_j | \psi_i \rangle
    = \sum_{a,b} c^{(f)}_{i,a} c^{(w)}_{j,b}
      \langle 0^n | P_{j,b} Q_{i,a} | 0^n \rangle .
    \label{eq:overlap_expansion}
\end{equation}
The coefficients $c^{(f)}_{i,a}$ and $c^{(w)}_{j,b}$ are generally complex and
retain the real or imaginary phases generated by the Jordan--Wigner
expansions of the individual determinant operators.  In addition, multiplying
two phase-free Pauli strings can generate a further Pauli-product phase.  
We therefore perform the classical Pauli multiplication
\begin{equation}
P_{j,b}Q_{i,a}
=
\eta_{ab}^{(ij)}R_{ab}^{(ij)}.
\label{eq:doe_pauli-product-phase}
\end{equation}
Here, $\eta_{ab}^{(ij)}\in\{\pm1,\pm i\}$ is the phase generated by
the Pauli multiplication, while
$R_{ab}^{(ij)}\in\{I,X,Y,Z\}^{\otimes n}$ is written without
an overall phase. Consequently,
\begin{equation}
    \langle 0^n | P_{j,b}Q_{i,a} |0^n\rangle
    = \eta_{ab}^{(ij)}
      \langle 0^n |R_{ab}^{(ij)}|0^n\rangle .
    \label{eq:doe_phase_resolved_amplitude}
\end{equation}
The phase $\eta_{ab}^{(ij)}$ is obtained exactly by classical Pauli algebra;
it is a global phase of the circuit output and therefore cannot be recovered
from computational-basis probabilities alone.

\begin{figure}[t]
\centering

\begin{quantikz}[row sep=0.25cm, column sep=0.4cm]
\lstick{$|0\rangle$} 
    & \gate{P_{j,b}Q_{i,a}} & \meter{} \\
\lstick{$|0\rangle$} 
    & \gate{P_{j,b}Q_{i,a}} & \meter{} \\
\lstick{$\vdots$} & \vdots & \vdots \\
\lstick{$|0\rangle$} 
    & \gate{P_{j,b}Q_{i,a}} & \meter{}
\end{quantikz}
\caption{
Quantum circuit used to determine the vacuum-return probability associated
with a Pauli pair $P_{j,b}Q_{i,a}$.  The classically computed phase
$\eta_{ab}^{(ij)}$ in Eq.~\eqref{eq:doe_pauli-product-phase} does not affect the
measurement statistics and is restored during classical post-processing.
Since each Pauli string is a tensor product of single-qubit Clifford gates,
the circuit consists only of local operations followed by computational-basis
measurement.
}
\label{fig:overlap-circuit}
\end{figure}
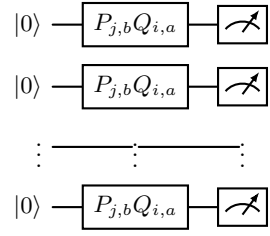

Our algorithm determines the remaining vacuum-return indicator using the
shallow Clifford circuit shown in Fig.~\ref{fig:overlap-circuit}:

\begin{enumerate}
    \item initialize the register in $|0\rangle^{\otimes n}$;
    \item apply all non-identity Pauli gates in $Q_{i,a}$;
    \item apply all non-identity Pauli gates in $P_{j,b}$;
    \item measure all qubits in the computational basis.
\end{enumerate}

Because Pauli operators act locally and require no entangling gates,
the circuit depth is $\mathcal{O}(1)$ and depends only on the number of 
non-identity Pauli factors in $P_{j,b} Q_{i,a}$.

\paragraph{Measurement rule.}
The measurement samples from the distribution generated by
$P_{j,b}Q_{i,a}|0^n\rangle$.  Because an $n$-qubit Pauli string maps the vacuum
to one computational-basis state up to a global phase, the all-zero
probability is
\begin{equation}
    p_{ab}^{(ij)}
    = \Pr\!\left(0^n\,\middle|\,P_{j,b}Q_{i,a}|0^n\rangle\right)
    = \begin{cases}
        1, & R_{ab}^{(ij)}\in\{I,Z\}^{\otimes n},\\[2pt]
        0, & \text{otherwise},
      \end{cases}
    \label{eq:doe_vacuum_return_probability}
\end{equation}
in the ideal noiseless limit.  Hence the probability determines whether the
phase-free Pauli product returns the register to the vacuum, whereas the
complex phase is supplied separately by $\eta_{ab}^{(ij)}$:
\begin{equation}
    \langle 0^n|P_{j,b}Q_{i,a}|0^n\rangle
    = \eta_{ab}^{(ij)}p_{ab}^{(ij)}.
    \label{eq:doe_amplitude_from_phase_probability}
\end{equation}
The phase-aware overlap estimator is therefore
\begin{equation}
    \widetilde{O}_{ij}
    = \sum_{a,b}
      c^{(f)}_{i,a}\,c^{(w)}_{j,b}\,
      \eta_{ab}^{(ij)}\,p_{ab}^{(ij)}.
    \label{eq:overlap_estimator}
\end{equation}
For example, $YX=-iZ$, so that $\langle0|YX|0\rangle=-i$ although the
all-zero measurement probability is unity.  In this case the missing factor
$-i$ is precisely the classically tracked Pauli-product phase $\eta$.
Thus, phases already present in the generally complex coefficients
$c^{(f)}_{i,a}$ and $c^{(w)}_{j,b}$ are retained, while the additional phase
created by multiplying $P_{j,b}$ and $Q_{i,a}$ is restored explicitly in
post-processing.  No ancilla qubits, controlled unitaries, or mid-circuit
measurements are required.

\paragraph{Computational scaling.}
Let $n_{\rm Pauli}^{(f)}$ and $n_{\rm Pauli}^{(w)}$ denote the number of Pauli
strings in $\hat{f}_i$ and $\hat{w}_j$, respectively.
Each pair $(Q_{i,a},P_{j,b})$ requires one shallow circuit execution.
Thus a single overlap evaluation scales as
\begin{equation}
    \mathcal{O}\!\left(n_{\rm Pauli}^{(f)} n_{\rm Pauli}^{(w)}\right),
\end{equation}
and computing all overlaps between $N_f$ ``bra'' and $N_w$ ``ket'' states scales as
\begin{equation}
    \mathcal{O}\!\left(N_f N_w n_{\rm Pauli}^{(f)} n_{\rm Pauli}^{(w)}\right).
\end{equation}
The procedure is embarrassingly parallel over $(i,j)$ and over Pauli-pair indices $(a,b)$.
The corresponding pseudocode is given in Algorithm~\ref{algo:overlap}.

\vspace{0.3cm}

\noindent\textbf{Algorithm 1. Phase-aware Determinant--Overlap Estimator (DOE) for evaluating
$\langle \psi_j | \psi_i \rangle$.}
\label{algo:overlap}
\begin{algorithmic}[1]
\Require Pauli expansions $\hat{f}_i=\sum_a c_{i,a}^{(f)}Q_{i,a}$,
        $\hat{w}_j=\sum_b c_{j,b}^{(w)}P_{j,b}$; number of shots $S$.
\Ensure Estimate $\widetilde{O}_{ij}$
\State $\widetilde{O}_{ij} \gets 0$
\For{each Pauli string $Q_{i,a}$ in $\hat{f}_i$}
    \For{each Pauli string $P_{j,b}$ in $\hat{w}_j$}
        \State $(\eta_{ab}^{(ij)},R_{ab}^{(ij)})
        \gets \operatorname{MultiplyPaulis}(P_{j,b},Q_{i,a})$
        \Comment{$P_{j,b}Q_{i,a}=\eta_{ab}^{(ij)}R_{ab}^{(ij)}$}
        \State prepare $|0^n\rangle$
        \State apply $Q_{i,a}$ followed by $P_{j,b}$
        \Comment{equivalently apply $R_{ab}^{(ij)}$}
        \State measure all qubits in the computational basis $S$ times
        \State $p_{ab}^{(ij)} \gets \Pr(0^n)$ from measurement statistics
        \State $\widetilde{O}_{ij} \gets \widetilde{O}_{ij}
               + c_{i,a}^{(f)}c_{j,b}^{(w)}
                 \eta_{ab}^{(ij)}p_{ab}^{(ij)}$
    \EndFor
\EndFor
\State \Return $\widetilde{O}_{ij}$
\end{algorithmic}

\subsection{SCGVB Pauli-Grouped Hamiltonian Estimator (PGHE)}
\label{sec:hamiltonian-eval}
While determinant overlaps can be evaluated directly using Pauli applications on the vacuum, the evaluation of Hamiltonian matrix elements requires estimation of Pauli expectation values. This leads to a different circuit structure, detailed below. We evaluate matrix elements of the form
\begin{equation}
    H_{ij} \equiv 
    \langle \phi_0 |\, w_i \, \hat{H}\, f_j \,| \phi_0\rangle ,
\end{equation}
where $f_j$ and $w_i$ are the fermionic excitation and de-excitation
operators defining the SCGVB determinant basis, mapped to qubits through the
nonorthogonal Jordan--Wigner transformation.  
Each operator is expanded as a sparse linear combination of Pauli strings,
\begin{equation}
    f_j = \sum_{\alpha} c^{(f_j)}_\alpha P_\alpha, \qquad 
    w_i = \sum_{\beta}  c^{(w_i)}_\beta P_\beta, \qquad
    \hat{H} = \sum_{\gamma} h_\gamma P_\gamma .
\end{equation}

The operator whose expectation value must be measured is
\begin{equation}
    \hat{O}_{ij} = w_i\, \hat{H}\, f_j
    = \sum_{k} d_k^{(ij)} P_k ,
    \label{eq:Oij-pauli}
\end{equation}
with coefficients $d_k^{(ij)}$ obtained by symbolic multiplication of Pauli terms.
The reference state used throughout is the computational-basis vacuum,
$|\phi_0\rangle = |0\ldots 0\rangle$.

\paragraph{Measurement strategy.}
Each Pauli operator $P_k$ appearing in Eq.~\eqref{eq:Oij-pauli}
is measured using a standard basis-rotation protocol.
For a single-qubit Pauli,
\begin{align}
    X &\rightarrow H Z H, \\
    Y &\rightarrow S^\dagger H Z H S, \\
    Z &\rightarrow Z, \qquad 
    I \rightarrow I ,
\end{align}
so that the composite rotation $U_k$ diagonalizes the full Pauli string.
After applying $U_k$, all qubits are measured in the computational ($Z$) basis.
For a measurement outcome $b\in\{0,1\}^n$, the eigenvalue is reconstructed via
\[
    \lambda_k(b) = (-1)^{\sum_{q\in \mathrm{supp}(P_k)} b_q}.
\]

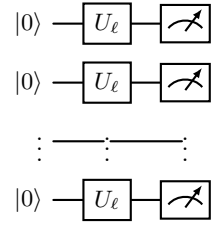
\begin{figure}[t]
\centering
\begin{quantikz}[row sep=0.25cm, column sep=0.4cm]
\lstick{$|0\rangle$} 
    & \gate{U_\ell} & \meter{} \\
\lstick{$|0\rangle$} 
    & \gate{U_\ell} & \meter{} \\
\lstick{$\vdots$} & \vdots & \vdots \\
\lstick{$|0\rangle$} 
    & \gate{U_\ell} & \meter{}
\end{quantikz}
\caption{
Quantum circuit used to measure all Pauli operators $P_k \in G_\ell$
belonging to a qubit-wise commuting group.
A single local basis rotation $U_\ell$ simultaneously diagonalizes every
operator in $G_\ell$, after which all qubits are measured in the
computational basis.
The resulting bitstring statistics are reused to obtain 
$\langle P_k\rangle_0$ for all $P_k\in G_\ell$, which are then assembled
to recover $H_{ij}$.
}
\label{fig:H-circuit}
\end{figure}

\paragraph{Qubit-wise commuting (QWC) grouping.}
To reduce measurement overhead, the Pauli strings $\{P_k\}$ are partitioned into
qubit-wise commuting groups,
\[
    \{P_k\} = G_1 \cup G_2 \cup \dots \cup G_M ,
\qquad 
[P_k, P_{k'}] = 0 \;\; \forall\, P_k,P_{k'}\in G_\ell.
\]
All operators in the same group share the same diagonalizing rotation.
Thus each $G_\ell$ requires only one circuit execution.
This measurement procedure is summarized in the Pauli-Grouped Hamiltonian Estimator (PGHE), shown in
\textbf{Algorithm~\ref{alg:H-eval}}, and the corresponding quantum circuit
structure is shown in \textbf{Figure~\ref{fig:H-circuit}}.

\paragraph{Reconstruction of the matrix element.}
From repeated measurements (shots) of each QWC group $G_\ell$, we compute
$\langle P_k\rangle_0$ for all $P_k \in G_\ell$ and reconstruct
\begin{equation}
    H_{ij}
    = \langle \phi_0|\, \hat{O}_{ij}\,|\phi_0\rangle
    = \sum_{k} d_k^{(ij)} \, \langle P_k \rangle_0 .
\end{equation}
This process is repeated for all $i,j\in\{1,\ldots,N_{\rm det}\}$,
where $N_{\rm det}=6$ in the SCGVB space used here.

\vspace{0.3cm}
\noindent\textbf{Algorithm 2. Pauli--Grouped Hamiltonian Estimator (PGHE) for evaluating 
$H_{ij} = \langle \phi_0 | w_i H f_j | \phi_0 \rangle$.}
\label{alg:H-eval}

\begin{algorithmic}[1]
\Require Pauli expansion $\hat{O}_{ij}=\sum_k d_k^{(ij)} P_k$,
         QWC grouping routine $\mathrm{GroupQWC}(\{P_k\})$,
         number of shots $S$.
\Ensure Estimate $\widetilde{H}_{ij}$
\State $\widetilde{H}_{ij} \gets 0$
\State $\{G_1,\dots,G_M\} \gets \mathrm{GroupQWC}(\{P_k\})$
\For{each group $G_\ell$}
    \State choose representative Pauli $P_{\mathrm{rep}} \in G_\ell$
    \State prepare $|\phi_0\rangle = |0^n\rangle$
    \State apply basis-rotation $U_\ell$ diagonalizing every $P_k \in G_\ell$
    \State measure all qubits in the $Z$ basis $S$ times
    \For{each $P_k\in G_\ell$}
        \State compute eigenvalues $\lambda_k(b)$ from bitstrings $b$
        \State $\langle P_k \rangle_0 = \frac{1}{S}\sum_b \lambda_k(b)p(b)$
        \State $\widetilde{H}_{ij} \gets \widetilde{H}_{ij} + d_k^{(ij)}\langle P_k\rangle_0$
    \EndFor
\EndFor
\State \Return $\widetilde{H}_{ij}$
\end{algorithmic}

\paragraph{Comparison with the overlap estimator.}
The Hamiltonian estimator described above is structurally similar to the
overlap-evaluation circuit of Sec.~\ref{alg:overlap_1}, but with two key
differences.  
(i)~The overlap $\langle \psi_j|\psi_i\rangle$ requires applying the Pauli
operators associated with the determinant creation/annihilation operators
$(f_i,w_j)$ directly to the vacuum and measuring the probability of the
all-zero outcome.  
No basis rotations or grouping are needed, and the resulting circuits consist
only of single-qubit Clifford gates with constant depth.
(ii)~The Hamiltonian estimator, instead, measures expectation values of the
Pauli operators appearing in $\hat{O}_{ij} = w_i H f_j$.
Because these Pauli strings must be measured in their eigenbases, a basis
rotation is required for each qubit-wise commuting (QWC) group.  
Consequently, Hamiltonian estimation involves more circuit executions, but
each circuit remains extremely shallow (a single layer of local rotations
followed by measurement).  
Both estimators are ancilla-free, deterministic, and use only single-qubit
gates. Their shallow circuit structure minimizes coherent-control requirements, although the present Clifford implementation remains efficiently classically simulable.

\paragraph{Computational scaling.}
Let $n^{(f)}_P$, $n^{(w)}_P$, and $n^{(H)}_P$ denote the number of Pauli terms
in the expansions of $f_j$, $w_i$, and $\hat{H}$, respectively.
Forming the operator $\hat{O}_{ij}=w_i H f_j$ involves symbolic multiplication
of Pauli strings, yielding
\[
    n^{(O)}_P(i,j)
    \;\sim\;
    \mathcal{O}\!\left(
        n^{(f)}_P \, n^{(w)}_P \, n^{(H)}_P
    \right),
\]
after term-combination and cancellation.  
These $n^{(O)}_P$ Pauli operators must be partitioned into qubit-wise
commuting groups $\{G_\ell\}_{\ell=1}^{M_{ij}}$.
Since each group requires one circuit execution, the number of quantum
circuits for element $(i,j)$ is
\[
    M_{ij}
    \;\le\;
    n^{(O)}_P(i,j),
\]
with equality in the worst case (no commuting structure) and typically
$M_{ij}\ll n^{(O)}_P(i,j)$ for electronic-structure Hamiltonians.

Each circuit has constant depth and uses only $n$ single-qubit gates,
so the total quantum cost per matrix element is
\[
    \mathcal{Q}_{ij}
    \;=\;
    \mathcal{O}\!\left(
        M_{ij} \cdot S
    \right),
\]
where $S$ is the number of measurement shots.
For a determinant space of size $N_{\rm det}$, the full Hamiltonian requires
\[
    \mathcal{O}\!\left(
        N_{\rm det}^2 \;
        M_{\rm avg} \;
        S
    \right)
\]
quantum operations, where $M_{\rm avg}$ is the average number of QWC groups per
matrix element. Because all $(i,j)$ evaluations and all QWC groups are independent, the entire
procedure parallelizes naturally over GPUs, CPUs, or distributed quantum backends.

{
\begin{table*}[t]

\small
\caption{Consolidated scaling summary for the present direct implementation. $M$ denotes the number of spin orbitals, $N_{\rm det}$ the determinant-space size, $n_P^{(f)}$, $n_P^{(w)}$, and $n_P^{(H)}$ the Pauli counts entering the determinant and Hamiltonian expansions, $M_{ij}$ the number of QWC measurement groups for a determinant pair, and $S$ the shot count. These expressions are algorithmic upper bounds or resource scalings; the H$_4$ and C$_2$ benchmarks are not used to fit an empirical asymptotic exponent.}
\label{tab:scaling_summary}
\centering
\begin{tabular}{ll}
\toprule
Quantity & Scaling / upper bound \\
\midrule
Unique determinant pairs & $N_{\rm det}(N_{\rm det}+1)/2$ \\
NO--JW one-body Pauli growth & $\mathcal{O}(M^3)$, worst case \\
NO--JW two-body Pauli growth & $\mathcal{O}(M^6)$, worst case \\
DOE per determinant pair & $\mathcal{O}\!\left(n_P^{(f)}n_P^{(w)}\right)$ \\
PGHE symbolic product per pair & $\mathcal{O}\!\left(n_P^{(f)}n_P^{(w)}n_P^{(H)}\right)$ before compression \\
Measurement groups per pair & $M_{ij}\le n_P^{(O)}(i,j)$ \\
Shot cost per pair & $\mathcal{O}(M_{ij}S)$ \\
Full determinant-space measurement cost & $\mathcal{O}(N_{\rm det}^{2}M_{\rm avg}S)$ \\
\bottomrule
\end{tabular}
\end{table*}
}

Although the two-electron Hamiltonian typically contains a significantly larger number of Pauli terms than the one-electron part, the number of required quantum circuits is governed by the number of qubit-wise commuting (QWC) groups rather than by the total number of Pauli strings. In practice, many two-body Pauli terms share identical local support patterns and commute qubit-wise, allowing them to be measured within the same circuit. As a result, the effective number of circuits required for the two-body contribution can be comparable to that of the one-body term, despite the much larger Pauli expansion. For larger systems, related measurement ideas inspired by local quantum overlapping tomography ~\cite{araujo2022lqot} may also prove useful.

It is worth noting that, although the Hadamard test and modified Hadamard-test schemes~\cite{Kitaev1997,NielsenChuang,Cleve1998,Baek2023NO} can in principle be used to evaluate the same matrix elements, such techniques require ancilla qubits, controlled operations, and deeper circuits.

The approach introduced here avoids controlled unitaries and uses only single-qubit Clifford rotations, thereby reducing circuit-level depth, ancilla, and controlled-operation requirements relative to those interference-based constructions. This circuit-level reduction is not an overall computational advantage because the nonorthogonal determinant algebra and Pauli preprocessing remain classical.

The circuits used in the present estimators contain only single-qubit Clifford operations acting on computational-basis states and therefore fall within the class efficiently simulable according to the Gottesman--Knill theorem. The present circuit family, by itself, does not demonstrate a computational need for quantum hardware.

The present study instead isolates the matrix-element measurement layer of the SCGVB problem and quantifies the accompanying symbolic and measurement costs. The combinatorial Pauli expansion and determinant-pair growth remain important classical bottlenecks, as summarized in Table~\ref{tab:scaling_summary}.

Accordingly, the current circuits should be interpreted as a circuit-compatible measurement formulation and benchmarking primitive rather than as evidence of quantum speedup. Their value in the present work is to expose the resource trade-off obtained when ancilla-assisted transition measurements are replaced by local measurement circuits.

The estimators can serve as modular building blocks in broader coherent workflows involving nontrivial state preparation. Whether such an end-to-end construction provides quantum utility requires a separate resource analysis and is not inferred from the present Clifford measurement circuits.

\section{Materials and Methods}
\label{details}
Molecular geometries, atomic-orbital bases, and one- and two-electron integrals
were generated using PySCF \cite{pyscf}. The resulting nonorthogonal second-quantized
Hamiltonian was constructed and mapped to qubit operators within Qiskit \cite{qiskit}. All
quantum estimators described in Sec.~\ref{sec:SCGVB-NOQA} were implemented
directly at the Pauli-string level, without the use of ancilla qubits or
controlled operations. For completeness and reproducibility, a step-by-step
pseudocode description of the Qiskit implementation of the nonorthogonal
Jordan--Wigner mapping~\cite{marruzzo2025nojw} is provided in Appendix A of the Supporting Information.

\paragraph{Implementation refinements for larger determinant spaces.}
The estimator definitions above are independent of the molecular system and were used without modification for both the H$_4$ and C$_2$ applications. For larger determinant spaces, however, the direct symbolic expansion of the nonorthogonal Jordan--Wigner products can generate a very large number of raw Pauli products before simplification. In the practical implementation, we therefore exploit two algebra-preserving reductions. First, independent determinant-pair matrix elements are evaluated in parallel, since each pair $(I,J)$ can be reconstructed independently. Second, after the full operator product has been formed and simplified, we apply the exact vacuum-selection rule
\begin{equation}
\langle 0|P|0\rangle = 0
\quad
\text{unless } P \text{ contains only } I \text{ and } Z ,
\end{equation}
so that Pauli strings containing $X$ or $Y$ in the final product are discarded before circuit execution. This vacuum filtering does not change the value of the vacuum expectation estimator; it only avoids sampling terms that are known analytically to vanish. These reductions were used as implementation-level optimizations for the larger C$_2$ determinant-space validation discussed in this text.

\subsection{SCGVB wavefunction of H$_4$ in a Singlet State}
\label{H4-SCGVB_wf}

To make the preceding formalism more concrete, consider the H$_4$ cluster in its
singlet ground state ($S=0$, $M=0$).  
For $N=4$ electrons and $S=0$, there are $f_0^4 = 2$ linearly independent spin
eigenfunctions, corresponding to two distinct spin-coupling patterns
(``pairings'') of the four electrons~\cite{cooper_introVB,dunning2021_scgvb}.
The SCGVB wavefunction can therefore be written as
\begin{eqnarray}
  \Psi_{\mathrm{SCGVB}} 
  &=& \mathcal{N}\;\Bigl[c_1 \; \psi_{0,0;1}^4 \,+ \, c_2 \; \psi_{0,0;2}^4 \Bigr]
  \nonumber \\
  &=& \mathcal{N} \; \Bigl[
  c_1 \; \hat{A}\Bigl(\phi_1 \,\phi_2\, \phi_3\, \phi_4\, \Theta_{0,0;1}^4\Bigr)
  \,+ \nonumber \\[4pt]
  && \quad
  c_2 \; \hat{A}\Bigl(\phi_1 \,\phi_2\, \phi_3\, \phi_4\, \Theta_{0,0;2}^4\Bigr)
  \Bigr],
  \label{eq:scgvb}
\end{eqnarray}
where $c_1$ and $c_2$ are spin-coupling coefficients that will be determined by
minimizing Eq.~(\ref{eq13}), and $\phi_1,\dots,\phi_4$ are the (generally
nonorthogonal) spin-coupled molecular orbitals. Adopting the Rumer basis, the two
spin eigenfunctions for $S=0$ can be written as products of singlet pairs: in one
case the pairs are (1,2) and (3,4), in the other (1,4) and (2,3).  

Explicitly,
\begin{equation}
 \Theta_{0,0;1}^4=
 \frac{1}{\sqrt{2}}\Bigl(\alpha_1\,\beta_2-\beta_1\,\alpha_2\Bigr)\;
 \frac{1}{\sqrt{2}}\Bigl(\alpha_3\,\beta_4-\beta_3\,\alpha_4\Bigr),
\end{equation}
\begin{equation}
 \Theta_{0,0;2}^4=
 \frac{1}{\sqrt{2}}\Bigl(\alpha_1\,\beta_4-\beta_1\,\alpha_4\Bigr)\;
 \frac{1}{\sqrt{2}}\Bigl(\alpha_2\,\beta_3-\beta_2\,\alpha_3\Bigr),
\end{equation}
where $\alpha_i \equiv \alpha(\sigma_i)$ and $\beta_i \equiv \beta(\sigma_i)$.
Expanding these products in primitive spin functions and applying the
antisymmetrizer yields the corresponding spin-coupled structures as linear
combinations of Slater determinants. Using the shorthand
$\phi_i(\mathbf{x})=\phi_i(\mathbf{r})\,\alpha(\sigma)$ and
$\overline{\phi}_i(\mathbf{x})=\phi_i(\mathbf{r})\,\beta(\sigma)$, the two
spin-coupled structures can be compactly expressed as
\begin{widetext}
\begin{eqnarray}
  \psi_{0,0;1}^4 
  &=& \frac{1}{2\sqrt{4!}} \; \Biggl[
  \Big|\phi_1 \,\overline{\phi}_2\, \phi_3\, \overline{\phi}_4\Big| 
  - \Big|\phi_1 \,\overline{\phi}_2\, \overline{\phi}_3\, \phi_4\Big|
  - \Big|\overline{\phi}_1 \,\phi_2\, \phi_3\, \overline{\phi}_4\Big|
  + \Big|\overline{\phi}_1 \,\phi_2\, \overline{\phi}_3\, \phi_4\Big|
  \Biggr],
  \label{eq21}
\\
  \psi_{0,0;2}^4 
  &=& \frac{1}{2\sqrt{4!}} \; \Biggl[
  \Big|\phi_1 \,\phi_2\, \overline{\phi}_3\, \overline{\phi}_4\Big|
  - \Big|\phi_1 \,\overline{\phi}_2\, \phi_3\, \overline{\phi}_4\Big|
  - \Big|\overline{\phi}_1 \,\phi_2\, \overline{\phi}_3\, \phi_4\Big|
  + \Big|\overline{\phi}_1 \,\overline{\phi}_2\, \phi_3\, \phi_4\Big|
  \Biggr].
  \label{eq22}
\end{eqnarray}
\end{widetext}

Thus, for H$_4$ in a singlet state, the SCGVB wavefunction is written as a
two-structure expansion in a basis of four spin-coupled orbitals. The
variational optimization of $c_1$, $c_2$, and the orbitals $\{\phi_i\}$ proceeds
by minimizing Eq.~(\ref{eq13}), which in practice requires the evaluation of
Hamiltonian and overlap matrix elements between the determinants appearing in
Eqs.~(\ref{eq21})--(\ref{eq22})~\cite{thorsteinsson1995,dunning2021_scgvb}.

A useful simplification, particularly in the context of testing quantum
algorithms or approximate treatments, is to freeze the spin-coupled orbitals and
identify them with a minimal set of atomic basis functions for H$_4$, e.g.\ the
four $1s$ functions of an STO-3G basis. Denoting these fixed orbitals by
$\chi_1,\chi_2,\chi_3,\chi_4$, the approximate SCGVB wavefunction becomes
\begin{widetext}
\begin{eqnarray}
  \tilde{\Psi}_{\mathrm{SCGVB}} 
  &=& c_1\Biggl[
  \Big|\chi_1 \,\overline{\chi}_2\, \chi_3\, \overline{\chi}_4\Big| 
  - \Big|\chi_1 \,\overline{\chi}_2\, \overline{\chi}_3\, \chi_4\Big|
  - \Big|\overline{\chi}_1 \,\chi_2\, \chi_3\, \overline{\chi}_4\Big|
  + \Big|\overline{\chi}_1 \,\chi_2\, \overline{\chi}_3\, \chi_4\Big|
  \Biggr] \nonumber
\\
  &+& c_2 \Biggl[
  \Big|\chi_1 \,\chi_2\, \overline{\chi}_3\, \overline{\chi}_4\Big|
  - \Big|\chi_1 \,\overline{\chi}_2\, \chi_3\, \overline{\chi}_4\Big|
  - \Big|\overline{\chi}_1 \,\chi_2\, \overline{\chi}_3\, \chi_4\Big|
  + \Big|\overline{\chi}_1 \,\overline{\chi}_2\, \chi_3\, \chi_4\Big|
  \Biggr],
  \label{eq24}
\end{eqnarray}
\end{widetext}
which is formally a traditional valence bond wavefunction built from a fixed set
of localized orbitals. In this simplified setting, one may focus exclusively on
the evaluation (classically or on a quantum device) of the Hamiltonian and
overlap matrix elements between the determinants in
Eqs.~(\ref{eq21})--(\ref{eq22}), use them to construct the $2\times2$ Hamiltonian
and overlap matrices in the $\{\psi_{0,0;1}^4,\psi_{0,0;2}^4\}$ basis, and finally
solve the generalized eigenvalue problem to obtain the SCGVB coefficients
$(c_1,c_2)$ and the ground-state energy of H$_4$.

This H$_4$ model provides a minimal yet nontrivial testbed for SCGVB theory, for
Chirgwin--Coulson weight analysis, and for exploring alternative strategies
(including quantum computing) to handle the nonorthogonal determinant algebra
that underpins modern valence bond methods.

For the larger C$_2$ benchmark, the same DOE/PGHE estimator workflow was used, but the determinant-pair evaluations were parallelized and the final Pauli expansions were simplified using the vacuum-selection rule described in Sec. \ref{sec:SCGVB-NOQA}. This distinction is important: the C$_2$ calculation does not introduce a different estimator, but rather uses the same estimator with implementation-level reductions that remove zero-expectation Pauli strings and distribute independent matrix-element evaluations across the computing cluster.

{\section{Results and Discussion} \label{Results}}
 
To assess the operational feasibility of the proposed SCGVB quantum estimators, we consider the $H_{4}$ cluster.

\subsection{$H_{4}$ cluster}
The $H_{4}$ system serves as a minimal yet nontrivial test case, featuring strong static correlation and qualitative changes in the bonding pattern along the dissociation coordinate. For this system, the full overlap and Hamiltonian matrices were evaluated using the Pauli-grouped Hamiltonian estimator, demonstrating the complete ancilla-free and shallow-circuit workflow of the method.

\paragraph{Accuracy of the quantum-estimated Hamiltonian.}
To assess the accuracy of the proposed measurement-only quantum algorithm,
we compared the complete SCGVB Hamiltonian matrix 
$H^{\mathrm{QA}}$ obtained on the quantum circuit emulator with the 
classical reference matrix $H^{\mathrm{ref}}$. The absolute elemental deviation is defined as
$|\Delta H_{ij}| = |H^{\mathrm{QA}}_{ij} - H^{\mathrm{ref}}_{ij}|$.
 The full deviation matrix for the square H$_4$ geometry is reported in 
Table~\ref{tab:deviations}. Complete data for the other geometries are provided in Section S4 of the Supporting Information.
Quantitative error assessment is reported at several complementary levels in this work: Table~\ref{tab:h4_geo2_shot_convergence} gives shot-dependent mean, RMS, maximum, and propagated statistical errors for representative H$_4$ Hamiltonian elements; Table~\ref{tab:hamiltonian_deviation_summary} summarizes matrix-level mean, RMS, and maximum deviations over all H$_4$ geometries; Table~\ref{tab:c2_matrix_validation} reports maximum, mean, and Frobenius-norm deviations for the full C$_2$ matrices; and Table~\ref{tab:c2_unfiltered_shot_scaling} isolates the finite-shot component for representative C$_2$ pairs. 

\begin{table*}[t!]
\small
\caption{Hamiltonian matrix elements (in Hartree) between nonorthogonal SCGVB Slater determinants. Columns show the quantum-estimated matrix elements (PGHE quantum circuit), the classical L\"{o}wdin reference, and the absolute deviation $|\Delta H_{ij}|$. The target system is a square $H_{4}$ molecule $0.7414 \times 0.7414$ (\AA). Only unique elements ($i \le j$) are reported since the Hamiltonian matrix is Hermitian.}
\label{tab:deviations}
\vspace{2mm}
\centering
\begin{threeparttable}
\begin{tabular}{|l|c|c|c|}
\hline
$\mathbf{H_{ij}}$ & \textbf{Quantum Circuit} & \textbf{L\"{o}wdin reference} & $\mathbf{\left|\Delta H_{ij}\right|}$ \\
\hline

$\langle \psi_1 | H | \psi_1 \rangle$ & -1.4174356874 & -1.447215508 & 0.0297798206 \\ \hline
$\langle \psi_1 | H | \psi_2 \rangle$ & 0.3223085002 & 0.318808565 & 0.0034999352 \\ \hline
$\langle \psi_1 | H | \psi_3 \rangle$ & 0.3265732036 & 0.318808565 & 0.0077646386 \\ \hline
$\langle \psi_1 | H | \psi_4 \rangle$ & 0.0111227864 & 0.01720101851 & 0.0060782321 \\ \hline
$\langle \psi_1 | H | \psi_5 \rangle$ & 0.3185223847 & 0.318808565 & 0.0002861803 \\ \hline
$\langle \psi_1 | H | \psi_6 \rangle$ & 0.3302189815 & 0.318808565 & 0.0114104160 \\ \hline

$\langle \psi_2 | H | \psi_2 \rangle$ & -0.6430841052 & -0.6331848711 & 0.0098992341 \\ \hline
$\langle \psi_2 | H | \psi_3 \rangle$ & -0.1505910757 & -0.1478321999 & 0.0027588758 \\ \hline
$\langle \psi_2 | H | \psi_4 \rangle$ & 0.3091913045 & 0.318808565 & 0.0096172605 \\ \hline
$\langle \psi_2 | H | \psi_5 \rangle$ & -0.0090499437 & -0.005690144139 & 0.0033597996 \\ \hline
$\langle \psi_2 | H | \psi_6 \rangle$ & -0.0075174089 & -0.005690144139 & 0.0018272648 \\ \hline

$\langle \psi_3 | H | \psi_3 \rangle$ & -0.6594546921 & -0.6331848711 & 0.0262698210 \\ \hline
$\langle \psi_3 | H | \psi_4 \rangle$ & 0.3111067327 & 0.318808565 & 0.0077018323 \\ \hline
$\langle \psi_3 | H | \psi_5 \rangle$ & -0.0079179479 & -0.005690144139 & 0.0022278038 \\ \hline
$\langle \psi_3 | H | \psi_6 \rangle$ & -0.0028908252 & -0.005690144139 & 0.0027993189 \\ \hline

$\langle \psi_4 | H | \psi_4 \rangle$ & -1.4142364353 & -1.447215508 & 0.0329790727 \\ \hline
$\langle \psi_4 | H | \psi_5 \rangle$ & 0.3245050937 & 0.318808565 & 0.0056965287 \\ \hline
$\langle \psi_4 | H | \psi_6 \rangle$ & 0.3281355396 & 0.318808565 & 0.0093269747 \\ \hline

$\langle \psi_5 | H | \psi_5 \rangle$ & -0.6364821722 & -0.6331848711 & 0.0032973011 \\ \hline
$\langle \psi_5 | H | \psi_6 \rangle$ & -0.1488527680 & -0.1478321999 & 0.0010205681 \\ \hline

$\langle \psi_6 | H | \psi_6 \rangle$ & -0.6417711786 & -0.6331848711 & 0.0085863075 \\ \hline

\end{tabular}
\end{threeparttable}
\end{table*}

\begin{table*}[t]
\small
\caption{Overlap matrix elements between the six SCGVB determinants.
The column ``QA (DOE)'' lists the values obtained from the quantum circuit,
while ``L\"{o}wdin rules'' contains the classical reference.
Absolute discrepancies are $|\Delta S_{ij}| = |S^{\mathrm{QA}}_{ij} - S^{\mathrm{ref}}_{ij}|$.
Only unique elements ($i \le j$) are reported since the overlap matrix is symmetric.}
\label{tab:overlap_unique}
\centering
\begin{threeparttable}
\begin{tabular}{|l|c|c|c|}
\hline
$\mathbf{S_{ij}}$ & \textbf{QA (DOE)} & \textbf{L\"{o}wdin rules} & $\mathbf{|\Delta S_{ij}|}$ \\
\hline

$\langle \psi_1|\psi_1\rangle$ & 0.6093766053 & 0.6093766115 & $6.2\times10^{-9}$ \\ \hline
$\langle \psi_1|\psi_2\rangle$ & -0.1227232093 & -0.1227232103 & $1.0\times10^{-9}$ \\ \hline
$\langle \psi_1|\psi_3\rangle$ & -0.1227232093 & -0.1227232103 & $1.0\times10^{-9}$ \\ \hline
$\langle \psi_1|\psi_4\rangle$ & 0.0000000000 & 0.0000000000 & $0$ \\ \hline
$\langle \psi_1|\psi_5\rangle$ & -0.1227232093 & -0.1227232103 & $1.0\times10^{-9}$ \\ \hline
$\langle \psi_1|\psi_6\rangle$ & -0.1227232093 & -0.1227232103 & $1.0\times10^{-9}$ \\ \hline

$\langle \psi_2|\psi_2\rangle$ & 0.3201019318 & 0.3201019355 & $3.7\times10^{-9}$ \\ \hline
$\langle \psi_2|\psi_3\rangle$ & 0.0461606361 & 0.0461606364 & $3.0\times10^{-10}$ \\ \hline
$\langle \psi_2|\psi_4\rangle$ & -0.1227232093 & -0.1227232103 & $1.0\times10^{-9}$ \\ \hline
$\langle \psi_2|\psi_5\rangle$ & -0.0011661906 & -0.0011661905 & $1.0\times10^{-10}$ \\ \hline
$\langle \psi_2|\psi_6\rangle$ & -0.0011661906 & -0.0011661905 & $1.0\times10^{-10}$ \\ \hline

$\langle \psi_3|\psi_3\rangle$ & 0.3201019318 & 0.3201019355 & $3.7\times10^{-9}$ \\ \hline
$\langle \psi_3|\psi_4\rangle$ & -0.1227232093 & -0.1227232103 & $1.0\times10^{-9}$ \\ \hline
$\langle \psi_3|\psi_5\rangle$ & -0.0011661906 & -0.0011661905 & $1.0\times10^{-10}$ \\ \hline
$\langle \psi_3|\psi_6\rangle$ & -0.0011661906 & -0.0011661905 & $1.0\times10^{-10}$ \\ \hline

$\langle \psi_4|\psi_4\rangle$ & 0.6093766053 & 0.6093766115 & $6.2\times10^{-9}$ \\ \hline
$\langle \psi_4|\psi_5\rangle$ & -0.1227232093 & -0.1227232103 & $1.0\times10^{-9}$ \\ \hline
$\langle \psi_4|\psi_6\rangle$ & -0.1227232093 & -0.1227232103 & $1.0\times10^{-9}$ \\ \hline

$\langle \psi_5|\psi_5\rangle$ & 0.3201019318 & 0.3201019355 & $3.7\times10^{-9}$ \\ \hline
$\langle \psi_5|\psi_6\rangle$ & 0.0461606361 & 0.0461606364 & $3.0\times10^{-10}$ \\ \hline

$\langle \psi_6|\psi_6\rangle$ & 0.3201019318 & 0.3201019355 & $3.7\times10^{-9}$ \\ \hline

\end{tabular}
\end{threeparttable}
\end{table*}

Importantly, the observed deviations do not preferentially affect any specific class of matrix elements, indicating that the measurement-only estimator preserves the relative structure of the nonorthogonal Hamiltonian rather than introducing systematic bias. This property is essential for the stability of generalized eigenvalue problems involving strongly overlapping determinant bases.

\paragraph{Accuracy of the quantum-estimated overlap matrix.}

Because a minimal basis is employed, the present results do not aim at chemical accuracy with respect to the complete basis set limit, but rather at validating the accuracy of the quantum algorithm within a fixed orbital representation.
Table~\ref{tab:overlap_unique} reports the full overlap matrix between the six nonorthogonal SCGVB determinants for the square H$_4$ geometry.
The agreement between the quantum-estimated overlaps obtained from the
Determinant--Overlap Estimator (DOE) and the classical values computed using L\"owdin’s nonorthogonal determinant rules is essentially exact, with absolute discrepancies on the order of $10^{-9}$ or smaller.
This level of agreement validates both the nonorthogonal Jordan--Wigner mapping and the measurement protocol used to reconstruct determinant overlaps from vacuum-projected Pauli strings.

From a physical perspective, the overlap matrix clearly reflects the strongly nonorthogonal character of the SCGVB determinant basis at the square geometry. Several off-diagonal elements are sizable, indicating significant interference between distinct valence-bond pairing patterns enforced by the high symmetry of the cluster. This nonorthogonality is intrinsic to the SCGVB description and cannot be eliminated without altering the physical content of the wavefunction. The ability of the quantum algorithm to reproduce these overlap patterns with
high fidelity is therefore a nontrivial requirement for the stability of the generalized eigenvalue problem.

Overall, the agreement between quantum and classical matrix elements is excellent. The maximum deviation observed is $\max_{i,j} |\Delta H_{ij}| = 3.30 \times 10^{-2}\, \text{Ha}$, which occurs on a diagonal element. This behavior is expected, as diagonal matrix elements involve larger
absolute Hamiltonian expectation values and accumulate contributions from
a greater number of Pauli terms, rendering them statistically more
sensitive to finite-shot noise than off-diagonal couplings.
Most off-diagonal contributions exhibit significantly smaller deviations,
typically in the range $10^{-3}$--$10^{-2}\,\text{Ha}$, and several elements fall below $3 \times 10^{-4}\,\text{Ha}$. Such fluctuations are fully consistent with statistical noise due to finite sampling:
each matrix element corresponds to a weighted sum of thousands of Pauli-term expectation values, each estimated using $5.24 \times 10^5$ measurement shots.
The deviations appear randomly distributed across diagonal and off-diagonal sectors, with no discernible systematic trend, suggesting that the quantum reconstruction does not introduce a dominant bias beyond statistical noise.  From a physical perspective, the magnitude of these errors is sufficiently small
to guarantee a stable recovery of the SCGVB ground-state energy via the generalized eigenvalue problem $Hc = E\,Sc$.
The largest deviation (0.033 Ha), although prominent at the matrix-element level, represents roughly 0.2\% of the diagonal entry's magnitude, and its effect is moderated when solving the full nonorthogonal eigenvalue problem. These results demonstrate that the measurement-only architecture---comprising only single-qubit basis rotations and measurements, and requiring no entangling gates--- is capable of accurately reconstructing a correlated many-electron Hamiltonian.

\paragraph{Shot-convergence analysis for the Hamiltonian estimator.}

To further verify that the deviations observed in the Hamiltonian matrix are dominated by finite-shot sampling rather than by a systematic bias of the estimator, we performed an explicit shot-convergence analysis for the total Hamiltonian,
\[
H_{\mathrm{full}} = H^{(1)}+H^{(2)},
\]
at the representative rectangular H$_4$ geometry \(0.8800\times0.7414\)~\AA. This geometry is part of the H$_4\rightarrow\mathrm{H}_2+\mathrm{H}_2\) dissociation pathway and was chosen as a representative non-square geometry for testing the statistical behavior of the measurement protocol. The analysis was restricted to the Hamiltonian estimator, since the finite-shot deviations discussed above arise primarily from the reconstruction of Hamiltonian matrix elements. Seven representative matrix elements were selected, including both diagonal and off-diagonal entries,
\[
(1,1),\ (1,2),\ (2,2),\ (3,3),\ (4,4),\ (5,6),\ (6,6),
\]
and each element was recomputed with
\[
N_{\rm shots}=4096,\ 16384,\ 65536,\ 262144,\ 524288,
\]
using five independent repetitions at each shot count.

For each measured matrix element, the statistical uncertainty was estimated by propagating the variances of the Pauli measurements entering the grouped Hamiltonian estimator. For an element written as
\[
H_{ij} = \sum_{\alpha} c_{\alpha}^{(ij)} \langle P_{\alpha}\rangle,
\]
we used the independent-Pauli estimate
\[
\sigma(H_{ij})
=
\left[
\sum_{\alpha}
|c_{\alpha}^{(ij)}|^2
\frac{1-\langle P_{\alpha}\rangle^2}{N_{\rm shots}}
\right]^{1/2}.
\]
This estimate neglects covariance between grouped Pauli terms and is therefore used as an operational uncertainty diagnostic rather than as a strict confidence interval.

The shot-convergence results are summarized in Table~\ref{tab:h4_geo2_shot_convergence}. The mean absolute deviation between the measured estimator and the analytic Pauli estimator decreases from \(1.34\times10^{-2}\)~Ha at \(4096\) shots to \(9.07\times10^{-4}\)~Ha at \(524288\) shots. The corresponding RMS deviation decreases from \(1.55\times10^{-2}\)~Ha to \(1.14\times10^{-3}\)~Ha, while the largest observed deviation decreases from \(4.66\times10^{-2}\)~Ha to \(2.90\times10^{-3}\)~Ha. The empirical standard deviations are generally of the same order as the propagated uncertainty estimates and decrease overall with increasing shot count. These results support the interpretation that the observed Hamiltonian-matrix deviations are finite-sampling fluctuations rather than a systematic error of the ancilla-free measurement-only estimator.

\begin{table*}[t!]
\small
\centering
\caption{Shot-convergence analysis for representative total-Hamiltonian matrix elements of rectangular H$_4$ at \(0.8800\times0.7414\)~\AA. Statistics are computed over the selected elements \((1,1),(1,2),(2,2),(3,3),(4,4),(5,6),(6,6)\), using five independent repetitions for each shot count.}
\label{tab:h4_geo2_shot_convergence}
\begin{tabular}{c c c c c}
\hline
\(N_{\rm shots}\) &
Mean \(|\Delta H|\) (Ha) &
RMS \(|\Delta H|\) (Ha) &
Max \(|\Delta H|\) (Ha) &
Mean \(\sigma(H)\) (Ha) \\
\hline
\(4096\)   & \(1.34\times10^{-2}\) & \(1.55\times10^{-2}\) & \(4.66\times10^{-2}\) & \(1.78\times10^{-2}\) \\
\(16384\)  & \(5.69\times10^{-3}\) & \(6.68\times10^{-3}\) & \(1.64\times10^{-2}\) & \(8.90\times10^{-3}\) \\
\(65536\)  & \(2.00\times10^{-3}\) & \(2.52\times10^{-3}\) & \(8.99\times10^{-3}\) & \(4.45\times10^{-3}\) \\
\(262144\) & \(1.33\times10^{-3}\) & \(1.64\times10^{-3}\) & \(4.55\times10^{-3}\) & \(2.22\times10^{-3}\) \\
\(524288\) & \(9.07\times10^{-4}\) & \(1.14\times10^{-3}\) & \(2.90\times10^{-3}\) & \(1.57\times10^{-3}\) \\
\hline
\end{tabular}
\end{table*}

The same calculations confirm that the statistical convergence is obtained without increasing circuit depth. All Hamiltonian measurement circuits in this test have maximum depth equal to 2 and contain only single-qubit basis rotations and computational-basis measurements, with no CNOT or other entangling gates. Thus, the dominant numerical limitation of the present H$_4$ Hamiltonian reconstruction is measurement sampling, not coherent circuit depth.

To make the propagation of the matrix-element errors to the final SCGVB observable explicit, we projected the six-determinant matrices onto the two spin-coupled structures of Eqs.~(\ref{eq21})--(\ref{eq22}) and solved the corresponding generalized eigenvalue problem for both the circuit-estimated and L\"owdin-reference matrices. Table~\ref{tab:h4_energy_error} reports the ground-state energies and their absolute deviations. The resulting errors range from approximately 1.99 to 6.93~mHa over the geometries considered.

{
\begin{table}[t]

\small
\caption{Propagation of the reconstructed H$_4$ matrix-element errors to the final SCGVB ground-state energy. $E_0^{\rm est}$ is obtained from the circuit-estimated determinant matrices and $E_0^{\rm ref}$ from the L\"owdin-reference matrices, after projection onto the two spin-coupled structures and solution of the generalized eigenvalue problem.}
\label{tab:h4_energy_error}
\centering
\begin{tabular}{lrrr}
\hline
Geometry (\AA) & $E_0^{\rm est}$ (Ha) & $E_0^{\rm ref}$ (Ha) & $|\Delta E_0|$ (mHa) \\
\hline
$0.7414\times0.7414$ & $-2.41350956$ & $-2.41878324$ & $5.274$ \\
$0.8800\times0.7414$ & $-2.38689169$ & $-2.39128376$ & $4.392$ \\
$0.92675\times0.7414$ & $-2.39267599$ & $-2.39466770$ & $1.992$ \\
$1.2000\times0.7414$ & $-2.42991061$ & $-2.42297868$ & $6.932$ \\
$1.2600\times0.7414$ & $-2.43383632$ & $-2.42793949$ & $5.897$ \\
\hline
\end{tabular}
\end{table}
}

\paragraph{Hardware-like noise sensitivity and propagation to the SCGVB energy.}
To complement the finite-shot convergence analysis, we performed a representative hardware-like noise study for the rectangular H$_4$ geometry $0.8800\times0.7414$~\AA\ using Qiskit Aer density-matrix simulations.  The test used 8192 shots per circuit and compared the original unfiltered Pauli reconstruction with the algebraically equivalent vacuum-filtered estimator.  Single-qubit depolarizing noise with $p_{\rm gate}=10^{-3}$ was applied to the nontrivial one-qubit measurement operations, and symmetric independent readout-assignment error with $p_{\rm ro}=10^{-2}$ was applied at measurement.  No two-qubit noise was included because the present measurement circuits contain no entangling gates.  This is a simulator sensitivity analysis and not a hardware experiment.

Vacuum filtering produces a substantial circuit reduction for the representative pairs.  The overlap estimator decreases from 2704 final Pauli terms and 36 measurement circuits to 16 surviving terms measured in one circuit, while the Hamiltonian estimator decreases from approximately 11632--12080 final Pauli terms and 1297--1369 QWC measurement circuits to 176 surviving $I/Z$ terms measured in one circuit.  In the present noise model, this filtering removes the finite-shot contribution from analytically zero vacuum-expectation terms and also removes sensitivity to the modeled basis-rotation gate channel.  Readout error, however, remains because it directly perturbs the computational-basis outcomes.

\begin{table}[t]

\small
\caption{Hardware-like H$_4$ Hamiltonian-noise sensitivity for representative determinant pairs at $0.8800\times0.7414$~\AA.  Values are MAEs in Hartree over five repetitions with 8192 shots per circuit.  The gate-only case uses $p_{\rm gate}=10^{-3}$; cases containing readout noise use $p_{\rm ro}=10^{-2}$.}
\label{tab:h4_hw_noise_filtering}
\centering
\begin{tabular}{lcc}
\hline
Noise scenario & Unfiltered MAE & Vacuum-filtered MAE \\
\hline
Ideal finite-shot & $5.71\times10^{-3}$ & $<10^{-14}$ \\
Gate only & $7.50\times10^{-3}$ & $<10^{-14}$ \\
Readout only & $7.78\times10^{-2}$ & $7.55\times10^{-2}$ \\
Gate + readout & $7.66\times10^{-2}$ & $7.52\times10^{-2}$ \\
\hline
\end{tabular}
\end{table}

Because the filtered estimator is dominated by readout noise in this test, we next isolated the readout channel and varied $p_{\rm ro}=0,0.001,0.005,$ and $0.01$ using ten independent repetitions.  For representative matrix elements, the overlap MAE increases from $3.07\times10^{-3}$ at $p_{\rm ro}=0.001$ to $2.89\times10^{-2}$ at $p_{\rm ro}=0.01$, whereas the Hamiltonian MAE increases from $7.69\times10^{-3}$ to $7.57\times10^{-2}$~Ha over the same interval.  Descriptive linear fits over this low-noise range give $R^2=0.999847$ for $S$ and $R^2=0.999989$ for $H$; these fits are used only to summarize the tested interval and are not asserted as asymptotic scaling laws.

Finally, to quantify the effect on the derived SCGVB observable, all 21 independent determinant pairs were evaluated at each readout-error probability, the complete $6\times6$ determinant-space matrices were reconstructed, and the two-structure matrices were formed as $H_{\rm SC}=B^{\rm T}H_{\rm det}B$ and $S_{\rm SC}=B^{\rm T}S_{\rm det}B$ before solving $H_{\rm SC}c=ES_{\rm SC}c$.  The noiseless vacuum-filtered reference is $E_0=-2.391283762$~Ha in the electronic-Hamiltonian convention used for this H$_4$ benchmark.  The mean absolute energy deviations are 1.89, 13.41, and 27.21~mHa for $p_{\rm ro}=0.001$, 0.005, and 0.01, respectively.  The projected overlap matrix remains positive definite throughout the tested range, indicating that the energy degradation is caused by systematic readout bias in the reconstructed matrix elements rather than numerical collapse of the generalized eigenvalue problem.

\begin{table*}[t]

\small
\caption{Propagation of symmetric single-qubit readout error to the vacuum-filtered H$_4$ SCGVB ground-state energy at $0.8800\times0.7414$~\AA.  Complete $6\times6$ determinant matrices were reconstructed from all 21 independent pairs using 8192 shots per circuit and ten repetitions.}
\label{tab:h4_readout_energy}
\centering
\begin{tabular}{cccccc}
\hline
$p_{\rm ro}$ & $\langle E_0\rangle$ (Ha) & $\sigma(E_0)$ (Ha) & MAE($\Delta E_0$) (mHa) & Max.$|\Delta E_0|$ (mHa) & $\min\lambda(S_{\rm SC})$ \\
\hline
0       & -2.391283762 & 0 & 0.000 & 0.000 & 0.8158 \\
0.001   & -2.389395433 & $9.80\times10^{-4}$ & 1.888 & 3.603 & 0.8087 \\
0.005   & -2.377877297 & $3.13\times10^{-3}$ & 13.406 & 17.423 & 0.7821 \\
0.010   & -2.364077808 & $2.54\times10^{-3}$ & 27.206 & 31.462 & 0.7513 \\
\hline
\end{tabular}
\end{table*}

\paragraph{Scope of quantum advantage and comparison with Hadamard tests.}
The present algorithm is not intended to demonstrate an asymptotic quantum
advantage in the complexity-theoretic sense. Instead, its objective is to formulate the SCGVB Hamiltonian and overlap matrix-element problem in terms of shallow, ancilla-free measurement circuits and to quantify the resulting circuit, sampling, and classical-preprocessing requirements.
In particular, the proposed measurement-only estimators avoid the use of
ancilla qubits, controlled unitaries, and deep circuits that are characteristic of standard Hadamard-test and modified Hadamard-test approaches.
While Hadamard-based schemes can in principle evaluate the same matrix elements, they require circuit depths that scale with the length of Pauli strings and the use of multi-qubit controlled operations, making them poorly suited for NISQ-era hardware. By contrast, the present approach relies exclusively on single-qubit Clifford gates and projective measurements, yielding constant-depth circuits independent of system size. This reduces circuit-level depth, controlled-gate, and ancillary-qubit overhead relative to Hadamard-test-based constructions; however, it does not imply an overall computational advantage once the classical SCGVB algebra, Pauli expansion, and post-processing costs are included.

\paragraph{Quantum resource requirements.}
A detailed resource estimate for the Hamiltonian reconstruction is presented in Table~\ref{tab:resources}.
\begin{table}[h!]
\caption{Quantum resource summary for the evaluation of the SCGVB Hamiltonian 
matrix elements of square H$_4$.}
\begin{tabular}{l c}
\hline
\textbf{Quantity} & \textbf{Value} \\
\hline
Number of spatial orbitals & 4 \\
Number of spin-orbitals & 8 \\
Number of qubits & 8 \\
Total circuits (one-body Hamiltonian) & 42\,102 \\
Total circuits (two-body Hamiltonian) & 44\,804 \\
Total circuits (all contributions) & 86\,906 \\
Average circuit depth & 2 \\
Maximum circuit depth & 2 \\
Average gate count per circuit & 13--14 \\
Maximum gate count per circuit & 16 \\
Total H gates (one-body) & 120\,372 \\
Total U2 gates (one-body) & 120\,372 \\
Total measurements (one-body) & 336\,816 \\
Total H gates (two-body) & 125\,488 \\
Total U2 gates (two-body) & 125\,488 \\
Total measurements (two-body) & 358\,432 \\
Total simulation time (one-body) & 19\,558.46 s \\
Total simulation time (two-body) & 22\,756.78 s \\
Peak memory (tracemalloc) & 9.95 GB \\
Shots per circuit & 524\,288 \\
\hline
\end{tabular}
\label{tab:resources}
\end{table}
Each Pauli group is measured using shallow circuits of depth~2, containing only single-qubit gates (H and S$^\dagger$) and projective measurements.
No CNOT or other entangling gates are required, a consequence of the 
qubit-wise commuting grouping of Pauli operators.
In total, 86{,}906 circuits were executed, corresponding to 
42{,}102 one-body and 44{,}804 two-body contributions.
Across all circuits, the peak memory footprint of the simulation was approximately 9.95\,GB, and the total runtime was roughly 42{,}315\,s.
These data provide a realistic estimate of the classical resources required to simulate the measurement-based quantum algorithm for nonorthogonal SCGVB states, and establish a clear baseline for the scaling of the method beyond eight qubits.
\paragraph{Resource profile of the shallow measurement implementation}
A key conceptual aspect of the present approach is that the SCGVB wavefunction itself is never prepared on the quantum device. All nonorthogonality, spin coupling, and configuration interaction coefficients are handled classically, while the quantum register is used solely to realize the measurement circuits associated with determinant overlap and Hamiltonian matrix elements. As a result, the present circuits act only on the computational vacuum and never generate entanglement.

The shallow circuit depths reported in Table~\ref{tab:resources}, together with the absence of entangling gates, show that coherent-control depth is not the limiting resource in the present implementation. Instead, the dominant costs are the classical Pauli preprocessing, the number of determinant-pair tasks and measurement groups, and finite-shot sampling. This distinction is important when assessing the method as an end-to-end computational workflow.

\paragraph{Chirgwin--Coulson weights and physical interpretation}
Because the Chirgwin--Coulson (CC) weights depend explicitly on both the generalized eigenvector coefficients and the nonorthogonal overlap matrix, they provide a chemically meaningful diagnostic of the relative importance of the two spin-coupled structures.

{The Chirgwin--Coulson weights are evaluated using the full nonorthogonal definition in Eq.~(\ref{eq:cc_two_structure_full}).  The structural basis used for H$_4$ is not individually normalized, so retaining the $S_{11}$ and $S_{22}$ factors is essential.  Table~\ref{tab:H4_SCGVB} reports the L\"owdin-reference energies, generalized-eigenvector coefficients, and CC weights for the five geometries used in the manuscript.}

\begin{table}[h!]

\small
\caption{Reference SCGVB ground-state energy, coefficients, and Chirgwin--Coulson weights for the H$_4\rightarrow$H$_2$+H$_2$ dissociation path.  The weights are evaluated from the full nonorthogonal expression $W_i=C_i\sum_j S_{ij}C_j$ after solving the generalized eigenvalue problem with the L\"owdin-reference determinant matrices.}
\label{tab:H4_SCGVB}
\centering
\setlength{\tabcolsep}{3.2pt}
\begin{tabular}{c r r r r r}
\hline
Geometry (\AA) & $E_0$ (Ha) & $c_1$ & $c_2$ & $W_1$ & $W_2$ \\
\hline
$0.7414\times0.7414$  & -2.4187 & -0.3122 & 0.3122 & 0.5000 & 0.5000 \\
$0.8800\times0.7414$  & -2.3912 & -0.0887 & 0.4482 & 0.1202 & 0.8798 \\
$0.92675\times0.7414$ & -2.3946 & -0.0560 & 0.4518 & 0.0738 & 0.9262 \\
$1.2000\times0.7414$  & -2.4229 & -0.0018 & 0.4103 & 0.0024 & 0.9976 \\
$1.2600\times0.7414$  & -2.4279 &  0.0002 & 0.4013 & 0.0000 & 1.0000 \\
\hline
\end{tabular}
\end{table}

\begin{figure}[h!]
    \includegraphics[width=1\linewidth]{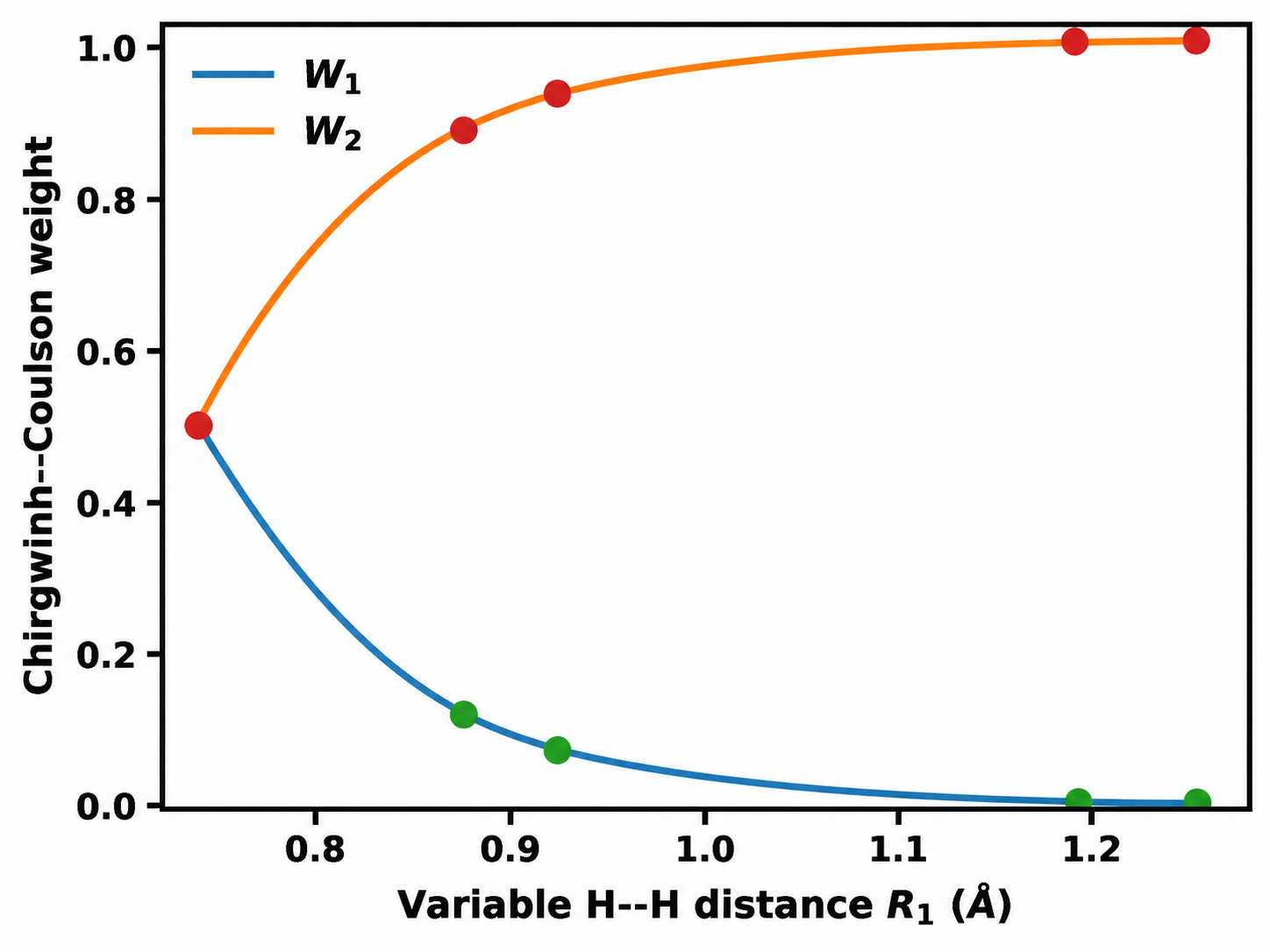}
    \caption{{Chirgwin--Coulson reference weights $W_1$ and $W_2$ for the two SCGVB structures of H$_4$ along the H$_4\rightarrow$H$_2$+H$_2$ dissociation coordinate, evaluated with the full nonorthogonal weight definition.}}
    \label{fig:H4_weights}
\end{figure}

{As shown in Table~\ref{tab:H4_SCGVB} and Fig.~\ref{fig:H4_weights}, the CC analysis preserves the expected chemical evolution: the two structures contribute equally at the square geometry, whereas the second structure rapidly becomes dominant as one H--H separation is elongated.  At the most elongated geometry reported here, the residual deviation of the CC weights from the limiting values 0 and 1 is at the level of the numerical post-processing precision; accordingly, the values are reported as $W_1=0.0000$ and $W_2=1.0000$.  The qualitative bonding assignment is consistent with the expected dissociation behavior.}

{The circuit-estimated determinant matrices are assessed independently through the direct matrix-element deviations and through the propagated ground-state-energy errors in Table~\ref{tab:h4_energy_error}.  Thus, the CC table provides the chemically interpretable L\"owdin-reference description of the dissociation path, while the estimator accuracy is quantified directly against that reference.}

\begin{table*}[t]
\small
\caption{Summary of absolute deviations between quantum-circuit and L\"{o}wdin-reference
Hamiltonian matrix elements for square and rectangular H$_4$ geometries.
Statistics are computed over all $36$ matrix elements:
Mean$\,|\Delta H|=\frac{1}{36}\sum_{ij}|\Delta H_{ij}|$,
RMS$\,|\Delta H|=\sqrt{\frac{1}{36}\sum_{ij}|\Delta H_{ij}|^2}$,
and Max$\,|\Delta H|=\max_{ij}|\Delta H_{ij}|$.}
\label{tab:hamiltonian_deviation_summary}
\begin{tabular}{|c|c|c|c|c|}
\hline
\textbf{Geometry (\AA)} &
\textbf{Mean $|\Delta H|$ (Ha)} &
\textbf{RMS $|\Delta H|$ (Ha)} &
\textbf{Max $|\Delta H|$ (Ha)} &
\textbf{Argmax $(i,j)$} \\
\hline
$0.7414\times0.7414$  & 0.0079 & 0.0115 & 0.03298 & (4,4) \\ \hline
$0.7414\times0.88$            & 0.0075 & 0.0109 & 0.03336 & (1,1) \\ \hline
$0.7414\times0.92675$         & 0.0066 & 0.0095 & 0.03111 & (4,4) \\ \hline
$0.7414\times1.20$            & 0.0055 & 0.0079 & 0.02338 & (3,3) \\ \hline
$0.7414\times1.26$            & 0.0060 & 0.0085 & 0.02739 & (4,4) \\ \hline
\end{tabular}
\end{table*}
Overall, Table \ref{tab:hamiltonian_deviation_summary} shows that the quantum-estimated Hamiltonian reproduces the classical L\"owdin reference with sub-$10^{-2}$ Hartree mean absolute error across all geometries considered. The largest deviations consistently arising
from the diagonal matrix elements corresponding to strongly interacting
configurations.

{The Chirgwin--Coulson weights provide the chemically interpretable reference for the evolution of the two spin-coupled structures, while the circuit-estimated matrices are tested directly through their element-wise deviations and the generalized-eigenvalue errors in Table~\ref{tab:h4_energy_error}.  This separation avoids attributing the reference CC analysis itself to finite-shot circuit data.}

Taken together, these results show that the proposed estimator provides a conceptually transparent, low-depth circuit formulation of nonorthogonal SCGVB matrix elements and reproduces the L\"owdin-rule references with quantified finite-shot errors, without implying an overall computational advantage.

Although the present work is motivated by near-term hardware considerations, the proposed measurement-only estimators are fully compatible with fault-tolerant quantum architectures, as they rely exclusively on single-qubit Clifford operations and projective measurements.

Taken together, the H$_4$ results demonstrate that the proposed
Measurement-only estimators reliably reproduce both the numerical and physical structure of the SCGVB problem across distinct bonding regimes, using only shallow, ancilla-free quantum circuits.

The implementation details for this system are provided in Appendices A--D of the Supporting Information.

\subsection{$\mathrm{C}_2$ determinant-space validation in the $X^1\Sigma_g^+$ state}

As an additional validation beyond the H$_4$ benchmark, we applied the same determinant-space estimator to the $X^1\Sigma_g^+$ state of C$_2$ at an internuclear distance of $R_{\mathrm{C-C}}=1.20$~\AA. The active-valence determinant space contains 70 primitive $M_S=0$ determinants in the fixed $N_\alpha=N_\beta=4$ sector in 16 spin orbitals, leading to a $70\times70$ overlap matrix and a corresponding determinant-space Hamiltonian matrix. The vacuum-estimator matrices reconstructed after the exact selection and simplification steps described above were compared directly against an independent determinant-space L\"owdin/(PySCF/Python) reference.

For the full 70-determinant C$_2$ space, the maximum absolute deviation is $6.66\times10^{-15}$ for the overlap matrix and $4.32\times10^{-11}$ Ha for the total Hamiltonian matrix. Thus, the circuit-sampled Pauli estimator reproduces the independent L\"owdin/PySCF determinant-space matrix elements to numerical precision. This confirms that the estimator algebra is not specific to the six-determinant H$_4$ case, but also applies to a substantially larger nonorthogonal determinant basis.

\begin{table}[h]
\centering
\caption{Direct comparison between the circuit-sampled Pauli estimator and the independent L\"owdin/PySCF determinant-space reference for the full 70-determinant C$_2$ space at $R_{\mathrm{C-C}}=1.20$~\AA. The reported deviation is $\Delta=O^{\rm circuit}-O^{\mathrm{L\text{"o}wdin/(PySCF/Python)}}$.}
\label{tab:c2_matrix_validation}
\begin{tabular}{lccc}
\hline
Matrix & $\max|\Delta|$ & $\langle |\Delta| \rangle$ & $\|\Delta\|_F$ \\
\hline
$S$ & $6.66\times10^{-15}$ & $7.92\times10^{-17}$ & $3.13\times10^{-14}$ \\
$H$ & $4.32\times10^{-11}$ & $1.93\times10^{-13}$ & $1.07\times10^{-10}$ \\
\hline
\end{tabular}
\end{table}

The small deviations in Table~\ref{tab:c2_matrix_validation} are at the level of numerical roundoff and confirm the correctness of the nonorthogonal Jordan--Wigner algebra, Pauli reconstruction, and determinant-space reference comparison. At the same time, the C$_2$ calculation exposes the main practical bottleneck of the present implementation: the direct symbolic expansion of the nonorthogonal Jordan--Wigner products generates a very large number of raw Pauli products before simplification and vacuum selection. Therefore, the C$_2$ benchmark should be interpreted as an algebraic and determinant-space validation, not as evidence of scalable quantum advantage.

\paragraph{Algebraic validation versus finite-shot measurement.}
The full filtered C$_2$ production calculation and the finite-shot C$_2$ test serve different purposes and should not be conflated. After the complete operator products are simplified, the exact vacuum-selection rule removes terms whose vacuum expectation value is known to vanish; consequently, the near-machine-precision agreement in Table~\ref{tab:c2_matrix_validation} primarily validates the nonorthogonal Pauli reconstruction, determinant-space bookkeeping, and classical preprocessing. By contrast, Sec.~\ref{sec:c2_unfiltered_shot_noise} retains representative raw products before final vacuum filtering and explicitly samples them at finite shot counts. Table~\ref{tab:c2_unfiltered_shot_scaling} therefore provides the stochastic measurement test, whereas Table~\ref{tab:c2_matrix_validation} provides the algebraic validation.

\subsection{Unfiltered shot-noise and resource analysis for C$_2$}
\label{sec:c2_unfiltered_shot_noise}

To assess the statistical behavior of the estimator before applying the final vacuum-filtering shortcut, we performed an additional unfiltered raw-product sampling analysis for selected C$_2$ determinant pairs. This test was designed as a representative shot-noise and resource estimate, rather than as the final production estimator for all $70\times70$ matrix elements. In this analysis, raw Pauli products were sampled before discarding terms with $X$ or $Y$ support, so that finite-shot measurement fluctuations appear explicitly.

Each matrix element is written as a weighted sum of Pauli expectation values,
\begin{equation}
O_{IJ}=\sum_{\alpha} c_{\alpha}^{(IJ)}\langle P_{\alpha}\rangle ,
\end{equation}
where $O_{IJ}$ denotes either an overlap or Hamiltonian matrix element. For a Pauli measurement estimated with $N_{\alpha}$ shots, the variance satisfies
\begin{equation}
{\rm Var}(\langle P_{\alpha}\rangle)
=
\frac{1-\langle P_{\alpha}\rangle^2}{N_{\alpha}} .
\end{equation}
Assuming independent measurement groups, the propagated variance of the reconstructed matrix element is
\begin{equation}
{\rm Var}(O_{IJ})
\approx
\sum_{\alpha}
|c_{\alpha}^{(IJ)}|^2
\frac{1-\langle P_{\alpha}\rangle^2}{N_{\alpha}} .
\end{equation}
Equivalently, the conservative upper bound
\begin{equation}
\sigma(O_{IJ})
\leq
\frac{\sum_{\alpha}|c_{\alpha}^{(IJ)}|}{\sqrt{N_{\rm shots}}}
\end{equation}
follows from ${\rm Var}(P_{\alpha})\leq 1$.

The unfiltered analysis was performed for 13 representative determinant pairs:
$(1,1)$, $(1,2)$, $(1,5)$, $(2,2)$, $(2,5)$, $(3,3)$, $(4,4)$,
$(5,5)$, $(10,10)$, $(20,20)$, $(35,35)$, $(50,50)$, and $(70,70)$.
For these pairs, we considered the observables $S$, $H_1$, and $H_2$.
For each case, $5.0\times10^4$ raw Pauli products were sampled with importance
probabilities proportional to the product of coefficient magnitudes.
The shot counts were varied as
\begin{equation}
N_{\rm shots}=1024,2048,4096,8192,16384,32768,65536,
\end{equation}
with 10 independent repetitions for each shot count.

The observed finite-shot component was quantified by comparing the measured raw-product estimator with the corresponding analytic sampled estimator. Table~\ref{tab:c2_unfiltered_shot_scaling} shows that the mean absolute difference decreases approximately as $1/\sqrt{N_{\rm shots}}$. A log--log fit gives slopes close to $-1/2$ for all three observables: approximately $-0.49$ for $S$, $-0.51$ for $H_1$, and $-0.49$ for $H_2$.

\begin{table}[h]
\centering
\caption{Unfiltered C$_2$ raw-product shot-noise analysis for representative determinant pairs. The reported quantity is the mean absolute difference between the measured sampled estimator and the corresponding analytic sampled estimator, averaged over the selected determinant pairs and 10 repetitions.}
\label{tab:c2_unfiltered_shot_scaling}
\begin{tabular}{cccc}
\hline
$N_{\rm shots}$ & $S$ & $H_1$ & $H_2$ \\
\hline
$1024$  & $1.60\times10^{-3}$ & $1.80\times10^{-1}$ & $2.87\times10^{-1}$ \\
$4096$  & $7.69\times10^{-4}$ & $1.05\times10^{-1}$ & $1.32\times10^{-1}$ \\
$16384$ & $4.83\times10^{-4}$ & $4.06\times10^{-2}$ & $7.37\times10^{-2}$ \\
$65536$ & $1.96\times10^{-4}$ & $2.30\times10^{-2}$ & $3.85\times10^{-2}$ \\
\hline
\end{tabular}
\end{table}

The resource statistics for the unfiltered raw-product sampling experiment are summarized in Table~\ref{tab:c2_unfiltered_resources}. Across all sampled cases, the analysis involved $1.365\times10^8$ sampled circuits and $2.536\times10^{12}$ total shots. The circuits remain extremely shallow, with average depth about 2.82 and no CNOT gates. Therefore, the dominant cost is not circuit depth, but the large number of raw Pauli products and sampled measurement tasks generated by the uncompressed nonorthogonal Jordan--Wigner expansion.

\begin{table}[h]
\centering
\caption{Aggregate resource statistics for the unfiltered C$_2$ raw-product shot-noise analysis. This calculation was used as a representative resource and statistical-convergence test, not as the final production estimator for the full $70\times70$ matrices.}
\label{tab:c2_unfiltered_resources}
\begin{tabular}{lc}
\hline
Quantity & Value \\
\hline
Number of qubits & $16$ \\
Representative determinant pairs & $13$ \\
Observables & $S$, $H_1$, $H_2$ \\
Raw samples per case & $5.0\times10^4$ \\
Shot counts & $1024$--$65536$ \\
Repetitions per shot count & $10$ \\
Total sampled circuits & $1.365\times10^8$ \\
Total shots & $2.536\times10^{12}$ \\
Wall time & $4.26\times10^4$ s \\
Peak memory & $9.58\times10^2$ MB \\
Average circuit depth & $2.82$ \\
Average gate count & $22.85$ \\
Total CNOT gates & $0$ \\
\hline
\end{tabular}
\end{table}

Finally, in the production estimator used for Table~\ref{tab:c2_matrix_validation}, the full operator products are simplified by the exact vacuum-selection rule
\begin{equation}
\langle 0|P|0\rangle = 0
\quad
\text{unless } P \text{ contains only } I \text{ and } Z .
\end{equation}
Thus, Pauli strings containing $X$ or $Y$ after the complete operator product has been formed are discarded before circuit execution. This vacuum filtering substantially reduces the measurement burden while preserving the exact vacuum expectation value. The unfiltered analysis above shows the expected shot-noise scaling when these zero-expectation raw products are not removed at the measurement stage, whereas the filtered production estimator is the one used for the final C$_2$ matrix validation.

For completeness and reproducibility, all details of the C$_2$ implementation are given in the Supporting Information. Appendix E of the Supporting Information reports the construction of the 70-determinant active-valence space, including the spin-orbital ordering, determinant labels, bitstring representation, and creation/annihilation strings. Appendix F provides the complete C$_2$ determinant-space overlap and Hamiltonian matrix elements used in the comparison with the L\"owdin rules implemented by PySCF/Python.

\section{Conclusion and Outlook}

We have presented a quantum measurement framework for evaluating overlap and Hamiltonian matrix elements in nonorthogonal SCGVB wavefunctions. By recasting the required quantities as vacuum expectation values of Pauli-string operators, the method replaces direct manipulation of nonorthogonal wavefunctions on hardware with ancilla-free, shallow measurement circuits built from local Clifford rotations and computational-basis readout. The present Clifford circuits are efficiently classically simulable; accordingly, the contribution is a circuit-compatible formulation and resource analysis of the SCGVB matrix-element layer rather than a demonstration of quantum computational advantage.

Tests on square and rectangular H$_4$ show that the quantum-estimated overlap and Hamiltonian matrices agree well with classical L\"owdin-based references across the geometries considered. {Using the full nonorthogonal Chirgwin--Coulson definition, the reference weights retain the expected dissociation trend, while the additional hardware-like simulations identify readout assignment error as the dominant residual noise source after vacuum filtering.} Together with the direct ground-state-energy errors in Table~\ref{tab:h4_energy_error} and the readout-noise propagation in Table~\ref{tab:h4_readout_energy}, these results establish the numerical behavior of the measurement formulation without relying on a claim of overall quantum advantage.

Beyond H$_4$, we also tested the estimator on a larger active-valence C$_2$ determinant-space benchmark in the $X^1\Sigma_g^+$ state at an internuclear distance of $R_{\mathrm{C-C}}=1.20$~\AA. This benchmark involves 16 spin orbitals and 70 primitive $M_S=0$ determinants in the fixed $N_\alpha=N_\beta=4$ sector, leading to full $70\times70$ determinant-space overlap and Hamiltonian matrices. In the original unoptimized implementation, this calculation exposed a severe symbolic-preprocessing bottleneck associated with the direct nonorthogonal Jordan--Wigner Pauli expansion. However, after parallelizing the determinant-pair workflow and introducing vacuum filtering to discard Pauli strings with zero vacuum expectation value before circuit execution, the full C$_2$ overlap and Hamiltonian matrix reconstruction was completed on a computing cluster in approximately 14 hours. The resulting circuit-sampled Pauli estimator reproduces the independent L\"owdin/PySCF determinant-space reference to numerical precision, with maximum absolute deviations of $6.66\times10^{-15}$ for the overlap matrix and $4.32\times10^{-11}$ Ha for the total Hamiltonian matrix.

At the same time, the scope of the present contribution should be stated clearly. The method does not establish an asymptotic quantum advantage, and the classical preprocessing associated with SCGVB algebra remains essential. Its present contribution is methodological: it quantifies the trade-off obtained by avoiding full nonorthogonal state preparation and controlled-interference circuits at the measurement layer, while explicitly retaining the associated classical preprocessing and measurement-resource costs.

The C$_2$ calculation therefore changes the interpretation of the scalability limitation. The estimator algebra is not restricted to the six-determinant H$_4$ example and remains valid for a substantially larger nonorthogonal determinant basis. Nevertheless, the calculation also confirms that scalability is controlled by the growth of the symbolic NO-JW Pauli expansion, determinant-pair count, and measurement-resource organization. Future implementations should therefore combine vacuum filtering with early Pauli-term compression, determinant-pair screening, integral screening, low-rank or density-fitting decompositions of the two-electron tensor, reuse of qubit-wise commuting measurement groups, adaptive shot allocation, and tensor-network formulations for compressed quantum circuit constructions. These strategies are necessary for extending the present estimator framework to larger molecular systems.

{
\subsection*{Outlook: coherent nonorthogonal state preparation}
The measurement-only formulation studied here was deliberately restricted to a shallow circuit layer. A conceptually different extension is to encode each normalized nonorthogonal determinant as a coherently prepared state
\begin{equation}
|\Phi_I\rangle = U_I |0\rangle^{\otimes n},
\end{equation}
where the state-preparation unitary $U_I$ represents the determinant amplitudes in an orthonormal computational basis. Transition matrix elements can then be accessed through an ancilla-assisted coherent superposition
\begin{equation}
|\Psi_{IJ}\rangle =
\frac{|0\rangle_a|\Phi_I\rangle+|1\rangle_a|\Phi_J\rangle}{\sqrt{2}},
\end{equation}
for which a Pauli operator $P$ obeys
\begin{align}
\mathrm{Re}\langle\Phi_I|P|\Phi_J\rangle
&=\langle X_a\otimes P\rangle,\\
\mathrm{Im}\langle\Phi_I|P|\Phi_J\rangle
&=\langle Y_a\otimes P\rangle.
\end{align}
Such a construction introduces coherent state preparation, ancilla-mediated interference, controlled operations, and entangling gates. Efficient preparation of general nonorthogonal SCGVB states is itself a separate algorithmic problem and is therefore outside the scope of the present low-depth measurement study. We are investigating this extension separately, including its resource requirements in a fault-tolerant-oriented setting. Importantly, the present results should not be interpreted as establishing the resource advantage of that future coherent formulation.
}

Within this restricted scope, the present results establish and benchmark the matrix-element measurement layer that such a broader coherent SCGVB workflow would need to replace or incorporate.

\section*{Acknowledgments}

The Italian Research Center on High Performance Computing, Big Data, and Quantum Computing (ICSC) is acknowledged for financial support under the Innovation Grant Project “Molecular Energy Landscapes by Quantum Computing.
The author gratefully acknowledges Professors Guido Raos, Piero Macchi, Mosè Casalegno, and Alessandro Genoni for the opportunity to be part of this project and for all that I have learned from them in quantum chemistry. A special thanks to Prof. Alessandro Mariani for the fruitful discussions.

\section*{Supporting Information}

\textbf{Supporting Information.} General nonorthogonal Jordan--Wigner mapping and determinant-operator definitions; H$_4$ spin-coupled structures and extended matrix-element data; C$_2$ active-valence determinant-space construction and benchmark matrices; resource estimates and statistical-convergence analyses.

\section*{Data and Software Availability}

The data underlying this study are provided in the manuscript and in the Supporting Information. Specifically, the Supporting Information includes the determinant definitions, spin-coupled SCGVB expansions, bitstring representations, creation and annihilation strings, the general nonorthogonal Jordan--Wigner mapping without redundant operator-by-operator listings, overlap and Hamiltonian matrix elements, resource estimates, statistical-convergence analyses, and the H$_4$ and C$_2$ benchmark data needed to reproduce the results reported in the manuscript.  Additional scripts used to generate and analyze the results are available from the author upon reasonable request.

\clearpage
\onecolumngrid

\begin{center}
\textbf{For Table of Contents Only}

\vspace{0.5cm}

\includegraphics[
    width=3.25in,
    height=1.75in,
    keepaspectratio
]{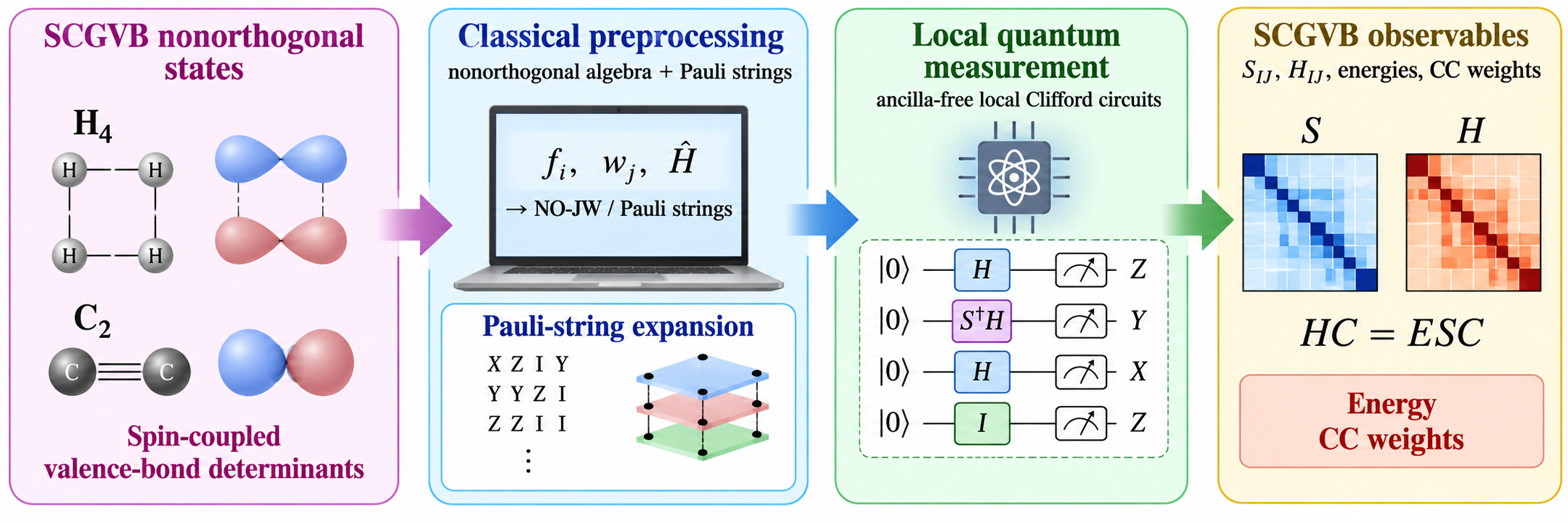}
\end{center}


\clearpage


\onecolumngrid
\begin{center}
{\Large \textbf{Supporting Information}}\\[1.2em]
{\large \textbf{Quantum Matrix-Element Estimators for Spin-Coupled Generalized Valence Bond (SCGVB) Wavefunctions}}\\[1.2em]
Bruna G. M. Ara\'{u}jo\\[0.6em]
Department of Chemistry, Materials and Chemical Engineering ``Giulio Natta'',\\
Politecnico di Milano, Via Bassini 6, 20133 Milano, Italy\\[0.6em]
Current address: Davidson School of Chemical Engineering, Purdue University, Forney Hall of Chemical Engineering, 480 Stadium Mall Drive, West Lafayette, IN 47907-2100, USA; and Instituto de Física, Universidade Federal da Bahia, Campus Universitário de Ondina, Salvador, 40210-340, BA, Brazil.\\[0.6em]
Email: bruna.gabriellyma@gmail.com
\end{center}

\vspace{1em}
This document contains additional implementation details, determinant definitions, extended H$_4$ and C$_2$ numerical data, resource estimates, and statistical-convergence analyses supporting the main manuscript.

\clearpage
\appendix
\setcounter{table}{0}
\setcounter{figure}{0}

\renewcommand{\thetable}{S\arabic{table}}
\renewcommand{\thefigure}{S\arabic{figure}}
\section{Implementation details of the nonorthogonal Jordan--Wigner encoding}
\label{implementation}
Here we implemented the NOJW mapping proposed by \cite{marruzzo2025nojw_SI} using Qiskit.
To assess the correctness of the proposed Jordan–Wigner encoding in the nonorthogonal case, we examined the mappings of the annihilation operators $\left\{\hat{\overline{a}}_p\right\}$. The implementation is conceptually straightforward. In our case, the challenge stands in the mapping of the $\left\{\hat{a}_p\right\}$ operators corresponding to nonorthogonal spin orbitals. To illustrate this concretely, below we present a Python routine snippet that uses the \texttt{SparsePauliOp} object from \texttt{Qiskit} to represent a nonorthogonal operator. 

\begin{lstlisting}[style=pythonstyle, caption={General NO-JW Mapping  }]
from qiskit.quantum_info import SparsePauliOp
import numpy as np

def nonorthogonal_annihilation_jw(p: int, S: np.ndarray) -> SparsePauliOp:
    """
    Returns the SparsePauliOp corresponding to the nonorthogonal JW mapping
    of the annihilation operator a_p, using the overlap matrix S.
    """
    M = S.shape[0]
    terms = []

    for q in range(M):
        coeff = S[p, q] / 2  # (X + iY)/2 has two parts
        pauli_x = ['I'] * M
        pauli_y = ['I'] * M
        for k in range(q):
            pauli_x[k] = 'Z'
            pauli_y[k] = 'Z'
        pauli_x[q] = 'X'
        pauli_y[q] = 'Y'
        # Qiskit Pauli labels are ordered as q_{M-1} ... q_0.
        label_x = ''.join(reversed(pauli_x))
        label_y = ''.join(reversed(pauli_y))
        # SparsePauliOp.from_list expects (Pauli label, coefficient).
        terms.append((label_x, coeff))
        terms.append((label_y, 1j * coeff))

    return SparsePauliOp.from_list(terms)
# ------------------------
#Example
# ------------------------
# Define a 4-orbital overlap matrix S (symmetric, real)
S = np.array([
    [1.0, 0.2, 0.1, 0.05],
    [0.2, 1.0, 0.3, 0.15],
    [0.1, 0.3, 1.0, 0.25],
    [0.05, 0.15, 0.25, 1.0]
])
# Choose the index p for the annihilation operator a_p
p = 2  # for example, a_2 (third orbital)
# Get SparsePauliOp corresponding to a_2 under the NO JW mapping
a2_operator = nonorthogonal_annihilation_jw(p, S)
# Print the output
print("Nonorthogonal JW mapping for annihilation operator a_2:\n")
for label, coeff in a2_operator.to_list():
    print(f"{coeff.real:.6f}{'+' if coeff.imag >= 0 else ''}{coeff.imag:.6f}j * {label}")
\end{lstlisting}
In particular, to obtain the nonorthogonal Jordan-Wigner (JW) mapping for the fermionic operators $\left\{\hat{a}_p\right\}$, we implemented a general function that takes as input the index \( p \) of the operator \( \hat{a}_p \) and the spin orbital overlap matrix $\mathbf{S}$. The mapping constructs the Pauli operator representation of \( \hat{a}_p \)  with $S_{pq}$ being the weight. In the example above, we defined a 4$\times$4 symmetric spin orbital overlap matrix and determined the mapped operator for \( \hat{a}_2 \). The output consists of the complete Pauli decomposition, expressed as a sum of complex-weighted Pauli strings, which can be directly used in quantum simulations of nonorthogonal wavefunctions. This example highlights the general applicability of the proposed encoding beyond system-specific implementations.

\section{Supplementary material about the implementation of SCGVB wavefunction for H$_4$ cluster}
\label{app:scgvb_expansions}
This section provides explicit operator-level details that are not required for understanding the conceptual workflow of the algorithm, but are included to ensure full reproducibility.
\subsection{Spin-coupled SCGVB structures in second quantization}
\label{app:scgvb_second_quant}

We provide explicit second-quantized expressions for the two spin-adapted SCGVB
structures used in the main text. Let $\ket{\phi_0} \equiv \ket{00000000}$ denote
the computational vacuum.

\begin{align}
\left|\psi_{0,0;1}^{4}\right\rangle &=
\left(
\hat{a}_{1}^{\dagger}\hat{a}_{3}^{\dagger}\hat{a}_{6}^{\dagger}\hat{a}_{8}^{\dagger}
-\hat{a}_{1}^{\dagger}\hat{a}_{4}^{\dagger}\hat{a}_{6}^{\dagger}\hat{a}_{7}^{\dagger}
-\hat{a}_{2}^{\dagger}\hat{a}_{3}^{\dagger}\hat{a}_{5}^{\dagger}\hat{a}_{8}^{\dagger}
+\hat{a}_{2}^{\dagger}\hat{a}_{4}^{\dagger}\hat{a}_{5}^{\dagger}\hat{a}_{7}^{\dagger}
\right)\ket{\phi_0},
\label{eq:psi001}
\\[2mm]
\left|\psi_{0,0;2}^{4}\right\rangle &=
\left(
\hat{a}_{1}^{\dagger}\hat{a}_{2}^{\dagger}\hat{a}_{7}^{\dagger}\hat{a}_{8}^{\dagger}
-\hat{a}_{1}^{\dagger}\hat{a}_{3}^{\dagger}\hat{a}_{6}^{\dagger}\hat{a}_{8}^{\dagger}
-\hat{a}_{2}^{\dagger}\hat{a}_{4}^{\dagger}\hat{a}_{5}^{\dagger}\hat{a}_{7}^{\dagger}
+\hat{a}_{3}^{\dagger}\hat{a}_{4}^{\dagger}\hat{a}_{5}^{\dagger}\hat{a}_{6}^{\dagger}
\right)\ket{\phi_0}.
\label{eq:psi002}
\end{align}

The SCGVB wavefunction is
\begin{equation}
\ket{\psi_{\mathrm{SCGVB}}} = c_1 \ket{\psi_{0,0;1}^{4}} + c_2 \ket{\psi_{0,0;2}^{4}}.
\label{eq:psi_scgvb}
\end{equation}

\subsection{Computational-basis (bitstring) representation}
\label{app:scgvb_bitstrings}

Using the occupation-number encoding in the main text, the same states can be written as
\begin{align}
\ket{\psi_{0,0;1}^{4}} &= \left(\ket{10100101} - \ket{10010110} - \ket{01101001} + \ket{01011010}\right),
\label{eq:psi001_bits}
\\
\ket{\psi_{0,0;2}^{4}} &= \left(\ket{11000011} - \ket{10100101} - \ket{01011010} + \ket{00111100}\right).
\label{eq:psi002_bits}
\end{align}

\subsection{Determinant labels}
\label{app:det_labels}

For compactness we define the six determinants:
\begin{align}
\ket{\psi_{1}} &= \ket{10100101}, &
\ket{\psi_{2}} &= \ket{10010110}, &
\ket{\psi_{3}} &= \ket{01101001}, \nonumber\\
\ket{\psi_{4}} &= \ket{01011010}, &
\ket{\psi_{5}} &= \ket{11000011}, &
\ket{\psi_{6}} &= \ket{00111100}.
\label{eq:psi1to6}
\end{align}
In this notation,
\begin{align}
\ket{\psi_{0,0;1}^{4}} &= \left(\ket{\psi_1} - \ket{\psi_2} - \ket{\psi_3} + \ket{\psi_4}\right), \\
\ket{\psi_{0,0;2}^{4}} &= \left(\ket{\psi_5} - \ket{\psi_1} - \ket{\psi_4} + \ket{\psi_6}\right).
\label{eq:psi_combo}
\end{align}

\section{Reduction of overlaps and Hamiltonian elements to vacuum expectation values}
\label{app:vacuum_reduction}

\subsection{Creation and annihilation strings}
\label{app:fandw}

Define creation strings $f_k$ and corresponding annihilation strings $w_k$ by
\begin{align}
f_{1}&=\hat{a}_{1}^{\dagger}\hat{a}_{3}^{\dagger}\hat{a}_{6}^{\dagger}\hat{a}_{8}^{\dagger}, &
w_{1}&=\hat{a}_{8}\hat{a}_{6}\hat{a}_{3}\hat{a}_{1}, \nonumber\\
f_{2}&=\hat{a}_{1}^{\dagger}\hat{a}_{4}^{\dagger}\hat{a}_{6}^{\dagger}\hat{a}_{7}^{\dagger}, &
w_{2}&=\hat{a}_{7}\hat{a}_{6}\hat{a}_{4}\hat{a}_{1}, \nonumber\\
f_{3}&=\hat{a}_{2}^{\dagger}\hat{a}_{3}^{\dagger}\hat{a}_{5}^{\dagger}\hat{a}_{8}^{\dagger}, &
w_{3}&=\hat{a}_{8}\hat{a}_{5}\hat{a}_{3}\hat{a}_{2}, \nonumber\\
f_{4}&=\hat{a}_{2}^{\dagger}\hat{a}_{4}^{\dagger}\hat{a}_{5}^{\dagger}\hat{a}_{7}^{\dagger}, &
w_{4}&=\hat{a}_{7}\hat{a}_{5}\hat{a}_{4}\hat{a}_{2}, \nonumber\\
f_{5}&=\hat{a}_{1}^{\dagger}\hat{a}_{2}^{\dagger}\hat{a}_{7}^{\dagger}\hat{a}_{8}^{\dagger}, &
w_{5}&=\hat{a}_{8}\hat{a}_{7}\hat{a}_{2}\hat{a}_{1}, \nonumber\\
f_{6}&=\hat{a}_{3}^{\dagger}\hat{a}_{4}^{\dagger}\hat{a}_{5}^{\dagger}\hat{a}_{6}^{\dagger}, &
w_{6}&=\hat{a}_{6}\hat{a}_{5}\hat{a}_{4}\hat{a}_{3}.
\label{eq:fandw}
\end{align}

Then each determinant is $\ket{\psi_k}=f_k\ket{\phi_0}$ and $\bra{\psi_k}=\bra{\phi_0}w_k$.

\subsection{Overlap matrix elements}
\label{app:overlaps_vacuum}

All overlaps reduce to
\begin{equation}
S_{ij}=\braket{\psi_i|\psi_j}=\bra{\phi_0} w_i f_j \ket{\phi_0}.
\label{eq:Sij_vacuum}
\end{equation}
This form is the starting point for the measurement-only overlap estimator used in the main text.

\subsection{Hamiltonian matrix elements}
\label{app:hij_vacuum}

Similarly,
\begin{equation}
H_{ij}=\bra{\psi_i}\hat{H}\ket{\psi_j}=\bra{\phi_0} w_i \hat{H} f_j \ket{\phi_0},
\label{eq:Hij_vacuum}
\end{equation}
which is evaluated by expressing $w_i \hat{H} f_j$ as a weighted sum of Pauli strings under the
(nonorthogonal) Jordan--Wigner mapping described in Section ~\ref{implementation}.

\subsection{NO-JW mapping of annihilation operators}
\label{app:nojw_annihilation}

For nonorthogonal spin-orbitals, the annihilation operator admits the expansion
\begin{equation}
\hat{a}_{p} = \sum_{q=1}^{N} S_{pq}\left(\prod_{k=1}^{q-1} Z_k\right)\frac{X_q+iY_q}{2},
\label{eq:nojw_general}
\end{equation}
where $S_{pq}$ are overlap matrix elements between spin-orbitals.

Equation~\eqref{eq:nojw_general} fully defines the mapping, so an explicit operator-by-operator listing of $\hat a_1,\ldots,\hat a_8$ is unnecessary.


\section{Matrix elements of Hamiltonian and overlap of the determinants of several H$_4$ geometries}
\label{H4_outputs}

Here we present all matrix elements of the Hamiltonian and overlap of the determinants for $H_{4}$ geometries used to obtain the CC weights presented in the Results Section of the main manuscript. All information is contained in Tables \ref{tab:deviations_rectangular}, \ref{tab:overlap_rect_unique}, \ref{tab:hamiltonian_rect_07414_092675}, \ref{tab:overlap_rect_07414_092675}, \ref{tab:hamiltonian_rect_07414_12}, \ref{tab:overlap_rect_07414_12}, \ref{tab:hamiltonian_rect_07414_126}, and \ref{tab:overlap_rect_071414_126}.

\begin{table}[t]
\small
\caption{Hamiltonian matrix elements (in Hartree) between nonorthogonal SCGVB Slater determinants for rectangular $H_{4}$ (0.88 $\times$ 0.7414 \AA). Columns show the quantum-estimated matrix elements (PGHE quantum circuit), the classical L\"{o}wdin reference, and the absolute deviation $|\Delta H_{ij}|$. Only unique elements ($i \le j$) are reported since the Hamiltonian matrix is Hermitian.}
\label{tab:deviations_rectangular}
\vspace{2mm}
\centering
\begin{threeparttable}
\begin{tabular}{|l|c|c|c|}
\hline
$\mathbf{H_{ij}}$ & \textbf{Quantum Circuit} & \textbf{L\"{o}wdin reference} & $\mathbf{\left|\Delta H_{ij}\right|}$ \\
\hline

$\langle \psi_1 | H | \psi_1 \rangle$ & -1.5541859494 & -1.5841446083 & 0.0299586589 \\ \hline
$\langle \psi_1 | H | \psi_2 \rangle$ & 0.2226076322 & 0.2190002768 & 0.0036073554 \\ \hline
$\langle \psi_1 | H | \psi_3 \rangle$ & 0.2250376771 & 0.2190002768 & 0.0060374003 \\ \hline
$\langle \psi_1 | H | \psi_4 \rangle$ & -0.0233166916 & -0.0233296753 & 0.0000129837 \\ \hline
$\langle \psi_1 | H | \psi_5 \rangle$ & 0.4748015342 & 0.4695619606 & 0.0052395736 \\ \hline
$\langle \psi_1 | H | \psi_6 \rangle$ & 0.4747209865 & 0.4695619606 & 0.0051590259 \\ \hline

$\langle \psi_2 | H | \psi_2 \rangle$ & -0.5933011243 & -0.5846331188 & 0.0086680055 \\ \hline
$\langle \psi_2 | H | \psi_3 \rangle$ & -0.0762366273 & -0.0738186226 & 0.0024180047 \\ \hline
$\langle \psi_2 | H | \psi_4 \rangle$ & 0.2118894648 & 0.2190002768 & 0.0071108120 \\ \hline
$\langle \psi_2 | H | \psi_5 \rangle$ & -0.0054926232 & -0.0049493105 & 0.0005433127 \\ \hline
$\langle \psi_2 | H | \psi_6 \rangle$ & -0.0041797147 & -0.0049493105 & 0.0007695958 \\ \hline

$\langle \psi_3 | H | \psi_3 \rangle$ & -0.6081644331 & -0.5846331188 & 0.0235313143 \\ \hline
$\langle \psi_3 | H | \psi_4 \rangle$ & 0.2125972017 & 0.2190002768 & 0.0064030751 \\ \hline
$\langle \psi_3 | H | \psi_5 \rangle$ & -0.0053082375 & -0.0049493105 & 0.0003589270 \\ \hline
$\langle \psi_3 | H | \psi_6 \rangle$ & -0.0037096120 & -0.0049493105 & 0.0012396985 \\ \hline

$\langle \psi_4 | H | \psi_4 \rangle$ & -1.5508408075 & -1.5841446083 & 0.0333038008 \\ \hline
$\langle \psi_4 | H | \psi_5 \rangle$ & 0.4719640968 & 0.4695619606 & 0.0024021362 \\ \hline
$\langle \psi_4 | H | \psi_6 \rangle$ & 0.4728043214 & 0.4695619606 & 0.0032423608 \\ \hline

$\langle \psi_5 | H | \psi_5 \rangle$ & -0.9475455136 & -0.9390349435 & 0.0085105701 \\ \hline
$\langle \psi_5 | H | \psi_6 \rangle$ & -0.2243404939 & -0.2205401655 & 0.0038003284 \\ \hline

$\langle \psi_6 | H | \psi_6 \rangle$ & -0.9494356410 & -0.9390349435 & 0.0104006975 \\ \hline

\end{tabular}
\end{threeparttable}
\end{table}

\begin{table}
\caption{Overlap matrix elements between the six SCGVB determinants for rectangular H$_4$.
The column ``QA (DOE)'' lists the values obtained from the quantum circuit, while
``L\"{o}wdin rules'' contains the classical reference. Absolute discrepancies are
$|\Delta S_{ij}| = |S_{ij}^{\mathrm{QA}}-S_{ij}^{\mathrm{ref}}|$. Only unique elements
$(i\leq j)$ are reported since the overlap matrix is symmetric.}
\label{tab:overlap_rect_unique}
\centering
\begin{threeparttable}
\begin{tabular}{|l|c|c|c|}
\hline
$\mathbf{S_{ij}}$ & \textbf{QA (DOE)} & \textbf{L\"{o}wdin rules} & $\mathbf{|\Delta S_{ij}|}$ \\
\hline

$\langle \psi_1|\psi_1\rangle$ & 0.6887973612 & 0.6887973612 & $0$ \\ \hline
$\langle \psi_1|\psi_2\rangle$ & -0.0887852187 & -0.0887852187 & $0$ \\ \hline
$\langle \psi_1|\psi_3\rangle$ & -0.0887852187 & -0.0887852187 & $0$ \\ \hline
$\langle \psi_1|\psi_4\rangle$ & 0.0120235015 & 0.0120235015 & $0$ \\ \hline
$\langle \psi_1|\psi_5\rangle$ & -0.1797893730 & -0.1797893730 & $0$ \\ \hline
$\langle \psi_1|\psi_6\rangle$ & -0.1797893730 & -0.1797893730 & $0$ \\ \hline

$\langle \psi_2|\psi_2\rangle$ & 0.3201019318 & 0.3201019318 & $0$ \\ \hline
$\langle \psi_2|\psi_3\rangle$ & 0.0238736260 & 0.0238736260 & $0$ \\ \hline
$\langle \psi_2|\psi_4\rangle$ & -0.0887852187 & -0.0887852187 & $0$ \\ \hline
$\langle \psi_2|\psi_5\rangle$ & -0.0013667206 & -0.0013667206 & $0$ \\ \hline
$\langle \psi_2|\psi_6\rangle$ & -0.0013667206 & -0.0013667206 & $0$ \\ \hline

$\langle \psi_3|\psi_3\rangle$ & 0.3201019318 & 0.3201019318 & $0$ \\ \hline
$\langle \psi_3|\psi_4\rangle$ & -0.0887852187 & -0.0887852187 & $0$ \\ \hline
$\langle \psi_3|\psi_5\rangle$ & -0.0013667206 & -0.0013667206 & $0$ \\ \hline
$\langle \psi_3|\psi_6\rangle$ & -0.0013667206 & -0.0013667206 & $0$ \\ \hline

$\langle \psi_4|\psi_4\rangle$ & 0.6887973612 & 0.6887973612 & $0$ \\ \hline
$\langle \psi_4|\psi_5\rangle$ & -0.1797893730 & -0.1797893730 & $0$ \\ \hline
$\langle \psi_4|\psi_6\rangle$ & -0.1797893730 & -0.1797893730 & $0$ \\ \hline

$\langle \psi_5|\psi_5\rangle$ & 0.4562019590 & 0.4562019590 & $0$ \\ \hline
$\langle \psi_5|\psi_6\rangle$ & 0.0697819075 & 0.0697819075 & $0$ \\ \hline

$\langle \psi_6|\psi_6\rangle$ & 0.4562019590 & 0.4562019590 & $0$ \\ \hline

\end{tabular}
\end{threeparttable}
\end{table}

\begin{table}[t]
\small
\caption{Hamiltonian matrix elements (in Hartree) between nonorthogonal SCGVB Slater determinants.
Columns show the quantum-estimated matrix elements (PGHE quantum circuit), the classical L\"{o}wdin reference,
and the absolute deviation $|\Delta H_{ij}|$.
The target system is a rectangular H$_4$ molecule ($0.7414 \times 0.92675$ \AA). Only unique elements ($i \le j$) are reported since the Hamiltonian matrix is Hermitian.}
\label{tab:hamiltonian_rect_07414_092675}
\vspace{2mm}
\centering
\begin{threeparttable}
\begin{tabular}{|l|c|c|c|}
\hline
$\mathbf{H_{ij}}$ & \textbf{Quantum Circuit} & \textbf{L\"{o}wdin reference} & $\mathbf{\left|\Delta H_{ij}\right|}$ \\
\hline

$\langle \psi_1|H|\psi_1\rangle$ & -1.6035884013 & -1.6296920868 & 0.0261036855 \\ \hline
$\langle \psi_1|H|\psi_2\rangle$ & 0.1861899770 & 0.1922546039 & 0.0060646269 \\ \hline
$\langle \psi_1|H|\psi_3\rangle$ & 0.2024651090 & 0.1922546039 & 0.0102105051 \\ \hline
$\langle \psi_1|H|\psi_4\rangle$ & -0.0503718171 & -0.0500258244 & 0.0003459927 \\ \hline
$\langle \psi_1|H|\psi_5\rangle$ & 0.5200942739 & 0.5202726591 & 0.0001783852 \\ \hline
$\langle \psi_1|H|\psi_6\rangle$ & 0.5296038729 & 0.5202726591 & 0.0093312138 \\ \hline

$\langle \psi_2|H|\psi_2\rangle$ & -0.5832003218 & -0.5722451592 & 0.0109551626 \\ \hline
$\langle \psi_2|H|\psi_3\rangle$ & -0.0601484921 & -0.0576903182 & 0.0024581739 \\ \hline
$\langle \psi_2|H|\psi_4\rangle$ & 0.1857931916 & 0.1922546039 & 0.0064614123 \\ \hline
$\langle \psi_2|H|\psi_5\rangle$ & -0.0068580786 & -0.0046185579 & 0.0022395207 \\ \hline
$\langle \psi_2|H|\psi_6\rangle$ & -0.0032248990 & -0.0046185579 & 0.0013936589 \\ \hline

$\langle \psi_3|H|\psi_3\rangle$ & -0.5933468267 & -0.5722451592 & 0.0211016675 \\ \hline
$\langle \psi_3|H|\psi_4\rangle$ & 0.1858430859 & 0.1922546039 & 0.0064115180 \\ \hline
$\langle \psi_3|H|\psi_5\rangle$ & -0.0029459046 & -0.0046185579 & 0.0016726533 \\ \hline
$\langle \psi_3|H|\psi_6\rangle$ & -0.0066381811 & -0.0046185579 & 0.0020196232 \\ \hline

$\langle \psi_4|H|\psi_4\rangle$ & -1.5985847263 & -1.6296920868 & 0.0311073605 \\ \hline
$\langle \psi_4|H|\psi_5\rangle$ & 0.5249557894 & 0.5202726591 & 0.0046831303 \\ \hline
$\langle \psi_4|H|\psi_6\rangle$ & 0.5252628050 & 0.5202726591 & 0.0049901459 \\ \hline

$\langle \psi_5|H|\psi_5\rangle$ & -1.0517394094 & -1.0404865434 & 0.0112528660 \\ \hline
$\langle \psi_5|H|\psi_6\rangle$ & -0.2456620275 & -0.2454850444 & 0.0001769831 \\ \hline

$\langle \psi_6|H|\psi_6\rangle$ & -1.0460970146 & -1.0404865434 & 0.0056104712 \\ \hline

\end{tabular}
\end{threeparttable}
\end{table}

\begin{table}[t]
\small
\caption{Overlap matrix elements between the six SCGVB determinants for rectangular H$_4$
($0.7414 \times 0.92675$ \AA). The column ``QA (DOE)'' lists the values obtained from the quantum
circuit, while ``L\"{o}wdin rules'' contains the classical reference.
Absolute discrepancies are $|\Delta S_{ij}| = |S^{\mathrm{QA}}_{ij} - S^{\mathrm{ref}}_{ij}|$.
Only unique elements ($i \le j$) are reported since the overlap matrix is symmetric.}
\label{tab:overlap_rect_07414_092675}
\centering
\begin{threeparttable}
\begin{tabular}{|l|c|c|c|}
\hline
$\mathbf{S_{ij}}$ & \textbf{QA (DOE)} & \textbf{L\"{o}wdin rules} & $\mathbf{|\Delta S_{ij}|}$ \\
\hline
$\langle \psi_1|\psi_1\rangle$ & 0.7140258214 & 0.7140258214 & $0$ \\ \hline
$\langle \psi_1|\psi_2\rangle$ & -0.0790672997 & -0.0790672997 & $0$ \\ \hline
$\langle \psi_1|\psi_3\rangle$ & -0.0790672997 & -0.0790672997 & $0$ \\ \hline
$\langle \psi_1|\psi_4\rangle$ & 0.0201509724 & 0.0201509724 & $0$ \\ \hline
$\langle \psi_1|\psi_5\rangle$ & -0.1990186006 & -0.1990186006 & $0$ \\ \hline
$\langle \psi_1|\psi_6\rangle$ & -0.1990186006 & -0.1990186006 & $0$ \\ \hline

$\langle \psi_2|\psi_2\rangle$ & 0.3201019318 & 0.3201019318 & $0$ \\ \hline
$\langle \psi_2|\psi_3\rangle$ & 0.0188432867 & 0.0188432867 & $0$ \\ \hline
$\langle \psi_2|\psi_4\rangle$ & -0.0790672997 & -0.0790672997 & $0$ \\ \hline
$\langle \psi_2|\psi_5\rangle$ & -0.0014028130 & -0.0014028130 & $0$ \\ \hline
$\langle \psi_2|\psi_6\rangle$ & -0.0014028130 & -0.0014028130 & $0$ \\ \hline

$\langle \psi_3|\psi_3\rangle$ & 0.3201019318 & 0.3201019318 & $0$ \\ \hline
$\langle \psi_3|\psi_4\rangle$ & -0.0790672997 & -0.0790672997 & $0$ \\ \hline
$\langle \psi_3|\psi_5\rangle$ & -0.0014028130 & -0.0014028130 & $0$ \\ \hline
$\langle \psi_3|\psi_6\rangle$ & -0.0014028130 & -0.0014028130 & $0$ \\ \hline

$\langle \psi_4|\psi_4\rangle$ & 0.7140258214 & 0.7140258214 & $0$ \\ \hline
$\langle \psi_4|\psi_5\rangle$ & -0.1990186006 & -0.1990186006 & $0$ \\ \hline
$\langle \psi_4|\psi_6\rangle$ & -0.1990186006 & -0.1990186006 & $0$ \\ \hline

$\langle \psi_5|\psi_5\rangle$ & 0.5008812352 & 0.5008812352 & $0$ \\ \hline
$\langle \psi_5|\psi_6\rangle$ & 0.0779665852 & 0.0779665852 & $0$ \\ \hline

$\langle \psi_6|\psi_6\rangle$ & 0.5008812352 & 0.5008812352 & $0$ \\ \hline
\end{tabular}
\end{threeparttable}
\end{table}

\begin{table}[t]
\small
\caption{Hamiltonian matrix elements (in Hartree) between nonorthogonal SCGVB Slater determinants.
Columns show the quantum-estimated matrix elements (PGHE quantum circuit), the classical L\"{o}wdin reference,
and the absolute deviation $|\Delta H_{ij}|$.
The target system is a rectangular H$_4$ molecule ($0.7414 \times 1.2$ \AA).
Only unique elements ($i \le j$) are reported since the Hamiltonian matrix is Hermitian.}
\label{tab:hamiltonian_rect_07414_12}
\vspace{2mm}
\centering
\begin{threeparttable}
\begin{tabular}{|l|c|c|c|}
\hline
$\mathbf{H_{ij}}$ & \textbf{Quantum Circuit} & \textbf{L\"{o}wdin reference} & $\mathbf{\left|\Delta H_{ij}\right|}$ \\
\hline

$\langle \psi_1|H|\psi_1\rangle$ & -1.8464182798 & -1.8640455684 & 0.0176272886 \\ \hline
$\langle \psi_1|H|\psi_2\rangle$ & 0.0936360199 & 0.0869394156 & 0.0066966043 \\ \hline
$\langle \psi_1|H|\psi_3\rangle$ & 0.0880244225 & 0.0869394156 & 0.0010850069 \\ \hline
$\langle \psi_1|H|\psi_4\rangle$ & -0.2430388722 & -0.2437304082 & 0.0006915360 \\ \hline
$\langle \psi_1|H|\psi_5\rangle$ & 0.7815945022 & 0.7817148523 & 0.0001203501 \\ \hline
$\langle \psi_1|H|\psi_6\rangle$ & 0.7914416464 & 0.7817148523 & 0.0097267941 \\ \hline

$\langle \psi_2|H|\psi_2\rangle$ & -0.5357262078 & -0.5268539625 & 0.0088722453 \\ \hline
$\langle \psi_2|H|\psi_3\rangle$ & -0.0124061816 & -0.0123280523 & 0.0000781293 \\ \hline
$\langle \psi_2|H|\psi_4\rangle$ & 0.0848025242 & 0.0869394156 & 0.0021368914 \\ \hline
$\langle \psi_2|H|\psi_5\rangle$ & -0.0048022157 & -0.0025821286 & 0.0022200871 \\ \hline
$\langle \psi_2|H|\psi_6\rangle$ & -0.0024764124 & -0.0025821286 & 0.0001057162 \\ \hline

$\langle \psi_3|H|\psi_3\rangle$ & -0.5502374058 & -0.5268539625 & 0.0233834433 \\ \hline
$\langle \psi_3|H|\psi_4\rangle$ & 0.0843623245 & 0.0869394156 & 0.0025770911 \\ \hline
$\langle \psi_3|H|\psi_5\rangle$ & -0.0012726857 & -0.0025821286 & 0.0013094429 \\ \hline
$\langle \psi_3|H|\psi_6\rangle$ & -0.0050026172 & -0.0025821286 & 0.0024204886 \\ \hline

$\langle \psi_4|H|\psi_4\rangle$ & -1.8455586554 & -1.8640455684 & 0.0184869130 \\ \hline
$\langle \psi_4|H|\psi_5\rangle$ & 0.7954009424 & 0.7817148523 & 0.0136860901 \\ \hline
$\langle \psi_4|H|\psi_6\rangle$ & 0.7864479610 & 0.7817148523 & 0.0047331087 \\ \hline

$\langle \psi_5|H|\psi_5\rangle$ & -1.5580916078 & -1.5506245415 & 0.0074670663 \\ \hline
$\langle \psi_5|H|\psi_6\rangle$ & -0.3809163008 & -0.3781083466 & 0.0028079542 \\ \hline

$\langle \psi_6|H|\psi_6\rangle$ & -1.5593865408 & -1.5506245415 & 0.0087619993 \\ \hline

\end{tabular}
\end{threeparttable}
\end{table}

\begin{table}[t]
\small
\caption{Overlap matrix elements between the six SCGVB determinants for rectangular H$_4$
($0.7414 \times 1.2$ \AA). The column ``QA (DOE)'' lists the values obtained from the quantum
circuit, while ``L\"{o}wdin rules'' contains the classical reference.
Absolute discrepancies are $|\Delta S_{ij}| = |S^{\mathrm{QA}}_{ij} - S^{\mathrm{ref}}_{ij}|$.
Only unique elements ($i \le j$) are reported since the overlap matrix is symmetric.}
\label{tab:overlap_rect_07414_12}
\centering
\begin{threeparttable}
\begin{tabular}{|l|c|c|c|}
\hline
$\mathbf{S_{ij}}$ & \textbf{QA (DOE)} & \textbf{L\"{o}wdin rules} & $\mathbf{|\Delta S_{ij}|}$ \\
\hline
$\langle \psi_1|\psi_1\rangle$ & 0.8373995335 & 0.8373995335 & $0$ \\ \hline
$\langle \psi_1|\psi_2\rangle$ & -0.0380339632 & -0.0380339632 & $0$ \\ \hline
$\langle \psi_1|\psi_3\rangle$ & -0.0380339632 & -0.0380339632 & $0$ \\ \hline
$\langle \psi_1|\psi_4\rangle$ & 0.0809005960 & 0.0809005960 & $0$ \\ \hline
$\langle \psi_1|\psi_5\rangle$ & -0.2983148141 & -0.2983148141 & $0$ \\ \hline
$\langle \psi_1|\psi_6\rangle$ & -0.2983148141 & -0.2983148141 & $0$ \\ \hline

$\langle \psi_2|\psi_2\rangle$ & 0.3201019318 & 0.3201019318 & $0$ \\ \hline
$\langle \psi_2|\psi_3\rangle$ & 0.0042106542 & 0.0042106542 & $0$ \\ \hline
$\langle \psi_2|\psi_4\rangle$ & -0.0380339632 & -0.0380339632 & $0$ \\ \hline
$\langle \psi_2|\psi_5\rangle$ & -0.0013210424 & -0.0013210424 & $0$ \\ \hline
$\langle \psi_2|\psi_6\rangle$ & -0.0013210424 & -0.0013210424 & $0$ \\ \hline

$\langle \psi_3|\psi_3\rangle$ & 0.3201019318 & 0.3201019318 & $0$ \\ \hline
$\langle \psi_3|\psi_4\rangle$ & -0.0380339632 & -0.0380339632 & $0$ \\ \hline
$\langle \psi_3|\psi_5\rangle$ & -0.0013210424 & -0.0013210424 & $0$ \\ \hline
$\langle \psi_3|\psi_6\rangle$ & -0.0013210424 & -0.0013210424 & $0$ \\ \hline

$\langle \psi_4|\psi_4\rangle$ & 0.8373995335 & 0.8373995335 & $0$ \\ \hline
$\langle \psi_4|\psi_5\rangle$ & -0.2983148141 & -0.2983148141 & $0$ \\ \hline
$\langle \psi_4|\psi_6\rangle$ & -0.2983148141 & -0.2983148141 & $0$ \\ \hline

$\langle \psi_5|\psi_5\rangle$ & 0.7228499269 & 0.7228499269 & $0$ \\ \hline
$\langle \psi_5|\psi_6\rangle$ & 0.1220243610 & 0.1220243610 & $0$ \\ \hline

$\langle \psi_6|\psi_6\rangle$ & 0.7228499269 & 0.7228499269 & $0$ \\ \hline
\end{tabular}
\end{threeparttable}
\end{table}

\begin{table}[t!]
\small
\caption{Hamiltonian matrix elements (in Hartree) between nonorthogonal SCGVB Slater determinants.
Columns show the quantum-estimated matrix elements (PGHE quantum circuit),
the classical L\"{o}wdin reference, and the absolute deviation $|\Delta H_{ij}|$.
The target system is a rectangular H$_4$ molecule ($0.7414 \times 1.26$ \AA).
Only unique elements ($i \le j$) are reported since the Hamiltonian matrix is Hermitian.}
\label{tab:hamiltonian_rect_07414_126}
\vspace{2mm}
\centering
\begin{threeparttable}
\begin{tabular}{|l|c|c|c|}
\hline
$\mathbf{H_{ij}}$ & \textbf{Quantum Circuit} & \textbf{L\"{o}wdin reference} & $\mathbf{\left|\Delta H_{ij}\right|}$ \\
\hline

$\langle \psi_1|H|\psi_1\rangle$ & -1.8942763210 & -1.9057425327 & 0.0114662117 \\ \hline
$\langle \psi_1|H|\psi_2\rangle$ & 0.0709018507 & 0.0725345274 & 0.0016326767 \\ \hline
$\langle \psi_1|H|\psi_3\rangle$ & 0.0771095406 & 0.0725345274 & 0.0045750132 \\ \hline
$\langle \psi_1|H|\psi_4\rangle$ & -0.2844217969 & -0.2840928654 & 0.0003289315 \\ \hline
$\langle \psi_1|H|\psi_5\rangle$ & 0.8311658625 & 0.8278481529 & 0.0033177096 \\ \hline
$\langle \psi_1|H|\psi_6\rangle$ & 0.8342336179 & 0.8278481529 & 0.0063854650 \\ \hline

$\langle \psi_2|H|\psi_2\rangle$ & -0.5289858414 & -0.5211398372 & 0.0078460042 \\ \hline
$\langle \psi_2|H|\psi_3\rangle$ & -0.0091121107 & -0.0086104169 & 0.0005016938 \\ \hline
$\langle \psi_2|H|\psi_4\rangle$ & 0.0693655891 & 0.0725345274 & 0.0031689383 \\ \hline
$\langle \psi_2|H|\psi_5\rangle$ & -0.0031278476 & -0.0021944191 & 0.0009334285 \\ \hline
$\langle \psi_2|H|\psi_6\rangle$ & -0.0028232805 & -0.0021944191 & 0.0006288614 \\ \hline

$\langle \psi_3|H|\psi_3\rangle$ & -0.5450522159 & -0.5211398372 & 0.0239123787 \\ \hline
$\langle \psi_3|H|\psi_4\rangle$ & 0.0698525691 & 0.0725345274 & 0.0026819583 \\ \hline
$\langle \psi_3|H|\psi_5\rangle$ & -0.0014004555 & -0.0021944191 & 0.0007939636 \\ \hline
$\langle \psi_3|H|\psi_6\rangle$ & -0.0009935743 & -0.0021944191 & 0.0012008448 \\ \hline

$\langle \psi_4|H|\psi_4\rangle$ & -1.8783532902 & -1.9057425327 & 0.0273892425 \\ \hline
$\langle \psi_4|H|\psi_5\rangle$ & 0.8387496303 & 0.8278481529 & 0.0109014774 \\ \hline
$\langle \psi_4|H|\psi_6\rangle$ & 0.8363918543 & 0.8278481529 & 0.0085437014 \\ \hline

$\langle \psi_5|H|\psi_5\rangle$ & -1.6470579099 & -1.6381039770 & 0.0089539329 \\ \hline
$\langle \psi_5|H|\psi_6\rangle$ & -0.3984998909 & -0.4022735280 & 0.0037736371 \\ \hline

$\langle \psi_6|H|\psi_6\rangle$ & -1.6519666078 & -1.6381039770 & 0.0138626308 \\ \hline

\end{tabular}
\end{threeparttable}
\end{table}

\begin{table}[t]
\small
\caption{Overlap matrix elements between the six SCGVB determinants for rectangular H$_4$
($0.7414 \times 1.26$ \AA). The column ``QA (DOE)'' lists the values obtained from the quantum
circuit, while ``L\"{o}wdin rules'' contains the classical reference.
Absolute discrepancies are $|\Delta S_{ij}| = |S^{\mathrm{QA}}_{ij} - S^{\mathrm{ref}}_{ij}|$.
Only unique elements ($i \le j$) are reported since the overlap matrix is symmetric.}
\label{tab:overlap_rect_071414_126}
\centering
\begin{threeparttable}
\begin{tabular}{|l|c|c|c|}
\hline
$\mathbf{S_{ij}}$ & \textbf{QA (DOE)} & \textbf{L\"{o}wdin rules} & $\mathbf{|\Delta S_{ij}|}$ \\
\hline
$\langle \psi_1|\psi_1\rangle$ & 0.8584263458 & 0.8584263458 & $0$ \\ \hline
$\langle \psi_1|\psi_2\rangle$ & -0.0320637598 & -0.0320637598 & $0$ \\ \hline
$\langle \psi_1|\psi_3\rangle$ & -0.0320637598 & -0.0320637598 & $0$ \\ \hline
$\langle \psi_1|\psi_4\rangle$ & 0.0938058736 & 0.0938058736 & $0$ \\ \hline
$\langle \psi_1|\psi_5\rangle$ & -0.3158337956 & -0.3158337956 & $0$ \\ \hline
$\langle \psi_1|\psi_6\rangle$ & -0.3158337956 & -0.3158337956 & $0$ \\ \hline

$\langle \psi_2|\psi_2\rangle$ & 0.3201019318 & 0.3201019318 & $0$ \\ \hline
$\langle \psi_2|\psi_3\rangle$ & 0.0029658965 & 0.0029658965 & $0$ \\ \hline
$\langle \psi_2|\psi_4\rangle$ & -0.0320637598 & -0.0320637598 & $0$ \\ \hline
$\langle \psi_2|\psi_5\rangle$ & -0.0012516001 & -0.0012516001 & $0$ \\ \hline
$\langle \psi_2|\psi_6\rangle$ & -0.0012516001 & -0.0012516001 & $0$ \\ \hline

$\langle \psi_3|\psi_3\rangle$ & 0.3201019318 & 0.3201019318 & $0$ \\ \hline
$\langle \psi_3|\psi_4\rangle$ & -0.0320637598 & -0.0320637598 & $0$ \\ \hline
$\langle \psi_3|\psi_5\rangle$ & -0.0012516001 & -0.0012516001 & $0$ \\ \hline
$\langle \psi_3|\psi_6\rangle$ & -0.0012516001 & -0.0012516001 & $0$ \\ \hline

$\langle \psi_4|\psi_4\rangle$ & 0.8584263458 & 0.8584263458 & $0$ \\ \hline
$\langle \psi_4|\psi_5\rangle$ & -0.3158337956 & -0.3158337956 & $0$ \\ \hline
$\langle \psi_4|\psi_6\rangle$ & -0.3158337956 & -0.3158337956 & $0$ \\ \hline

$\langle \psi_5|\psi_5\rangle$ & 0.7604763634 & 0.7604763633 & $1.0\times10^{-10}$ \\ \hline
$\langle \psi_5|\psi_6\rangle$ & 0.1301315366 & 0.1301315367 & $1.0\times10^{-10}$ \\ \hline

$\langle \psi_6|\psi_6\rangle$ & 0.7604763634 & 0.7604763633 & $1.0\times10^{-10}$ \\ \hline
\end{tabular}
\end{threeparttable}
\end{table}

\FloatBarrier
\clearpage

\section{SCGVB wavefunction of C$_2$ in the X$^1\Sigma_g^+$ State}
\label{C2-SCGVB_wf}

To make the preceding formalism concrete for a larger molecule, we consider the
C$_2$ molecule in its singlet ground state, X$^1\Sigma_g^+$. The label
X$^1\Sigma_g^+$ denotes the ground electronic state of this symmetry: the
superscript ``1'' indicates total spin $S=0$, $\Sigma$ indicates zero projection
of the electronic orbital angular momentum along the molecular axis, the
subscript $g$ denotes gerade inversion symmetry, and the $+$ label denotes
reflection symmetry with respect to a plane containing the molecular axis.

The bonding pattern of C$_2$ is known to be unusually subtle, and
spin-coupled/generalized-valence-bond analyses have shown that it cannot be
described satisfactorily by a single classical bonding picture
\cite{XuDunning_C2_GVB_SI,cooper2015C2_SI,RaosGerrattCooperRaimondi1993_spinbases_SI}.
In the full spin-coupled description of the X$^1\Sigma_g^+$ state, the eight
valence electrons are accommodated in eight generally nonorthogonal
spin-coupled orbitals,
\begin{equation}
  \phi_1,\phi_2,\phi_3,\phi_4,\phi_5,\phi_6,\phi_7,\phi_8.
\end{equation}
The two carbon core pairs may be treated as doubly occupied inactive orbitals.
Since the present quantum estimator acts on the active determinant space, we
write explicitly below the active-valence part of the wavefunction.

For $N=8$ active electrons and $S=0$, the number of linearly independent spin
eigenfunctions is
\begin{equation}
  f_0^8
  =
  \frac{(2S+1)N!}
  {\left(\frac{N}{2}+S+1\right)!\left(\frac{N}{2}-S\right)!}
  =
  \frac{8!}{5!\,4!}
  =
  14.
  \label{eq:C2_f08}
\end{equation}
Therefore, in direct analogy with the H$_4$ singlet case, the active SCGVB
wavefunction of C$_2$ can be written as
\begin{eqnarray}
  \Psi_{\mathrm{SCGVB}}^{\mathrm{act}}
  &=&
  \mathcal{N}
  \Bigl[
    c_{0,1}\psi^8_{0,0;1}
    +c_{0,2}\psi^8_{0,0;2}
    +\cdots
    +c_{0,14}\psi^8_{0,0;14}
  \Bigr]
  \nonumber\\
  &=&
  \mathcal{N}
  \sum_{k=1}^{14}
  c_{0,k}\,
  \psi^8_{0,0;k},
  \label{eq:C2_total_SCGVB}
\end{eqnarray}
where $\mathcal{N}$ is a normalization factor, $c_{0,k}$ are spin-coupling
coefficients, and the fourteen spin-coupled branches are
\begin{equation}
  \psi^8_{0,0;k}
  =
  \hat A
  \Bigl(
    \phi_1\,\phi_2\,\phi_3\,\phi_4\,
    \phi_5\,\phi_6\,\phi_7\,\phi_8\,
    \Theta^8_{0,0;k}
  \Bigr),
  \qquad k=1,\ldots,14.
  \label{eq:C2_branch_def}
\end{equation}
Here $\Theta^8_{0,0;k}$ are the fourteen independent eight-electron singlet spin
eigenfunctions. We choose a Rumer-type singlet-pair basis, which spans the same
$S=0$ spin space and is convenient for writing the branches explicitly. Each
branch is defined by four singlet-coupled pairs. For a pair $(p,q)$ we define
\begin{equation}
  [pq]
  =
  \frac{1}{\sqrt{2}}
  \left(
    \alpha_p\beta_q-\beta_p\alpha_q
  \right),
  \label{eq:C2_singlet_pair}
\end{equation}
where $\alpha_i\equiv\alpha(\sigma_i)$ and
$\beta_i\equiv\beta(\sigma_i)$. The fourteen Rumer branches for eight electrons
are then
\begin{eqnarray}
\Theta^8_{0,0;1}  &=& [12][34][56][78],
\nonumber\\
\Theta^8_{0,0;2}  &=& [12][34][58][67],
\nonumber\\
\Theta^8_{0,0;3}  &=& [12][36][45][78],
\nonumber\\
\Theta^8_{0,0;4}  &=& [12][38][45][67],
\nonumber\\
\Theta^8_{0,0;5}  &=& [12][38][47][56],
\nonumber\\
\Theta^8_{0,0;6}  &=& [14][23][56][78],
\nonumber\\
\Theta^8_{0,0;7}  &=& [14][23][58][67],
\nonumber\\
\Theta^8_{0,0;8}  &=& [16][23][45][78],
\nonumber\\
\Theta^8_{0,0;9}  &=& [16][25][34][78],
\nonumber\\
\Theta^8_{0,0;10} &=& [18][23][45][67],
\nonumber\\
\Theta^8_{0,0;11} &=& [18][23][47][56],
\nonumber\\
\Theta^8_{0,0;12} &=& [18][25][34][67],
\nonumber\\
\Theta^8_{0,0;13} &=& [18][27][34][56],
\nonumber\\
\Theta^8_{0,0;14} &=& [18][27][36][45].
\label{eq:C2_14_Rumer_branches}
\end{eqnarray}
Equations~(\ref{eq:C2_singlet_pair})--(\ref{eq:C2_14_Rumer_branches}) give the
explicit spin-coupling branches entering the C$_2$ singlet SCGVB wavefunction.
The first branch is the perfect-pairing branch in the chosen orbital ordering,
while the remaining branches correspond to the other independent singlet
recouplings of eight spin-$1/2$ electrons.

Expanding each product of four singlet pairs in primitive spin functions and
applying the antisymmetrizer gives the determinant expansion of each branch.
Using the shorthand
\begin{equation}
  \phi_i(\mathbf{x})=\phi_i(\mathbf{r})\,\alpha(\sigma),
  \qquad
  \overline{\phi}_i(\mathbf{x})=\phi_i(\mathbf{r})\,\beta(\sigma),
\end{equation}
each branch contains $2^4=16$ determinant terms before the different branches
are combined. To write all branches compactly, let the $k$th Rumer branch be
defined by the four pairs
\begin{equation}
  \mathcal{P}_k
  =
  \bigl\{
  (p_{k1},q_{k1}),
  (p_{k2},q_{k2}),
  (p_{k3},q_{k3}),
  (p_{k4},q_{k4})
  \bigr\}.
\end{equation}
For four binary variables
$\boldsymbol{\epsilon}=(\epsilon_1,\epsilon_2,\epsilon_3,\epsilon_4)$, with
$\epsilon_r=0,1$, define the spin orbital assigned to pair
$(p_{kr},q_{kr})$ as
\begin{equation}
  Y_{p_{kr}}(\epsilon_r)
  =
  \begin{cases}
    \phi_{p_{kr}}, & \epsilon_r=0,\\
    \overline{\phi}_{p_{kr}}, & \epsilon_r=1,
  \end{cases}
  \qquad
  Y_{q_{kr}}(\epsilon_r)
  =
  \begin{cases}
    \overline{\phi}_{q_{kr}}, & \epsilon_r=0,\\
    \phi_{q_{kr}}, & \epsilon_r=1.
  \end{cases}
  \label{eq:C2_pair_spinorbital_assignment}
\end{equation}
Then the determinant associated with branch $k$ and binary string
$\boldsymbol{\epsilon}$ is
\begin{equation}
  \left|
  \Omega_k(\boldsymbol{\epsilon})
  \right|
  =
  \left|
  Y_{p_{k1}}(\epsilon_1)\,
  Y_{q_{k1}}(\epsilon_1)\,
  Y_{p_{k2}}(\epsilon_2)\,
  Y_{q_{k2}}(\epsilon_2)\,
  Y_{p_{k3}}(\epsilon_3)\,
  Y_{q_{k3}}(\epsilon_3)\,
  Y_{p_{k4}}(\epsilon_4)\,
  Y_{q_{k4}}(\epsilon_4)
  \right|.
  \label{eq:C2_branch_determinant_general}
\end{equation}
With this convention, the explicit determinant expansion of the $k$th branch is
\begin{equation}
  \psi^8_{0,0;k}
  =
  \frac{1}{4\sqrt{8!}}
  \sum_{\epsilon_1,\epsilon_2,\epsilon_3,\epsilon_4=0}^{1}
  (-1)^{\epsilon_1+\epsilon_2+\epsilon_3+\epsilon_4}
  \left|
  \Omega_k(\epsilon_1,\epsilon_2,\epsilon_3,\epsilon_4)
  \right|,
  \qquad k=1,\ldots,14.
  \label{eq:C2_each_branch_explicit}
\end{equation}
Equation~(\ref{eq:C2_each_branch_explicit}), together with the fourteen pairing
patterns in Eq.~(\ref{eq:C2_14_Rumer_branches}), gives a complete explicit
definition of all determinant terms in the C$_2$ SCGVB branches.

The explicit determinant expansions are listed below.

\begin{eqnarray}
  \psi^8_{0,0;1}
  &=&
  \frac{1}{4\sqrt{8!}}
  \Bigl[
  \left|\phi_1\overline{\phi}_2\phi_3\overline{\phi}_4\phi_5\overline{\phi}_6\phi_7\overline{\phi}_8\right|
  - \left|\phi_1\overline{\phi}_2\phi_3\overline{\phi}_4\phi_5\overline{\phi}_6\overline{\phi}_7\phi_8\right|\nonumber\\
  &&- \left|\phi_1\overline{\phi}_2\phi_3\overline{\phi}_4\overline{\phi}_5\phi_6\phi_7\overline{\phi}_8\right|
  + \left|\phi_1\overline{\phi}_2\phi_3\overline{\phi}_4\overline{\phi}_5\phi_6\overline{\phi}_7\phi_8\right|\nonumber\\
  &&- \left|\phi_1\overline{\phi}_2\overline{\phi}_3\phi_4\phi_5\overline{\phi}_6\phi_7\overline{\phi}_8\right|
  + \left|\phi_1\overline{\phi}_2\overline{\phi}_3\phi_4\phi_5\overline{\phi}_6\overline{\phi}_7\phi_8\right|\nonumber\\
  &&+ \left|\phi_1\overline{\phi}_2\overline{\phi}_3\phi_4\overline{\phi}_5\phi_6\phi_7\overline{\phi}_8\right|
  - \left|\phi_1\overline{\phi}_2\overline{\phi}_3\phi_4\overline{\phi}_5\phi_6\overline{\phi}_7\phi_8\right|\nonumber\\
  &&- \left|\overline{\phi}_1\phi_2\phi_3\overline{\phi}_4\phi_5\overline{\phi}_6\phi_7\overline{\phi}_8\right|
  + \left|\overline{\phi}_1\phi_2\phi_3\overline{\phi}_4\phi_5\overline{\phi}_6\overline{\phi}_7\phi_8\right|\nonumber\\
  &&+ \left|\overline{\phi}_1\phi_2\phi_3\overline{\phi}_4\overline{\phi}_5\phi_6\phi_7\overline{\phi}_8\right|
  - \left|\overline{\phi}_1\phi_2\phi_3\overline{\phi}_4\overline{\phi}_5\phi_6\overline{\phi}_7\phi_8\right|\nonumber\\
  &&+ \left|\overline{\phi}_1\phi_2\overline{\phi}_3\phi_4\phi_5\overline{\phi}_6\phi_7\overline{\phi}_8\right|
  - \left|\overline{\phi}_1\phi_2\overline{\phi}_3\phi_4\phi_5\overline{\phi}_6\overline{\phi}_7\phi_8\right|\nonumber\\
  &&- \left|\overline{\phi}_1\phi_2\overline{\phi}_3\phi_4\overline{\phi}_5\phi_6\phi_7\overline{\phi}_8\right|
  + \left|\overline{\phi}_1\phi_2\overline{\phi}_3\phi_4\overline{\phi}_5\phi_6\overline{\phi}_7\phi_8\right|
  \Bigr].
  \label{eq:C2_branch1_explicit}
\end{eqnarray}

\begin{eqnarray}
  \psi^8_{0,0;2}
  &=&
  \frac{1}{4\sqrt{8!}}
  \Bigl[
  \left|\phi_1\overline{\phi}_2\phi_3\overline{\phi}_4\phi_5\phi_6\overline{\phi}_7\overline{\phi}_8\right|
  - \left|\phi_1\overline{\phi}_2\phi_3\overline{\phi}_4\phi_5\overline{\phi}_6\phi_7\overline{\phi}_8\right|\nonumber\\
  &&- \left|\phi_1\overline{\phi}_2\phi_3\overline{\phi}_4\overline{\phi}_5\phi_6\overline{\phi}_7\phi_8\right|
  + \left|\phi_1\overline{\phi}_2\phi_3\overline{\phi}_4\overline{\phi}_5\overline{\phi}_6\phi_7\phi_8\right|\nonumber\\
  &&- \left|\phi_1\overline{\phi}_2\overline{\phi}_3\phi_4\phi_5\phi_6\overline{\phi}_7\overline{\phi}_8\right|
  + \left|\phi_1\overline{\phi}_2\overline{\phi}_3\phi_4\phi_5\overline{\phi}_6\phi_7\overline{\phi}_8\right|\nonumber\\
  &&+ \left|\phi_1\overline{\phi}_2\overline{\phi}_3\phi_4\overline{\phi}_5\phi_6\overline{\phi}_7\phi_8\right|
  - \left|\phi_1\overline{\phi}_2\overline{\phi}_3\phi_4\overline{\phi}_5\overline{\phi}_6\phi_7\phi_8\right|\nonumber\\
  &&- \left|\overline{\phi}_1\phi_2\phi_3\overline{\phi}_4\phi_5\phi_6\overline{\phi}_7\overline{\phi}_8\right|
  + \left|\overline{\phi}_1\phi_2\phi_3\overline{\phi}_4\phi_5\overline{\phi}_6\phi_7\overline{\phi}_8\right|\nonumber\\
  &&+ \left|\overline{\phi}_1\phi_2\phi_3\overline{\phi}_4\overline{\phi}_5\phi_6\overline{\phi}_7\phi_8\right|
  - \left|\overline{\phi}_1\phi_2\phi_3\overline{\phi}_4\overline{\phi}_5\overline{\phi}_6\phi_7\phi_8\right|\nonumber\\
  &&+ \left|\overline{\phi}_1\phi_2\overline{\phi}_3\phi_4\phi_5\phi_6\overline{\phi}_7\overline{\phi}_8\right|
  - \left|\overline{\phi}_1\phi_2\overline{\phi}_3\phi_4\phi_5\overline{\phi}_6\phi_7\overline{\phi}_8\right|\nonumber\\
  &&- \left|\overline{\phi}_1\phi_2\overline{\phi}_3\phi_4\overline{\phi}_5\phi_6\overline{\phi}_7\phi_8\right|
  + \left|\overline{\phi}_1\phi_2\overline{\phi}_3\phi_4\overline{\phi}_5\overline{\phi}_6\phi_7\phi_8\right|
  \Bigr].
  \label{eq:C2_branch2_explicit}
\end{eqnarray}

\begin{eqnarray}
  \psi^8_{0,0;3}
  &=&
  \frac{1}{4\sqrt{8!}}
  \Bigl[
  \left|\phi_1\overline{\phi}_2\phi_3\phi_4\overline{\phi}_5\overline{\phi}_6\phi_7\overline{\phi}_8\right|
  - \left|\phi_1\overline{\phi}_2\phi_3\phi_4\overline{\phi}_5\overline{\phi}_6\overline{\phi}_7\phi_8\right|\nonumber\\
  &&- \left|\phi_1\overline{\phi}_2\phi_3\overline{\phi}_4\phi_5\overline{\phi}_6\phi_7\overline{\phi}_8\right|
  + \left|\phi_1\overline{\phi}_2\phi_3\overline{\phi}_4\phi_5\overline{\phi}_6\overline{\phi}_7\phi_8\right|\nonumber\\
  &&- \left|\phi_1\overline{\phi}_2\overline{\phi}_3\phi_4\overline{\phi}_5\phi_6\phi_7\overline{\phi}_8\right|
  + \left|\phi_1\overline{\phi}_2\overline{\phi}_3\phi_4\overline{\phi}_5\phi_6\overline{\phi}_7\phi_8\right|\nonumber\\
  &&+ \left|\phi_1\overline{\phi}_2\overline{\phi}_3\overline{\phi}_4\phi_5\phi_6\phi_7\overline{\phi}_8\right|
  - \left|\phi_1\overline{\phi}_2\overline{\phi}_3\overline{\phi}_4\phi_5\phi_6\overline{\phi}_7\phi_8\right|\nonumber\\
  &&- \left|\overline{\phi}_1\phi_2\phi_3\phi_4\overline{\phi}_5\overline{\phi}_6\phi_7\overline{\phi}_8\right|
  + \left|\overline{\phi}_1\phi_2\phi_3\phi_4\overline{\phi}_5\overline{\phi}_6\overline{\phi}_7\phi_8\right|\nonumber\\
  &&+ \left|\overline{\phi}_1\phi_2\phi_3\overline{\phi}_4\phi_5\overline{\phi}_6\phi_7\overline{\phi}_8\right|
  - \left|\overline{\phi}_1\phi_2\phi_3\overline{\phi}_4\phi_5\overline{\phi}_6\overline{\phi}_7\phi_8\right|\nonumber\\
  &&+ \left|\overline{\phi}_1\phi_2\overline{\phi}_3\phi_4\overline{\phi}_5\phi_6\phi_7\overline{\phi}_8\right|
  - \left|\overline{\phi}_1\phi_2\overline{\phi}_3\phi_4\overline{\phi}_5\phi_6\overline{\phi}_7\phi_8\right|\nonumber\\
  &&- \left|\overline{\phi}_1\phi_2\overline{\phi}_3\overline{\phi}_4\phi_5\phi_6\phi_7\overline{\phi}_8\right|
  + \left|\overline{\phi}_1\phi_2\overline{\phi}_3\overline{\phi}_4\phi_5\phi_6\overline{\phi}_7\phi_8\right|
  \Bigr].
  \label{eq:C2_branch3_explicit}
\end{eqnarray}

\begin{eqnarray}
  \psi^8_{0,0;4}
  &=&
  \frac{1}{4\sqrt{8!}}
  \Bigl[
  \left|\phi_1\overline{\phi}_2\phi_3\phi_4\overline{\phi}_5\phi_6\overline{\phi}_7\overline{\phi}_8\right|
  - \left|\phi_1\overline{\phi}_2\phi_3\phi_4\overline{\phi}_5\overline{\phi}_6\phi_7\overline{\phi}_8\right|\nonumber\\
  &&- \left|\phi_1\overline{\phi}_2\phi_3\overline{\phi}_4\phi_5\phi_6\overline{\phi}_7\overline{\phi}_8\right|
  + \left|\phi_1\overline{\phi}_2\phi_3\overline{\phi}_4\phi_5\overline{\phi}_6\phi_7\overline{\phi}_8\right|\nonumber\\
  &&- \left|\phi_1\overline{\phi}_2\overline{\phi}_3\phi_4\overline{\phi}_5\phi_6\overline{\phi}_7\phi_8\right|
  + \left|\phi_1\overline{\phi}_2\overline{\phi}_3\phi_4\overline{\phi}_5\overline{\phi}_6\phi_7\phi_8\right|\nonumber\\
  &&+ \left|\phi_1\overline{\phi}_2\overline{\phi}_3\overline{\phi}_4\phi_5\phi_6\overline{\phi}_7\phi_8\right|
  - \left|\phi_1\overline{\phi}_2\overline{\phi}_3\overline{\phi}_4\phi_5\overline{\phi}_6\phi_7\phi_8\right|\nonumber\\
  &&- \left|\overline{\phi}_1\phi_2\phi_3\phi_4\overline{\phi}_5\phi_6\overline{\phi}_7\overline{\phi}_8\right|
  + \left|\overline{\phi}_1\phi_2\phi_3\phi_4\overline{\phi}_5\overline{\phi}_6\phi_7\overline{\phi}_8\right|\nonumber\\
  &&+ \left|\overline{\phi}_1\phi_2\phi_3\overline{\phi}_4\phi_5\phi_6\overline{\phi}_7\overline{\phi}_8\right|
  - \left|\overline{\phi}_1\phi_2\phi_3\overline{\phi}_4\phi_5\overline{\phi}_6\phi_7\overline{\phi}_8\right|\nonumber\\
  &&+ \left|\overline{\phi}_1\phi_2\overline{\phi}_3\phi_4\overline{\phi}_5\phi_6\overline{\phi}_7\phi_8\right|
  - \left|\overline{\phi}_1\phi_2\overline{\phi}_3\phi_4\overline{\phi}_5\overline{\phi}_6\phi_7\phi_8\right|\nonumber\\
  &&- \left|\overline{\phi}_1\phi_2\overline{\phi}_3\overline{\phi}_4\phi_5\phi_6\overline{\phi}_7\phi_8\right|
  + \left|\overline{\phi}_1\phi_2\overline{\phi}_3\overline{\phi}_4\phi_5\overline{\phi}_6\phi_7\phi_8\right|
  \Bigr].
  \label{eq:C2_branch4_explicit}
\end{eqnarray}

\begin{eqnarray}
  \psi^8_{0,0;5}
  &=&
  \frac{1}{4\sqrt{8!}}
  \Bigl[
  \left|\phi_1\overline{\phi}_2\phi_3\phi_4\phi_5\overline{\phi}_6\overline{\phi}_7\overline{\phi}_8\right|
  - \left|\phi_1\overline{\phi}_2\phi_3\phi_4\overline{\phi}_5\phi_6\overline{\phi}_7\overline{\phi}_8\right|\nonumber\\
  &&- \left|\phi_1\overline{\phi}_2\phi_3\overline{\phi}_4\phi_5\overline{\phi}_6\phi_7\overline{\phi}_8\right|
  + \left|\phi_1\overline{\phi}_2\phi_3\overline{\phi}_4\overline{\phi}_5\phi_6\phi_7\overline{\phi}_8\right|\nonumber\\
  &&- \left|\phi_1\overline{\phi}_2\overline{\phi}_3\phi_4\phi_5\overline{\phi}_6\overline{\phi}_7\phi_8\right|
  + \left|\phi_1\overline{\phi}_2\overline{\phi}_3\phi_4\overline{\phi}_5\phi_6\overline{\phi}_7\phi_8\right|\nonumber\\
  &&+ \left|\phi_1\overline{\phi}_2\overline{\phi}_3\overline{\phi}_4\phi_5\overline{\phi}_6\phi_7\phi_8\right|
  - \left|\phi_1\overline{\phi}_2\overline{\phi}_3\overline{\phi}_4\overline{\phi}_5\phi_6\phi_7\phi_8\right|\nonumber\\
  &&- \left|\overline{\phi}_1\phi_2\phi_3\phi_4\phi_5\overline{\phi}_6\overline{\phi}_7\overline{\phi}_8\right|
  + \left|\overline{\phi}_1\phi_2\phi_3\phi_4\overline{\phi}_5\phi_6\overline{\phi}_7\overline{\phi}_8\right|\nonumber\\
  &&+ \left|\overline{\phi}_1\phi_2\phi_3\overline{\phi}_4\phi_5\overline{\phi}_6\phi_7\overline{\phi}_8\right|
  - \left|\overline{\phi}_1\phi_2\phi_3\overline{\phi}_4\overline{\phi}_5\phi_6\phi_7\overline{\phi}_8\right|\nonumber\\
  &&+ \left|\overline{\phi}_1\phi_2\overline{\phi}_3\phi_4\phi_5\overline{\phi}_6\overline{\phi}_7\phi_8\right|
  - \left|\overline{\phi}_1\phi_2\overline{\phi}_3\phi_4\overline{\phi}_5\phi_6\overline{\phi}_7\phi_8\right|\nonumber\\
  &&- \left|\overline{\phi}_1\phi_2\overline{\phi}_3\overline{\phi}_4\phi_5\overline{\phi}_6\phi_7\phi_8\right|
  + \left|\overline{\phi}_1\phi_2\overline{\phi}_3\overline{\phi}_4\overline{\phi}_5\phi_6\phi_7\phi_8\right|
  \Bigr].
  \label{eq:C2_branch5_explicit}
\end{eqnarray}

\begin{eqnarray}
  \psi^8_{0,0;6}
  &=&
  \frac{1}{4\sqrt{8!}}
  \Bigl[
  \left|\phi_1\phi_2\overline{\phi}_3\overline{\phi}_4\phi_5\overline{\phi}_6\phi_7\overline{\phi}_8\right|
  - \left|\phi_1\phi_2\overline{\phi}_3\overline{\phi}_4\phi_5\overline{\phi}_6\overline{\phi}_7\phi_8\right|\nonumber\\
  &&- \left|\phi_1\phi_2\overline{\phi}_3\overline{\phi}_4\overline{\phi}_5\phi_6\phi_7\overline{\phi}_8\right|
  + \left|\phi_1\phi_2\overline{\phi}_3\overline{\phi}_4\overline{\phi}_5\phi_6\overline{\phi}_7\phi_8\right|\nonumber\\
  &&- \left|\phi_1\overline{\phi}_2\phi_3\overline{\phi}_4\phi_5\overline{\phi}_6\phi_7\overline{\phi}_8\right|
  + \left|\phi_1\overline{\phi}_2\phi_3\overline{\phi}_4\phi_5\overline{\phi}_6\overline{\phi}_7\phi_8\right|\nonumber\\
  &&+ \left|\phi_1\overline{\phi}_2\phi_3\overline{\phi}_4\overline{\phi}_5\phi_6\phi_7\overline{\phi}_8\right|
  - \left|\phi_1\overline{\phi}_2\phi_3\overline{\phi}_4\overline{\phi}_5\phi_6\overline{\phi}_7\phi_8\right|\nonumber\\
  &&- \left|\overline{\phi}_1\phi_2\overline{\phi}_3\phi_4\phi_5\overline{\phi}_6\phi_7\overline{\phi}_8\right|
  + \left|\overline{\phi}_1\phi_2\overline{\phi}_3\phi_4\phi_5\overline{\phi}_6\overline{\phi}_7\phi_8\right|\nonumber\\
  &&+ \left|\overline{\phi}_1\phi_2\overline{\phi}_3\phi_4\overline{\phi}_5\phi_6\phi_7\overline{\phi}_8\right|
  - \left|\overline{\phi}_1\phi_2\overline{\phi}_3\phi_4\overline{\phi}_5\phi_6\overline{\phi}_7\phi_8\right|\nonumber\\
  &&+ \left|\overline{\phi}_1\overline{\phi}_2\phi_3\phi_4\phi_5\overline{\phi}_6\phi_7\overline{\phi}_8\right|
  - \left|\overline{\phi}_1\overline{\phi}_2\phi_3\phi_4\phi_5\overline{\phi}_6\overline{\phi}_7\phi_8\right|\nonumber\\
  &&- \left|\overline{\phi}_1\overline{\phi}_2\phi_3\phi_4\overline{\phi}_5\phi_6\phi_7\overline{\phi}_8\right|
  + \left|\overline{\phi}_1\overline{\phi}_2\phi_3\phi_4\overline{\phi}_5\phi_6\overline{\phi}_7\phi_8\right|
  \Bigr].
  \label{eq:C2_branch6_explicit}
\end{eqnarray}

\begin{eqnarray}
  \psi^8_{0,0;7}
  &=&
  \frac{1}{4\sqrt{8!}}
  \Bigl[
  \left|\phi_1\phi_2\overline{\phi}_3\overline{\phi}_4\phi_5\phi_6\overline{\phi}_7\overline{\phi}_8\right|
  - \left|\phi_1\phi_2\overline{\phi}_3\overline{\phi}_4\phi_5\overline{\phi}_6\phi_7\overline{\phi}_8\right|\nonumber\\
  &&- \left|\phi_1\phi_2\overline{\phi}_3\overline{\phi}_4\overline{\phi}_5\phi_6\overline{\phi}_7\phi_8\right|
  + \left|\phi_1\phi_2\overline{\phi}_3\overline{\phi}_4\overline{\phi}_5\overline{\phi}_6\phi_7\phi_8\right|\nonumber\\
  &&- \left|\phi_1\overline{\phi}_2\phi_3\overline{\phi}_4\phi_5\phi_6\overline{\phi}_7\overline{\phi}_8\right|
  + \left|\phi_1\overline{\phi}_2\phi_3\overline{\phi}_4\phi_5\overline{\phi}_6\phi_7\overline{\phi}_8\right|\nonumber\\
  &&+ \left|\phi_1\overline{\phi}_2\phi_3\overline{\phi}_4\overline{\phi}_5\phi_6\overline{\phi}_7\phi_8\right|
  - \left|\phi_1\overline{\phi}_2\phi_3\overline{\phi}_4\overline{\phi}_5\overline{\phi}_6\phi_7\phi_8\right|\nonumber\\
  &&- \left|\overline{\phi}_1\phi_2\overline{\phi}_3\phi_4\phi_5\phi_6\overline{\phi}_7\overline{\phi}_8\right|
  + \left|\overline{\phi}_1\phi_2\overline{\phi}_3\phi_4\phi_5\overline{\phi}_6\phi_7\overline{\phi}_8\right|\nonumber\\
  &&+ \left|\overline{\phi}_1\phi_2\overline{\phi}_3\phi_4\overline{\phi}_5\phi_6\overline{\phi}_7\phi_8\right|
  - \left|\overline{\phi}_1\phi_2\overline{\phi}_3\phi_4\overline{\phi}_5\overline{\phi}_6\phi_7\phi_8\right|\nonumber\\
  &&+ \left|\overline{\phi}_1\overline{\phi}_2\phi_3\phi_4\phi_5\phi_6\overline{\phi}_7\overline{\phi}_8\right|
  - \left|\overline{\phi}_1\overline{\phi}_2\phi_3\phi_4\phi_5\overline{\phi}_6\phi_7\overline{\phi}_8\right|\nonumber\\
  &&- \left|\overline{\phi}_1\overline{\phi}_2\phi_3\phi_4\overline{\phi}_5\phi_6\overline{\phi}_7\phi_8\right|
  + \left|\overline{\phi}_1\overline{\phi}_2\phi_3\phi_4\overline{\phi}_5\overline{\phi}_6\phi_7\phi_8\right|
  \Bigr].
  \label{eq:C2_branch7_explicit}
\end{eqnarray}

\begin{eqnarray}
  \psi^8_{0,0;8}
  &=&
  \frac{1}{4\sqrt{8!}}
  \Bigl[
  \left|\phi_1\phi_2\overline{\phi}_3\phi_4\overline{\phi}_5\overline{\phi}_6\phi_7\overline{\phi}_8\right|
  - \left|\phi_1\phi_2\overline{\phi}_3\phi_4\overline{\phi}_5\overline{\phi}_6\overline{\phi}_7\phi_8\right|\nonumber\\
  &&- \left|\phi_1\phi_2\overline{\phi}_3\overline{\phi}_4\phi_5\overline{\phi}_6\phi_7\overline{\phi}_8\right|
  + \left|\phi_1\phi_2\overline{\phi}_3\overline{\phi}_4\phi_5\overline{\phi}_6\overline{\phi}_7\phi_8\right|\nonumber\\
  &&- \left|\phi_1\overline{\phi}_2\phi_3\phi_4\overline{\phi}_5\overline{\phi}_6\phi_7\overline{\phi}_8\right|
  + \left|\phi_1\overline{\phi}_2\phi_3\phi_4\overline{\phi}_5\overline{\phi}_6\overline{\phi}_7\phi_8\right|\nonumber\\
  &&+ \left|\phi_1\overline{\phi}_2\phi_3\overline{\phi}_4\phi_5\overline{\phi}_6\phi_7\overline{\phi}_8\right|
  - \left|\phi_1\overline{\phi}_2\phi_3\overline{\phi}_4\phi_5\overline{\phi}_6\overline{\phi}_7\phi_8\right|\nonumber\\
  &&- \left|\overline{\phi}_1\phi_2\overline{\phi}_3\phi_4\overline{\phi}_5\phi_6\phi_7\overline{\phi}_8\right|
  + \left|\overline{\phi}_1\phi_2\overline{\phi}_3\phi_4\overline{\phi}_5\phi_6\overline{\phi}_7\phi_8\right|\nonumber\\
  &&+ \left|\overline{\phi}_1\phi_2\overline{\phi}_3\overline{\phi}_4\phi_5\phi_6\phi_7\overline{\phi}_8\right|
  - \left|\overline{\phi}_1\phi_2\overline{\phi}_3\overline{\phi}_4\phi_5\phi_6\overline{\phi}_7\phi_8\right|\nonumber\\
  &&+ \left|\overline{\phi}_1\overline{\phi}_2\phi_3\phi_4\overline{\phi}_5\phi_6\phi_7\overline{\phi}_8\right|
  - \left|\overline{\phi}_1\overline{\phi}_2\phi_3\phi_4\overline{\phi}_5\phi_6\overline{\phi}_7\phi_8\right|\nonumber\\
  &&- \left|\overline{\phi}_1\overline{\phi}_2\phi_3\overline{\phi}_4\phi_5\phi_6\phi_7\overline{\phi}_8\right|
  + \left|\overline{\phi}_1\overline{\phi}_2\phi_3\overline{\phi}_4\phi_5\phi_6\overline{\phi}_7\phi_8\right|
  \Bigr].
  \label{eq:C2_branch8_explicit}
\end{eqnarray}

\begin{eqnarray}
  \psi^8_{0,0;9}
  &=&
  \frac{1}{4\sqrt{8!}}
  \Bigl[
  \left|\phi_1\phi_2\phi_3\overline{\phi}_4\overline{\phi}_5\overline{\phi}_6\phi_7\overline{\phi}_8\right|
  - \left|\phi_1\phi_2\phi_3\overline{\phi}_4\overline{\phi}_5\overline{\phi}_6\overline{\phi}_7\phi_8\right|\nonumber\\
  &&- \left|\phi_1\phi_2\overline{\phi}_3\phi_4\overline{\phi}_5\overline{\phi}_6\phi_7\overline{\phi}_8\right|
  + \left|\phi_1\phi_2\overline{\phi}_3\phi_4\overline{\phi}_5\overline{\phi}_6\overline{\phi}_7\phi_8\right|\nonumber\\
  &&- \left|\phi_1\overline{\phi}_2\phi_3\overline{\phi}_4\phi_5\overline{\phi}_6\phi_7\overline{\phi}_8\right|
  + \left|\phi_1\overline{\phi}_2\phi_3\overline{\phi}_4\phi_5\overline{\phi}_6\overline{\phi}_7\phi_8\right|\nonumber\\
  &&+ \left|\phi_1\overline{\phi}_2\overline{\phi}_3\phi_4\phi_5\overline{\phi}_6\phi_7\overline{\phi}_8\right|
  - \left|\phi_1\overline{\phi}_2\overline{\phi}_3\phi_4\phi_5\overline{\phi}_6\overline{\phi}_7\phi_8\right|\nonumber\\
  &&- \left|\overline{\phi}_1\phi_2\phi_3\overline{\phi}_4\overline{\phi}_5\phi_6\phi_7\overline{\phi}_8\right|
  + \left|\overline{\phi}_1\phi_2\phi_3\overline{\phi}_4\overline{\phi}_5\phi_6\overline{\phi}_7\phi_8\right|\nonumber\\
  &&+ \left|\overline{\phi}_1\phi_2\overline{\phi}_3\phi_4\overline{\phi}_5\phi_6\phi_7\overline{\phi}_8\right|
  - \left|\overline{\phi}_1\phi_2\overline{\phi}_3\phi_4\overline{\phi}_5\phi_6\overline{\phi}_7\phi_8\right|\nonumber\\
  &&+ \left|\overline{\phi}_1\overline{\phi}_2\phi_3\overline{\phi}_4\phi_5\phi_6\phi_7\overline{\phi}_8\right|
  - \left|\overline{\phi}_1\overline{\phi}_2\phi_3\overline{\phi}_4\phi_5\phi_6\overline{\phi}_7\phi_8\right|\nonumber\\
  &&- \left|\overline{\phi}_1\overline{\phi}_2\overline{\phi}_3\phi_4\phi_5\phi_6\phi_7\overline{\phi}_8\right|
  + \left|\overline{\phi}_1\overline{\phi}_2\overline{\phi}_3\phi_4\phi_5\phi_6\overline{\phi}_7\phi_8\right|
  \Bigr].
  \label{eq:C2_branch9_explicit}
\end{eqnarray}

\begin{eqnarray}
  \psi^8_{0,0;10}
  &=&
  \frac{1}{4\sqrt{8!}}
  \Bigl[
  \left|\phi_1\phi_2\overline{\phi}_3\phi_4\overline{\phi}_5\phi_6\overline{\phi}_7\overline{\phi}_8\right|
  - \left|\phi_1\phi_2\overline{\phi}_3\phi_4\overline{\phi}_5\overline{\phi}_6\phi_7\overline{\phi}_8\right|\nonumber\\
  &&- \left|\phi_1\phi_2\overline{\phi}_3\overline{\phi}_4\phi_5\phi_6\overline{\phi}_7\overline{\phi}_8\right|
  + \left|\phi_1\phi_2\overline{\phi}_3\overline{\phi}_4\phi_5\overline{\phi}_6\phi_7\overline{\phi}_8\right|\nonumber\\
  &&- \left|\phi_1\overline{\phi}_2\phi_3\phi_4\overline{\phi}_5\phi_6\overline{\phi}_7\overline{\phi}_8\right|
  + \left|\phi_1\overline{\phi}_2\phi_3\phi_4\overline{\phi}_5\overline{\phi}_6\phi_7\overline{\phi}_8\right|\nonumber\\
  &&+ \left|\phi_1\overline{\phi}_2\phi_3\overline{\phi}_4\phi_5\phi_6\overline{\phi}_7\overline{\phi}_8\right|
  - \left|\phi_1\overline{\phi}_2\phi_3\overline{\phi}_4\phi_5\overline{\phi}_6\phi_7\overline{\phi}_8\right|\nonumber\\
  &&- \left|\overline{\phi}_1\phi_2\overline{\phi}_3\phi_4\overline{\phi}_5\phi_6\overline{\phi}_7\phi_8\right|
  + \left|\overline{\phi}_1\phi_2\overline{\phi}_3\phi_4\overline{\phi}_5\overline{\phi}_6\phi_7\phi_8\right|\nonumber\\
  &&+ \left|\overline{\phi}_1\phi_2\overline{\phi}_3\overline{\phi}_4\phi_5\phi_6\overline{\phi}_7\phi_8\right|
  - \left|\overline{\phi}_1\phi_2\overline{\phi}_3\overline{\phi}_4\phi_5\overline{\phi}_6\phi_7\phi_8\right|\nonumber\\
  &&+ \left|\overline{\phi}_1\overline{\phi}_2\phi_3\phi_4\overline{\phi}_5\phi_6\overline{\phi}_7\phi_8\right|
  - \left|\overline{\phi}_1\overline{\phi}_2\phi_3\phi_4\overline{\phi}_5\overline{\phi}_6\phi_7\phi_8\right|\nonumber\\
  &&- \left|\overline{\phi}_1\overline{\phi}_2\phi_3\overline{\phi}_4\phi_5\phi_6\overline{\phi}_7\phi_8\right|
  + \left|\overline{\phi}_1\overline{\phi}_2\phi_3\overline{\phi}_4\phi_5\overline{\phi}_6\phi_7\phi_8\right|
  \Bigr].
  \label{eq:C2_branch10_explicit}
\end{eqnarray}

\begin{eqnarray}
  \psi^8_{0,0;11}
  &=&
  \frac{1}{4\sqrt{8!}}
  \Bigl[
  \left|\phi_1\phi_2\overline{\phi}_3\phi_4\phi_5\overline{\phi}_6\overline{\phi}_7\overline{\phi}_8\right|
  - \left|\phi_1\phi_2\overline{\phi}_3\phi_4\overline{\phi}_5\phi_6\overline{\phi}_7\overline{\phi}_8\right|\nonumber\\
  &&- \left|\phi_1\phi_2\overline{\phi}_3\overline{\phi}_4\phi_5\overline{\phi}_6\phi_7\overline{\phi}_8\right|
  + \left|\phi_1\phi_2\overline{\phi}_3\overline{\phi}_4\overline{\phi}_5\phi_6\phi_7\overline{\phi}_8\right|\nonumber\\
  &&- \left|\phi_1\overline{\phi}_2\phi_3\phi_4\phi_5\overline{\phi}_6\overline{\phi}_7\overline{\phi}_8\right|
  + \left|\phi_1\overline{\phi}_2\phi_3\phi_4\overline{\phi}_5\phi_6\overline{\phi}_7\overline{\phi}_8\right|\nonumber\\
  &&+ \left|\phi_1\overline{\phi}_2\phi_3\overline{\phi}_4\phi_5\overline{\phi}_6\phi_7\overline{\phi}_8\right|
  - \left|\phi_1\overline{\phi}_2\phi_3\overline{\phi}_4\overline{\phi}_5\phi_6\phi_7\overline{\phi}_8\right|\nonumber\\
  &&- \left|\overline{\phi}_1\phi_2\overline{\phi}_3\phi_4\phi_5\overline{\phi}_6\overline{\phi}_7\phi_8\right|
  + \left|\overline{\phi}_1\phi_2\overline{\phi}_3\phi_4\overline{\phi}_5\phi_6\overline{\phi}_7\phi_8\right|\nonumber\\
  &&+ \left|\overline{\phi}_1\phi_2\overline{\phi}_3\overline{\phi}_4\phi_5\overline{\phi}_6\phi_7\phi_8\right|
  - \left|\overline{\phi}_1\phi_2\overline{\phi}_3\overline{\phi}_4\overline{\phi}_5\phi_6\phi_7\phi_8\right|\nonumber\\
  &&+ \left|\overline{\phi}_1\overline{\phi}_2\phi_3\phi_4\phi_5\overline{\phi}_6\overline{\phi}_7\phi_8\right|
  - \left|\overline{\phi}_1\overline{\phi}_2\phi_3\phi_4\overline{\phi}_5\phi_6\overline{\phi}_7\phi_8\right|\nonumber\\
  &&- \left|\overline{\phi}_1\overline{\phi}_2\phi_3\overline{\phi}_4\phi_5\overline{\phi}_6\phi_7\phi_8\right|
  + \left|\overline{\phi}_1\overline{\phi}_2\phi_3\overline{\phi}_4\overline{\phi}_5\phi_6\phi_7\phi_8\right|
  \Bigr].
  \label{eq:C2_branch11_explicit}
\end{eqnarray}

\begin{eqnarray}
  \psi^8_{0,0;12}
  &=&
  \frac{1}{4\sqrt{8!}}
  \Bigl[
  \left|\phi_1\phi_2\phi_3\overline{\phi}_4\overline{\phi}_5\phi_6\overline{\phi}_7\overline{\phi}_8\right|
  - \left|\phi_1\phi_2\phi_3\overline{\phi}_4\overline{\phi}_5\overline{\phi}_6\phi_7\overline{\phi}_8\right|\nonumber\\
  &&- \left|\phi_1\phi_2\overline{\phi}_3\phi_4\overline{\phi}_5\phi_6\overline{\phi}_7\overline{\phi}_8\right|
  + \left|\phi_1\phi_2\overline{\phi}_3\phi_4\overline{\phi}_5\overline{\phi}_6\phi_7\overline{\phi}_8\right|\nonumber\\
  &&- \left|\phi_1\overline{\phi}_2\phi_3\overline{\phi}_4\phi_5\phi_6\overline{\phi}_7\overline{\phi}_8\right|
  + \left|\phi_1\overline{\phi}_2\phi_3\overline{\phi}_4\phi_5\overline{\phi}_6\phi_7\overline{\phi}_8\right|\nonumber\\
  &&+ \left|\phi_1\overline{\phi}_2\overline{\phi}_3\phi_4\phi_5\phi_6\overline{\phi}_7\overline{\phi}_8\right|
  - \left|\phi_1\overline{\phi}_2\overline{\phi}_3\phi_4\phi_5\overline{\phi}_6\phi_7\overline{\phi}_8\right|\nonumber\\
  &&- \left|\overline{\phi}_1\phi_2\phi_3\overline{\phi}_4\overline{\phi}_5\phi_6\overline{\phi}_7\phi_8\right|
  + \left|\overline{\phi}_1\phi_2\phi_3\overline{\phi}_4\overline{\phi}_5\overline{\phi}_6\phi_7\phi_8\right|\nonumber\\
  &&+ \left|\overline{\phi}_1\phi_2\overline{\phi}_3\phi_4\overline{\phi}_5\phi_6\overline{\phi}_7\phi_8\right|
  - \left|\overline{\phi}_1\phi_2\overline{\phi}_3\phi_4\overline{\phi}_5\overline{\phi}_6\phi_7\phi_8\right|\nonumber\\
  &&+ \left|\overline{\phi}_1\overline{\phi}_2\phi_3\overline{\phi}_4\phi_5\phi_6\overline{\phi}_7\phi_8\right|
  - \left|\overline{\phi}_1\overline{\phi}_2\phi_3\overline{\phi}_4\phi_5\overline{\phi}_6\phi_7\phi_8\right|\nonumber\\
  &&- \left|\overline{\phi}_1\overline{\phi}_2\overline{\phi}_3\phi_4\phi_5\phi_6\overline{\phi}_7\phi_8\right|
  + \left|\overline{\phi}_1\overline{\phi}_2\overline{\phi}_3\phi_4\phi_5\overline{\phi}_6\phi_7\phi_8\right|
  \Bigr].
  \label{eq:C2_branch12_explicit}
\end{eqnarray}

\begin{eqnarray}
  \psi^8_{0,0;13}
  &=&
  \frac{1}{4\sqrt{8!}}
  \Bigl[
  \left|\phi_1\phi_2\phi_3\overline{\phi}_4\phi_5\overline{\phi}_6\overline{\phi}_7\overline{\phi}_8\right|
  - \left|\phi_1\phi_2\phi_3\overline{\phi}_4\overline{\phi}_5\phi_6\overline{\phi}_7\overline{\phi}_8\right|\nonumber\\
  &&- \left|\phi_1\phi_2\overline{\phi}_3\phi_4\phi_5\overline{\phi}_6\overline{\phi}_7\overline{\phi}_8\right|
  + \left|\phi_1\phi_2\overline{\phi}_3\phi_4\overline{\phi}_5\phi_6\overline{\phi}_7\overline{\phi}_8\right|\nonumber\\
  &&- \left|\phi_1\overline{\phi}_2\phi_3\overline{\phi}_4\phi_5\overline{\phi}_6\phi_7\overline{\phi}_8\right|
  + \left|\phi_1\overline{\phi}_2\phi_3\overline{\phi}_4\overline{\phi}_5\phi_6\phi_7\overline{\phi}_8\right|\nonumber\\
  &&+ \left|\phi_1\overline{\phi}_2\overline{\phi}_3\phi_4\phi_5\overline{\phi}_6\phi_7\overline{\phi}_8\right|
  - \left|\phi_1\overline{\phi}_2\overline{\phi}_3\phi_4\overline{\phi}_5\phi_6\phi_7\overline{\phi}_8\right|\nonumber\\
  &&- \left|\overline{\phi}_1\phi_2\phi_3\overline{\phi}_4\phi_5\overline{\phi}_6\overline{\phi}_7\phi_8\right|
  + \left|\overline{\phi}_1\phi_2\phi_3\overline{\phi}_4\overline{\phi}_5\phi_6\overline{\phi}_7\phi_8\right|\nonumber\\
  &&+ \left|\overline{\phi}_1\phi_2\overline{\phi}_3\phi_4\phi_5\overline{\phi}_6\overline{\phi}_7\phi_8\right|
  - \left|\overline{\phi}_1\phi_2\overline{\phi}_3\phi_4\overline{\phi}_5\phi_6\overline{\phi}_7\phi_8\right|\nonumber\\
  &&+ \left|\overline{\phi}_1\overline{\phi}_2\phi_3\overline{\phi}_4\phi_5\overline{\phi}_6\phi_7\phi_8\right|
  - \left|\overline{\phi}_1\overline{\phi}_2\phi_3\overline{\phi}_4\overline{\phi}_5\phi_6\phi_7\phi_8\right|\nonumber\\
  &&- \left|\overline{\phi}_1\overline{\phi}_2\overline{\phi}_3\phi_4\phi_5\overline{\phi}_6\phi_7\phi_8\right|
  + \left|\overline{\phi}_1\overline{\phi}_2\overline{\phi}_3\phi_4\overline{\phi}_5\phi_6\phi_7\phi_8\right|
  \Bigr].
  \label{eq:C2_branch13_explicit}
\end{eqnarray}

\begin{eqnarray}
  \psi^8_{0,0;14}
  &=&
  \frac{1}{4\sqrt{8!}}
  \Bigl[
  \left|\phi_1\phi_2\phi_3\phi_4\overline{\phi}_5\overline{\phi}_6\overline{\phi}_7\overline{\phi}_8\right|
  - \left|\phi_1\phi_2\phi_3\overline{\phi}_4\phi_5\overline{\phi}_6\overline{\phi}_7\overline{\phi}_8\right|\nonumber\\
  &&- \left|\phi_1\phi_2\overline{\phi}_3\phi_4\overline{\phi}_5\phi_6\overline{\phi}_7\overline{\phi}_8\right|
  + \left|\phi_1\phi_2\overline{\phi}_3\overline{\phi}_4\phi_5\phi_6\overline{\phi}_7\overline{\phi}_8\right|\nonumber\\
  &&- \left|\phi_1\overline{\phi}_2\phi_3\phi_4\overline{\phi}_5\overline{\phi}_6\phi_7\overline{\phi}_8\right|
  + \left|\phi_1\overline{\phi}_2\phi_3\overline{\phi}_4\phi_5\overline{\phi}_6\phi_7\overline{\phi}_8\right|\nonumber\\
  &&+ \left|\phi_1\overline{\phi}_2\overline{\phi}_3\phi_4\overline{\phi}_5\phi_6\phi_7\overline{\phi}_8\right|
  - \left|\phi_1\overline{\phi}_2\overline{\phi}_3\overline{\phi}_4\phi_5\phi_6\phi_7\overline{\phi}_8\right|\nonumber\\
  &&- \left|\overline{\phi}_1\phi_2\phi_3\phi_4\overline{\phi}_5\overline{\phi}_6\overline{\phi}_7\phi_8\right|
  + \left|\overline{\phi}_1\phi_2\phi_3\overline{\phi}_4\phi_5\overline{\phi}_6\overline{\phi}_7\phi_8\right|\nonumber\\
  &&+ \left|\overline{\phi}_1\phi_2\overline{\phi}_3\phi_4\overline{\phi}_5\phi_6\overline{\phi}_7\phi_8\right|
  - \left|\overline{\phi}_1\phi_2\overline{\phi}_3\overline{\phi}_4\phi_5\phi_6\overline{\phi}_7\phi_8\right|\nonumber\\
  &&+ \left|\overline{\phi}_1\overline{\phi}_2\phi_3\phi_4\overline{\phi}_5\overline{\phi}_6\phi_7\phi_8\right|
  - \left|\overline{\phi}_1\overline{\phi}_2\phi_3\overline{\phi}_4\phi_5\overline{\phi}_6\phi_7\phi_8\right|\nonumber\\
  &&- \left|\overline{\phi}_1\overline{\phi}_2\overline{\phi}_3\phi_4\overline{\phi}_5\phi_6\phi_7\phi_8\right|
  + \left|\overline{\phi}_1\overline{\phi}_2\overline{\phi}_3\overline{\phi}_4\phi_5\phi_6\phi_7\phi_8\right|
  \Bigr].
  \label{eq:C2_branch14_explicit}
\end{eqnarray}

A useful simplification, particularly in the context of testing quantum algorithms
or approximate treatments, is to freeze the spin-coupled orbitals and identify
them with a minimal set of localized valence functions for C$_2$. Denoting these
fixed orbitals by
\begin{equation}
  \chi_1,\chi_2,\chi_3,\chi_4,\chi_5,\chi_6,\chi_7,\chi_8,
\end{equation}
the approximate SCGVB wavefunction becomes
\begin{equation}
  \widetilde{\Psi}_{\mathrm{SCGVB}}^{\mathrm{act}}
  =
  \mathcal{N}
  \sum_{k=1}^{14}
  c_{0,k}\,
  \widetilde{\psi}^8_{0,0;k},
  \label{eq:C2_frozen_total}
\end{equation}
where each $\widetilde{\psi}^8_{0,0;k}$ is obtained from
Eq.~(\ref{eq:C2_each_branch_explicit}) by replacing
$\phi_j\rightarrow\chi_j$ and
$\overline{\phi}_j\rightarrow\overline{\chi}_j$.

Thus, for C$_2$ in the X$^1\Sigma_g^+$ ground state, the full singlet SCGVB
spin space is represented by fourteen spin-coupled branches built from eight
active spin-coupled orbitals. This is the direct analogue of the H$_4$ singlet
construction, but with $f_0^8=14$ instead of $f_0^4=2$. The variational
optimization of the coefficients $c_{0,1},\ldots,c_{0,14}$ and of the orbitals
$\{\phi_i\}$ proceeds by minimizing the SCGVB energy functional defined in the main manuscript. In the frozen-orbital
approximation, the same explicit branch structure provides a reproducible
larger testbed for the nonorthogonal determinant algebra underlying SCGVB
theory and its quantum matrix-element estimators.

\subsection{Explicit second-quantized and computational-basis representation of the $C_{2}$ SCGVB
wavefunction}
\label{app:C2_scgvb_expansions}

This section gives a fully explicit implementation of the active-valence
SCGVB determinant basis used for C$_2$ in the X$^1\Sigma_g^+$ state.  The notation
follows the H$_4$ appendix: $\ket{\phi_0}$ is the computational vacuum, $f_I$ is the
creation string that generates determinant $\ket{D_I}$, and $w_I$ is the corresponding
annihilation string acting on the bra side.

\subsection{Spin-orbital ordering and computational vacuum}
\label{app:C2_ordering}

Let
\begin{equation}
  \ket{\phi_0}\equiv \ket{0000000000000000}
\end{equation}
be the 16-qubit computational vacuum.  We use the spin-orbital ordering
\begin{equation}
  1,\ldots,8 \equiv \alpha_1,\ldots,\alpha_8,
  \qquad
  9,\ldots,16 \equiv \beta_1,\ldots,\beta_8.
  \label{eq:C2_spin_orbital_ordering_app}
\end{equation}
Thus
\begin{equation}
  \phi_i \longleftrightarrow \hat a_i^\dagger,
  \qquad
  \overline{\phi}_i \longleftrightarrow \hat a_{8+i}^\dagger,
  \qquad i=1,\ldots,8.
\end{equation}

\subsection{Compact determinant labels and bitstrings}
\label{app:C2_D_labels_bitstrings}

For an eight-electron singlet determinant, four spatial orbitals carry $\alpha$
spin and the complementary four carry $\beta$ spin.  We label each determinant
by the set $A$ of spatial orbitals occupied by $\alpha$ electrons:
\begin{equation}
  A=\{i_1<i_2<i_3<i_4\},\qquad
  \overline A=\{j_1<j_2<j_3<j_4\}.
\end{equation}
The associated creation and annihilation strings are
\begin{align}
  f_A
  &=
  \hat a_{i_1}^{\dagger}
  \hat a_{i_2}^{\dagger}
  \hat a_{i_3}^{\dagger}
  \hat a_{i_4}^{\dagger}
  \hat a_{8+j_1}^{\dagger}
  \hat a_{8+j_2}^{\dagger}
  \hat a_{8+j_3}^{\dagger}
  \hat a_{8+j_4}^{\dagger},
  \label{eq:C2_fA_general_app}
\\
  w_A
  &=
  \hat a_{8+j_4}
  \hat a_{8+j_3}
  \hat a_{8+j_2}
  \hat a_{8+j_1}
  \hat a_{i_4}
  \hat a_{i_3}
  \hat a_{i_2}
  \hat a_{i_1}.
  \label{eq:C2_wA_general_app}
\end{align}
Then
\begin{equation}
  \ket{D_A}=f_A\ket{\phi_0},
  \qquad
  \bra{D_A}=\bra{\phi_0}w_A.
\end{equation}

The following table \ref{tab:C2_complete_alpha_beta_bitstrings} gives only the compact labels, $\alpha$ sets, and bitstrings;
the explicit $f_I$ and $w_I$ strings are listed separately below to avoid page
overlap.

\begingroup
\scriptsize
\setlength{\LTleft}{0pt plus 1fill}
\setlength{\LTright}{0pt plus 1fill}
\setlength{\tabcolsep}{4pt}
\renewcommand{\arraystretch}{1.12}

\begin{longtable}{c c c c c}
\caption{Complete primitive $M=0$ determinant basis for the active C$_2$ singlet space. 
The set $A_\alpha$ lists the four spatial orbitals occupied by $\alpha$ electrons, while
$A_\beta=\overline{A_\alpha}$ lists the complementary four spatial orbitals occupied by
$\beta$ electrons. The occupied spin-orbitals are written using the ordering
$(\alpha_1,\ldots,\alpha_8,\beta_1,\ldots,\beta_8)$, so that $\beta_i$ corresponds to
spin-orbital $8+i$. The bitstring follows the same ordering.}
\label{tab:C2_complete_alpha_beta_bitstrings}
\\
\toprule
Label & $A_\alpha$ & $A_\beta=\overline{A_\alpha}$ & Occupied spin-orbitals & Bitstring \\
\midrule
\endfirsthead

\toprule
Label & $A_\alpha$ & $A_\beta=\overline{A_\alpha}$ & Occupied spin-orbitals & Bitstring \\
\midrule
\endhead

\midrule
\multicolumn{5}{r}{Continued on next page.}\\
\midrule
\endfoot

\bottomrule
\endlastfoot

$D_{1}$ & $\{1,2,3,4\}$ & $\{5,6,7,8\}$ & $\{1,2,3,4,13,14,15,16\}$ & \texttt{1111000000001111} \\
$D_{2}$ & $\{1,2,3,5\}$ & $\{4,6,7,8\}$ & $\{1,2,3,5,12,14,15,16\}$ & \texttt{1110100000010111} \\
$D_{3}$ & $\{1,2,3,6\}$ & $\{4,5,7,8\}$ & $\{1,2,3,6,12,13,15,16\}$ & \texttt{1110010000011011} \\
$D_{4}$ & $\{1,2,3,7\}$ & $\{4,5,6,8\}$ & $\{1,2,3,7,12,13,14,16\}$ & \texttt{1110001000011101} \\
$D_{5}$ & $\{1,2,3,8\}$ & $\{4,5,6,7\}$ & $\{1,2,3,8,12,13,14,15\}$ & \texttt{1110000100011110} \\
$D_{6}$ & $\{1,2,4,5\}$ & $\{3,6,7,8\}$ & $\{1,2,4,5,11,14,15,16\}$ & \texttt{1101100000100111} \\
$D_{7}$ & $\{1,2,4,6\}$ & $\{3,5,7,8\}$ & $\{1,2,4,6,11,13,15,16\}$ & \texttt{1101010000101011} \\
$D_{8}$ & $\{1,2,4,7\}$ & $\{3,5,6,8\}$ & $\{1,2,4,7,11,13,14,16\}$ & \texttt{1101001000101101} \\
$D_{9}$ & $\{1,2,4,8\}$ & $\{3,5,6,7\}$ & $\{1,2,4,8,11,13,14,15\}$ & \texttt{1101000100101110} \\
$D_{10}$ & $\{1,2,5,6\}$ & $\{3,4,7,8\}$ & $\{1,2,5,6,11,12,15,16\}$ & \texttt{1100110000110011} \\
$D_{11}$ & $\{1,2,5,7\}$ & $\{3,4,6,8\}$ & $\{1,2,5,7,11,12,14,16\}$ & \texttt{1100101000110101} \\
$D_{12}$ & $\{1,2,5,8\}$ & $\{3,4,6,7\}$ & $\{1,2,5,8,11,12,14,15\}$ & \texttt{1100100100110110} \\
$D_{13}$ & $\{1,2,6,7\}$ & $\{3,4,5,8\}$ & $\{1,2,6,7,11,12,13,16\}$ & \texttt{1100011000111001} \\
$D_{14}$ & $\{1,2,6,8\}$ & $\{3,4,5,7\}$ & $\{1,2,6,8,11,12,13,15\}$ & \texttt{1100010100111010} \\
$D_{15}$ & $\{1,2,7,8\}$ & $\{3,4,5,6\}$ & $\{1,2,7,8,11,12,13,14\}$ & \texttt{1100001100111100} \\
$D_{16}$ & $\{1,3,4,5\}$ & $\{2,6,7,8\}$ & $\{1,3,4,5,10,14,15,16\}$ & \texttt{1011100001000111} \\
$D_{17}$ & $\{1,3,4,6\}$ & $\{2,5,7,8\}$ & $\{1,3,4,6,10,13,15,16\}$ & \texttt{1011010001001011} \\
$D_{18}$ & $\{1,3,4,7\}$ & $\{2,5,6,8\}$ & $\{1,3,4,7,10,13,14,16\}$ & \texttt{1011001001001101} \\
$D_{19}$ & $\{1,3,4,8\}$ & $\{2,5,6,7\}$ & $\{1,3,4,8,10,13,14,15\}$ & \texttt{1011000101001110} \\
$D_{20}$ & $\{1,3,5,6\}$ & $\{2,4,7,8\}$ & $\{1,3,5,6,10,12,15,16\}$ & \texttt{1010110001010011} \\
$D_{21}$ & $\{1,3,5,7\}$ & $\{2,4,6,8\}$ & $\{1,3,5,7,10,12,14,16\}$ & \texttt{1010101001010101} \\
$D_{22}$ & $\{1,3,5,8\}$ & $\{2,4,6,7\}$ & $\{1,3,5,8,10,12,14,15\}$ & \texttt{1010100101010110} \\
$D_{23}$ & $\{1,3,6,7\}$ & $\{2,4,5,8\}$ & $\{1,3,6,7,10,12,13,16\}$ & \texttt{1010011001011001} \\
$D_{24}$ & $\{1,3,6,8\}$ & $\{2,4,5,7\}$ & $\{1,3,6,8,10,12,13,15\}$ & \texttt{1010010101011010} \\
$D_{25}$ & $\{1,3,7,8\}$ & $\{2,4,5,6\}$ & $\{1,3,7,8,10,12,13,14\}$ & \texttt{1010001101011100} \\
$D_{26}$ & $\{1,4,5,6\}$ & $\{2,3,7,8\}$ & $\{1,4,5,6,10,11,15,16\}$ & \texttt{1001110001100011} \\
$D_{27}$ & $\{1,4,5,7\}$ & $\{2,3,6,8\}$ & $\{1,4,5,7,10,11,14,16\}$ & \texttt{1001101001100101} \\
$D_{28}$ & $\{1,4,5,8\}$ & $\{2,3,6,7\}$ & $\{1,4,5,8,10,11,14,15\}$ & \texttt{1001100101100110} \\
$D_{29}$ & $\{1,4,6,7\}$ & $\{2,3,5,8\}$ & $\{1,4,6,7,10,11,13,16\}$ & \texttt{1001011001101001} \\
$D_{30}$ & $\{1,4,6,8\}$ & $\{2,3,5,7\}$ & $\{1,4,6,8,10,11,13,15\}$ & \texttt{1001010101101010} \\
$D_{31}$ & $\{1,4,7,8\}$ & $\{2,3,5,6\}$ & $\{1,4,7,8,10,11,13,14\}$ & \texttt{1001001101101100} \\
$D_{32}$ & $\{1,5,6,7\}$ & $\{2,3,4,8\}$ & $\{1,5,6,7,10,11,12,16\}$ & \texttt{1000111001110001} \\
$D_{33}$ & $\{1,5,6,8\}$ & $\{2,3,4,7\}$ & $\{1,5,6,8,10,11,12,15\}$ & \texttt{1000110101110010} \\
$D_{34}$ & $\{1,5,7,8\}$ & $\{2,3,4,6\}$ & $\{1,5,7,8,10,11,12,14\}$ & \texttt{1000101101110100} \\
$D_{35}$ & $\{1,6,7,8\}$ & $\{2,3,4,5\}$ & $\{1,6,7,8,10,11,12,13\}$ & \texttt{1000011101111000} \\
$D_{36}$ & $\{2,3,4,5\}$ & $\{1,6,7,8\}$ & $\{2,3,4,5,9,14,15,16\}$ & \texttt{0111100010000111} \\
$D_{37}$ & $\{2,3,4,6\}$ & $\{1,5,7,8\}$ & $\{2,3,4,6,9,13,15,16\}$ & \texttt{0111010010001011} \\
$D_{38}$ & $\{2,3,4,7\}$ & $\{1,5,6,8\}$ & $\{2,3,4,7,9,13,14,16\}$ & \texttt{0111001010001101} \\
$D_{39}$ & $\{2,3,4,8\}$ & $\{1,5,6,7\}$ & $\{2,3,4,8,9,13,14,15\}$ & \texttt{0111000110001110} \\
$D_{40}$ & $\{2,3,5,6\}$ & $\{1,4,7,8\}$ & $\{2,3,5,6,9,12,15,16\}$ & \texttt{0110110010010011} \\
$D_{41}$ & $\{2,3,5,7\}$ & $\{1,4,6,8\}$ & $\{2,3,5,7,9,12,14,16\}$ & \texttt{0110101010010101} \\
$D_{42}$ & $\{2,3,5,8\}$ & $\{1,4,6,7\}$ & $\{2,3,5,8,9,12,14,15\}$ & \texttt{0110100110010110} \\
$D_{43}$ & $\{2,3,6,7\}$ & $\{1,4,5,8\}$ & $\{2,3,6,7,9,12,13,16\}$ & \texttt{0110011010011001} \\
$D_{44}$ & $\{2,3,6,8\}$ & $\{1,4,5,7\}$ & $\{2,3,6,8,9,12,13,15\}$ & \texttt{0110010110011010} \\
$D_{45}$ & $\{2,3,7,8\}$ & $\{1,4,5,6\}$ & $\{2,3,7,8,9,12,13,14\}$ & \texttt{0110001110011100} \\
$D_{46}$ & $\{2,4,5,6\}$ & $\{1,3,7,8\}$ & $\{2,4,5,6,9,11,15,16\}$ & \texttt{0101110010100011} \\
$D_{47}$ & $\{2,4,5,7\}$ & $\{1,3,6,8\}$ & $\{2,4,5,7,9,11,14,16\}$ & \texttt{0101101010100101} \\
$D_{48}$ & $\{2,4,5,8\}$ & $\{1,3,6,7\}$ & $\{2,4,5,8,9,11,14,15\}$ & \texttt{0101100110100110} \\
$D_{49}$ & $\{2,4,6,7\}$ & $\{1,3,5,8\}$ & $\{2,4,6,7,9,11,13,16\}$ & \texttt{0101011010101001} \\
$D_{50}$ & $\{2,4,6,8\}$ & $\{1,3,5,7\}$ & $\{2,4,6,8,9,11,13,15\}$ & \texttt{0101010110101010} \\
$D_{51}$ & $\{2,4,7,8\}$ & $\{1,3,5,6\}$ & $\{2,4,7,8,9,11,13,14\}$ & \texttt{0101001110101100} \\
$D_{52}$ & $\{2,5,6,7\}$ & $\{1,3,4,8\}$ & $\{2,5,6,7,9,11,12,16\}$ & \texttt{0100111010110001} \\
$D_{53}$ & $\{2,5,6,8\}$ & $\{1,3,4,7\}$ & $\{2,5,6,8,9,11,12,15\}$ & \texttt{0100110110110010} \\
$D_{54}$ & $\{2,5,7,8\}$ & $\{1,3,4,6\}$ & $\{2,5,7,8,9,11,12,14\}$ & \texttt{0100101110110100} \\
$D_{55}$ & $\{2,6,7,8\}$ & $\{1,3,4,5\}$ & $\{2,6,7,8,9,11,12,13\}$ & \texttt{0100011110111000} \\
$D_{56}$ & $\{3,4,5,6\}$ & $\{1,2,7,8\}$ & $\{3,4,5,6,9,10,15,16\}$ & \texttt{0011110011000011} \\
$D_{57}$ & $\{3,4,5,7\}$ & $\{1,2,6,8\}$ & $\{3,4,5,7,9,10,14,16\}$ & \texttt{0011101011000101} \\
$D_{58}$ & $\{3,4,5,8\}$ & $\{1,2,6,7\}$ & $\{3,4,5,8,9,10,14,15\}$ & \texttt{0011100111000110} \\
$D_{59}$ & $\{3,4,6,7\}$ & $\{1,2,5,8\}$ & $\{3,4,6,7,9,10,13,16\}$ & \texttt{0011011011001001} \\
$D_{60}$ & $\{3,4,6,8\}$ & $\{1,2,5,7\}$ & $\{3,4,6,8,9,10,13,15\}$ & \texttt{0011010111001010} \\
$D_{61}$ & $\{3,4,7,8\}$ & $\{1,2,5,6\}$ & $\{3,4,7,8,9,10,13,14\}$ & \texttt{0011001111001100} \\
$D_{62}$ & $\{3,5,6,7\}$ & $\{1,2,4,8\}$ & $\{3,5,6,7,9,10,12,16\}$ & \texttt{0010111011010001} \\
$D_{63}$ & $\{3,5,6,8\}$ & $\{1,2,4,7\}$ & $\{3,5,6,8,9,10,12,15\}$ & \texttt{0010110111010010} \\
$D_{64}$ & $\{3,5,7,8\}$ & $\{1,2,4,6\}$ & $\{3,5,7,8,9,10,12,14\}$ & \texttt{0010101111010100} \\
$D_{65}$ & $\{3,6,7,8\}$ & $\{1,2,4,5\}$ & $\{3,6,7,8,9,10,12,13\}$ & \texttt{0010011111011000} \\
$D_{66}$ & $\{4,5,6,7\}$ & $\{1,2,3,8\}$ & $\{4,5,6,7,9,10,11,16\}$ & \texttt{0001111011100001} \\
$D_{67}$ & $\{4,5,6,8\}$ & $\{1,2,3,7\}$ & $\{4,5,6,8,9,10,11,15\}$ & \texttt{0001110111100010} \\
$D_{68}$ & $\{4,5,7,8\}$ & $\{1,2,3,6\}$ & $\{4,5,7,8,9,10,11,14\}$ & \texttt{0001101111100100} \\
$D_{69}$ & $\{4,6,7,8\}$ & $\{1,2,3,5\}$ & $\{4,6,7,8,9,10,11,13\}$ & \texttt{0001011111101000} \\
$D_{70}$ & $\{5,6,7,8\}$ & $\{1,2,3,4\}$ & $\{5,6,7,8,9,10,11,12\}$ & \texttt{0000111111110000} \\
\end{longtable}
\endgroup

\subsection{Explicit creation and annihilation strings}
\label{app:C2_explicit_f_w}

The following equations list all creation strings $f_I$ and annihilation strings
$w_I$ explicitly.  They are split into small blocks to avoid long tables and page
overlap.

\paragraph{Determinants $D_{1}$--$D_{5}$.}
\begin{align}
f_{1} &= \hat a_{1}^\dagger\hat a_{2}^\dagger\hat a_{3}^\dagger\hat a_{4}^\dagger\hat a_{13}^\dagger\hat a_{14}^\dagger\hat a_{15}^\dagger\hat a_{16}^\dagger, & w_{1} &= \hat a_{16}\hat a_{15}\hat a_{14}\hat a_{13}\hat a_{4}\hat a_{3}\hat a_{2}\hat a_{1} \\
f_{2} &= \hat a_{1}^\dagger\hat a_{2}^\dagger\hat a_{3}^\dagger\hat a_{5}^\dagger\hat a_{12}^\dagger\hat a_{14}^\dagger\hat a_{15}^\dagger\hat a_{16}^\dagger, & w_{2} &= \hat a_{16}\hat a_{15}\hat a_{14}\hat a_{12}\hat a_{5}\hat a_{3}\hat a_{2}\hat a_{1} \\
f_{3} &= \hat a_{1}^\dagger\hat a_{2}^\dagger\hat a_{3}^\dagger\hat a_{6}^\dagger\hat a_{12}^\dagger\hat a_{13}^\dagger\hat a_{15}^\dagger\hat a_{16}^\dagger, & w_{3} &= \hat a_{16}\hat a_{15}\hat a_{13}\hat a_{12}\hat a_{6}\hat a_{3}\hat a_{2}\hat a_{1} \\
f_{4} &= \hat a_{1}^\dagger\hat a_{2}^\dagger\hat a_{3}^\dagger\hat a_{7}^\dagger\hat a_{12}^\dagger\hat a_{13}^\dagger\hat a_{14}^\dagger\hat a_{16}^\dagger, & w_{4} &= \hat a_{16}\hat a_{14}\hat a_{13}\hat a_{12}\hat a_{7}\hat a_{3}\hat a_{2}\hat a_{1} \\
f_{5} &= \hat a_{1}^\dagger\hat a_{2}^\dagger\hat a_{3}^\dagger\hat a_{8}^\dagger\hat a_{12}^\dagger\hat a_{13}^\dagger\hat a_{14}^\dagger\hat a_{15}^\dagger, & w_{5} &= \hat a_{15}\hat a_{14}\hat a_{13}\hat a_{12}\hat a_{8}\hat a_{3}\hat a_{2}\hat a_{1} 
\end{align}
\paragraph{Determinants $D_{6}$--$D_{10}$.}
\begin{align}
f_{6} &= \hat a_{1}^\dagger\hat a_{2}^\dagger\hat a_{4}^\dagger\hat a_{5}^\dagger\hat a_{11}^\dagger\hat a_{14}^\dagger\hat a_{15}^\dagger\hat a_{16}^\dagger, & w_{6} &= \hat a_{16}\hat a_{15}\hat a_{14}\hat a_{11}\hat a_{5}\hat a_{4}\hat a_{2}\hat a_{1} \\
f_{7} &= \hat a_{1}^\dagger\hat a_{2}^\dagger\hat a_{4}^\dagger\hat a_{6}^\dagger\hat a_{11}^\dagger\hat a_{13}^\dagger\hat a_{15}^\dagger\hat a_{16}^\dagger, & w_{7} &= \hat a_{16}\hat a_{15}\hat a_{13}\hat a_{11}\hat a_{6}\hat a_{4}\hat a_{2}\hat a_{1} \\
f_{8} &= \hat a_{1}^\dagger\hat a_{2}^\dagger\hat a_{4}^\dagger\hat a_{7}^\dagger\hat a_{11}^\dagger\hat a_{13}^\dagger\hat a_{14}^\dagger\hat a_{16}^\dagger, & w_{8} &= \hat a_{16}\hat a_{14}\hat a_{13}\hat a_{11}\hat a_{7}\hat a_{4}\hat a_{2}\hat a_{1} \\
f_{9} &= \hat a_{1}^\dagger\hat a_{2}^\dagger\hat a_{4}^\dagger\hat a_{8}^\dagger\hat a_{11}^\dagger\hat a_{13}^\dagger\hat a_{14}^\dagger\hat a_{15}^\dagger, & w_{9} &= \hat a_{15}\hat a_{14}\hat a_{13}\hat a_{11}\hat a_{8}\hat a_{4}\hat a_{2}\hat a_{1} \\
f_{10} &= \hat a_{1}^\dagger\hat a_{2}^\dagger\hat a_{5}^\dagger\hat a_{6}^\dagger\hat a_{11}^\dagger\hat a_{12}^\dagger\hat a_{15}^\dagger\hat a_{16}^\dagger, & w_{10} &= \hat a_{16}\hat a_{15}\hat a_{12}\hat a_{11}\hat a_{6}\hat a_{5}\hat a_{2}\hat a_{1} 
\end{align}
\paragraph{Determinants $D_{11}$--$D_{15}$.}
\begin{align}
f_{11} &= \hat a_{1}^\dagger\hat a_{2}^\dagger\hat a_{5}^\dagger\hat a_{7}^\dagger\hat a_{11}^\dagger\hat a_{12}^\dagger\hat a_{14}^\dagger\hat a_{16}^\dagger, & w_{11} &= \hat a_{16}\hat a_{14}\hat a_{12}\hat a_{11}\hat a_{7}\hat a_{5}\hat a_{2}\hat a_{1} \\
f_{12} &= \hat a_{1}^\dagger\hat a_{2}^\dagger\hat a_{5}^\dagger\hat a_{8}^\dagger\hat a_{11}^\dagger\hat a_{12}^\dagger\hat a_{14}^\dagger\hat a_{15}^\dagger, & w_{12} &= \hat a_{15}\hat a_{14}\hat a_{12}\hat a_{11}\hat a_{8}\hat a_{5}\hat a_{2}\hat a_{1} \\
f_{13} &= \hat a_{1}^\dagger\hat a_{2}^\dagger\hat a_{6}^\dagger\hat a_{7}^\dagger\hat a_{11}^\dagger\hat a_{12}^\dagger\hat a_{13}^\dagger\hat a_{16}^\dagger, & w_{13} &= \hat a_{16}\hat a_{13}\hat a_{12}\hat a_{11}\hat a_{7}\hat a_{6}\hat a_{2}\hat a_{1} \\
f_{14} &= \hat a_{1}^\dagger\hat a_{2}^\dagger\hat a_{6}^\dagger\hat a_{8}^\dagger\hat a_{11}^\dagger\hat a_{12}^\dagger\hat a_{13}^\dagger\hat a_{15}^\dagger, & w_{14} &= \hat a_{15}\hat a_{13}\hat a_{12}\hat a_{11}\hat a_{8}\hat a_{6}\hat a_{2}\hat a_{1} \\
f_{15} &= \hat a_{1}^\dagger\hat a_{2}^\dagger\hat a_{7}^\dagger\hat a_{8}^\dagger\hat a_{11}^\dagger\hat a_{12}^\dagger\hat a_{13}^\dagger\hat a_{14}^\dagger, & w_{15} &= \hat a_{14}\hat a_{13}\hat a_{12}\hat a_{11}\hat a_{8}\hat a_{7}\hat a_{2}\hat a_{1} 
\end{align}
\paragraph{Determinants $D_{16}$--$D_{20}$.}
\begin{align}
f_{16} &= \hat a_{1}^\dagger\hat a_{3}^\dagger\hat a_{4}^\dagger\hat a_{5}^\dagger\hat a_{10}^\dagger\hat a_{14}^\dagger\hat a_{15}^\dagger\hat a_{16}^\dagger, & w_{16} &= \hat a_{16}\hat a_{15}\hat a_{14}\hat a_{10}\hat a_{5}\hat a_{4}\hat a_{3}\hat a_{1} \\
f_{17} &= \hat a_{1}^\dagger\hat a_{3}^\dagger\hat a_{4}^\dagger\hat a_{6}^\dagger\hat a_{10}^\dagger\hat a_{13}^\dagger\hat a_{15}^\dagger\hat a_{16}^\dagger, & w_{17} &= \hat a_{16}\hat a_{15}\hat a_{13}\hat a_{10}\hat a_{6}\hat a_{4}\hat a_{3}\hat a_{1} \\
f_{18} &= \hat a_{1}^\dagger\hat a_{3}^\dagger\hat a_{4}^\dagger\hat a_{7}^\dagger\hat a_{10}^\dagger\hat a_{13}^\dagger\hat a_{14}^\dagger\hat a_{16}^\dagger, & w_{18} &= \hat a_{16}\hat a_{14}\hat a_{13}\hat a_{10}\hat a_{7}\hat a_{4}\hat a_{3}\hat a_{1} \\
f_{19} &= \hat a_{1}^\dagger\hat a_{3}^\dagger\hat a_{4}^\dagger\hat a_{8}^\dagger\hat a_{10}^\dagger\hat a_{13}^\dagger\hat a_{14}^\dagger\hat a_{15}^\dagger, & w_{19} &= \hat a_{15}\hat a_{14}\hat a_{13}\hat a_{10}\hat a_{8}\hat a_{4}\hat a_{3}\hat a_{1} \\
f_{20} &= \hat a_{1}^\dagger\hat a_{3}^\dagger\hat a_{5}^\dagger\hat a_{6}^\dagger\hat a_{10}^\dagger\hat a_{12}^\dagger\hat a_{15}^\dagger\hat a_{16}^\dagger, & w_{20} &= \hat a_{16}\hat a_{15}\hat a_{12}\hat a_{10}\hat a_{6}\hat a_{5}\hat a_{3}\hat a_{1} 
\end{align}
\paragraph{Determinants $D_{21}$--$D_{25}$.}
\begin{align}
f_{21} &= \hat a_{1}^\dagger\hat a_{3}^\dagger\hat a_{5}^\dagger\hat a_{7}^\dagger\hat a_{10}^\dagger\hat a_{12}^\dagger\hat a_{14}^\dagger\hat a_{16}^\dagger, & w_{21} &= \hat a_{16}\hat a_{14}\hat a_{12}\hat a_{10}\hat a_{7}\hat a_{5}\hat a_{3}\hat a_{1} \\
f_{22} &= \hat a_{1}^\dagger\hat a_{3}^\dagger\hat a_{5}^\dagger\hat a_{8}^\dagger\hat a_{10}^\dagger\hat a_{12}^\dagger\hat a_{14}^\dagger\hat a_{15}^\dagger, & w_{22} &= \hat a_{15}\hat a_{14}\hat a_{12}\hat a_{10}\hat a_{8}\hat a_{5}\hat a_{3}\hat a_{1} \\
f_{23} &= \hat a_{1}^\dagger\hat a_{3}^\dagger\hat a_{6}^\dagger\hat a_{7}^\dagger\hat a_{10}^\dagger\hat a_{12}^\dagger\hat a_{13}^\dagger\hat a_{16}^\dagger, & w_{23} &= \hat a_{16}\hat a_{13}\hat a_{12}\hat a_{10}\hat a_{7}\hat a_{6}\hat a_{3}\hat a_{1} \\
f_{24} &= \hat a_{1}^\dagger\hat a_{3}^\dagger\hat a_{6}^\dagger\hat a_{8}^\dagger\hat a_{10}^\dagger\hat a_{12}^\dagger\hat a_{13}^\dagger\hat a_{15}^\dagger, & w_{24} &= \hat a_{15}\hat a_{13}\hat a_{12}\hat a_{10}\hat a_{8}\hat a_{6}\hat a_{3}\hat a_{1} \\
f_{25} &= \hat a_{1}^\dagger\hat a_{3}^\dagger\hat a_{7}^\dagger\hat a_{8}^\dagger\hat a_{10}^\dagger\hat a_{12}^\dagger\hat a_{13}^\dagger\hat a_{14}^\dagger, & w_{25} &= \hat a_{14}\hat a_{13}\hat a_{12}\hat a_{10}\hat a_{8}\hat a_{7}\hat a_{3}\hat a_{1} 
\end{align}
\paragraph{Determinants $D_{26}$--$D_{30}$.}
\begin{align}
f_{26} &= \hat a_{1}^\dagger\hat a_{4}^\dagger\hat a_{5}^\dagger\hat a_{6}^\dagger\hat a_{10}^\dagger\hat a_{11}^\dagger\hat a_{15}^\dagger\hat a_{16}^\dagger, & w_{26} &= \hat a_{16}\hat a_{15}\hat a_{11}\hat a_{10}\hat a_{6}\hat a_{5}\hat a_{4}\hat a_{1} \\
f_{27} &= \hat a_{1}^\dagger\hat a_{4}^\dagger\hat a_{5}^\dagger\hat a_{7}^\dagger\hat a_{10}^\dagger\hat a_{11}^\dagger\hat a_{14}^\dagger\hat a_{16}^\dagger, & w_{27} &= \hat a_{16}\hat a_{14}\hat a_{11}\hat a_{10}\hat a_{7}\hat a_{5}\hat a_{4}\hat a_{1} \\
f_{28} &= \hat a_{1}^\dagger\hat a_{4}^\dagger\hat a_{5}^\dagger\hat a_{8}^\dagger\hat a_{10}^\dagger\hat a_{11}^\dagger\hat a_{14}^\dagger\hat a_{15}^\dagger, & w_{28} &= \hat a_{15}\hat a_{14}\hat a_{11}\hat a_{10}\hat a_{8}\hat a_{5}\hat a_{4}\hat a_{1} \\
f_{29} &= \hat a_{1}^\dagger\hat a_{4}^\dagger\hat a_{6}^\dagger\hat a_{7}^\dagger\hat a_{10}^\dagger\hat a_{11}^\dagger\hat a_{13}^\dagger\hat a_{16}^\dagger, & w_{29} &= \hat a_{16}\hat a_{13}\hat a_{11}\hat a_{10}\hat a_{7}\hat a_{6}\hat a_{4}\hat a_{1} \\
f_{30} &= \hat a_{1}^\dagger\hat a_{4}^\dagger\hat a_{6}^\dagger\hat a_{8}^\dagger\hat a_{10}^\dagger\hat a_{11}^\dagger\hat a_{13}^\dagger\hat a_{15}^\dagger, & w_{30} &= \hat a_{15}\hat a_{13}\hat a_{11}\hat a_{10}\hat a_{8}\hat a_{6}\hat a_{4}\hat a_{1} 
\end{align}
\paragraph{Determinants $D_{31}$--$D_{35}$.}
\begin{align}
f_{31} &= \hat a_{1}^\dagger\hat a_{4}^\dagger\hat a_{7}^\dagger\hat a_{8}^\dagger\hat a_{10}^\dagger\hat a_{11}^\dagger\hat a_{13}^\dagger\hat a_{14}^\dagger, & w_{31} &= \hat a_{14}\hat a_{13}\hat a_{11}\hat a_{10}\hat a_{8}\hat a_{7}\hat a_{4}\hat a_{1} \\
f_{32} &= \hat a_{1}^\dagger\hat a_{5}^\dagger\hat a_{6}^\dagger\hat a_{7}^\dagger\hat a_{10}^\dagger\hat a_{11}^\dagger\hat a_{12}^\dagger\hat a_{16}^\dagger, & w_{32} &= \hat a_{16}\hat a_{12}\hat a_{11}\hat a_{10}\hat a_{7}\hat a_{6}\hat a_{5}\hat a_{1} \\
f_{33} &= \hat a_{1}^\dagger\hat a_{5}^\dagger\hat a_{6}^\dagger\hat a_{8}^\dagger\hat a_{10}^\dagger\hat a_{11}^\dagger\hat a_{12}^\dagger\hat a_{15}^\dagger, & w_{33} &= \hat a_{15}\hat a_{12}\hat a_{11}\hat a_{10}\hat a_{8}\hat a_{6}\hat a_{5}\hat a_{1} \\
f_{34} &= \hat a_{1}^\dagger\hat a_{5}^\dagger\hat a_{7}^\dagger\hat a_{8}^\dagger\hat a_{10}^\dagger\hat a_{11}^\dagger\hat a_{12}^\dagger\hat a_{14}^\dagger, & w_{34} &= \hat a_{14}\hat a_{12}\hat a_{11}\hat a_{10}\hat a_{8}\hat a_{7}\hat a_{5}\hat a_{1} \\
f_{35} &= \hat a_{1}^\dagger\hat a_{6}^\dagger\hat a_{7}^\dagger\hat a_{8}^\dagger\hat a_{10}^\dagger\hat a_{11}^\dagger\hat a_{12}^\dagger\hat a_{13}^\dagger, & w_{35} &= \hat a_{13}\hat a_{12}\hat a_{11}\hat a_{10}\hat a_{8}\hat a_{7}\hat a_{6}\hat a_{1} 
\end{align}
\paragraph{Determinants $D_{36}$--$D_{40}$.}
\begin{align}
f_{36} &= \hat a_{2}^\dagger\hat a_{3}^\dagger\hat a_{4}^\dagger\hat a_{5}^\dagger\hat a_{9}^\dagger\hat a_{14}^\dagger\hat a_{15}^\dagger\hat a_{16}^\dagger, & w_{36} &= \hat a_{16}\hat a_{15}\hat a_{14}\hat a_{9}\hat a_{5}\hat a_{4}\hat a_{3}\hat a_{2} \\
f_{37} &= \hat a_{2}^\dagger\hat a_{3}^\dagger\hat a_{4}^\dagger\hat a_{6}^\dagger\hat a_{9}^\dagger\hat a_{13}^\dagger\hat a_{15}^\dagger\hat a_{16}^\dagger, & w_{37} &= \hat a_{16}\hat a_{15}\hat a_{13}\hat a_{9}\hat a_{6}\hat a_{4}\hat a_{3}\hat a_{2} \\
f_{38} &= \hat a_{2}^\dagger\hat a_{3}^\dagger\hat a_{4}^\dagger\hat a_{7}^\dagger\hat a_{9}^\dagger\hat a_{13}^\dagger\hat a_{14}^\dagger\hat a_{16}^\dagger, & w_{38} &= \hat a_{16}\hat a_{14}\hat a_{13}\hat a_{9}\hat a_{7}\hat a_{4}\hat a_{3}\hat a_{2} \\
f_{39} &= \hat a_{2}^\dagger\hat a_{3}^\dagger\hat a_{4}^\dagger\hat a_{8}^\dagger\hat a_{9}^\dagger\hat a_{13}^\dagger\hat a_{14}^\dagger\hat a_{15}^\dagger, & w_{39} &= \hat a_{15}\hat a_{14}\hat a_{13}\hat a_{9}\hat a_{8}\hat a_{4}\hat a_{3}\hat a_{2} \\
f_{40} &= \hat a_{2}^\dagger\hat a_{3}^\dagger\hat a_{5}^\dagger\hat a_{6}^\dagger\hat a_{9}^\dagger\hat a_{12}^\dagger\hat a_{15}^\dagger\hat a_{16}^\dagger, & w_{40} &= \hat a_{16}\hat a_{15}\hat a_{12}\hat a_{9}\hat a_{6}\hat a_{5}\hat a_{3}\hat a_{2} 
\end{align}
\paragraph{Determinants $D_{41}$--$D_{45}$.}
\begin{align}
f_{41} &= \hat a_{2}^\dagger\hat a_{3}^\dagger\hat a_{5}^\dagger\hat a_{7}^\dagger\hat a_{9}^\dagger\hat a_{12}^\dagger\hat a_{14}^\dagger\hat a_{16}^\dagger, & w_{41} &= \hat a_{16}\hat a_{14}\hat a_{12}\hat a_{9}\hat a_{7}\hat a_{5}\hat a_{3}\hat a_{2} \\
f_{42} &= \hat a_{2}^\dagger\hat a_{3}^\dagger\hat a_{5}^\dagger\hat a_{8}^\dagger\hat a_{9}^\dagger\hat a_{12}^\dagger\hat a_{14}^\dagger\hat a_{15}^\dagger, & w_{42} &= \hat a_{15}\hat a_{14}\hat a_{12}\hat a_{9}\hat a_{8}\hat a_{5}\hat a_{3}\hat a_{2} \\
f_{43} &= \hat a_{2}^\dagger\hat a_{3}^\dagger\hat a_{6}^\dagger\hat a_{7}^\dagger\hat a_{9}^\dagger\hat a_{12}^\dagger\hat a_{13}^\dagger\hat a_{16}^\dagger, & w_{43} &= \hat a_{16}\hat a_{13}\hat a_{12}\hat a_{9}\hat a_{7}\hat a_{6}\hat a_{3}\hat a_{2} \\
f_{44} &= \hat a_{2}^\dagger\hat a_{3}^\dagger\hat a_{6}^\dagger\hat a_{8}^\dagger\hat a_{9}^\dagger\hat a_{12}^\dagger\hat a_{13}^\dagger\hat a_{15}^\dagger, & w_{44} &= \hat a_{15}\hat a_{13}\hat a_{12}\hat a_{9}\hat a_{8}\hat a_{6}\hat a_{3}\hat a_{2} \\
f_{45} &= \hat a_{2}^\dagger\hat a_{3}^\dagger\hat a_{7}^\dagger\hat a_{8}^\dagger\hat a_{9}^\dagger\hat a_{12}^\dagger\hat a_{13}^\dagger\hat a_{14}^\dagger, & w_{45} &= \hat a_{14}\hat a_{13}\hat a_{12}\hat a_{9}\hat a_{8}\hat a_{7}\hat a_{3}\hat a_{2} 
\end{align}
\paragraph{Determinants $D_{46}$--$D_{50}$.}
\begin{align}
f_{46} &= \hat a_{2}^\dagger\hat a_{4}^\dagger\hat a_{5}^\dagger\hat a_{6}^\dagger\hat a_{9}^\dagger\hat a_{11}^\dagger\hat a_{15}^\dagger\hat a_{16}^\dagger, & w_{46} &= \hat a_{16}\hat a_{15}\hat a_{11}\hat a_{9}\hat a_{6}\hat a_{5}\hat a_{4}\hat a_{2} \\
f_{47} &= \hat a_{2}^\dagger\hat a_{4}^\dagger\hat a_{5}^\dagger\hat a_{7}^\dagger\hat a_{9}^\dagger\hat a_{11}^\dagger\hat a_{14}^\dagger\hat a_{16}^\dagger, & w_{47} &= \hat a_{16}\hat a_{14}\hat a_{11}\hat a_{9}\hat a_{7}\hat a_{5}\hat a_{4}\hat a_{2} \\
f_{48} &= \hat a_{2}^\dagger\hat a_{4}^\dagger\hat a_{5}^\dagger\hat a_{8}^\dagger\hat a_{9}^\dagger\hat a_{11}^\dagger\hat a_{14}^\dagger\hat a_{15}^\dagger, & w_{48} &= \hat a_{15}\hat a_{14}\hat a_{11}\hat a_{9}\hat a_{8}\hat a_{5}\hat a_{4}\hat a_{2} \\
f_{49} &= \hat a_{2}^\dagger\hat a_{4}^\dagger\hat a_{6}^\dagger\hat a_{7}^\dagger\hat a_{9}^\dagger\hat a_{11}^\dagger\hat a_{13}^\dagger\hat a_{16}^\dagger, & w_{49} &= \hat a_{16}\hat a_{13}\hat a_{11}\hat a_{9}\hat a_{7}\hat a_{6}\hat a_{4}\hat a_{2} \\
f_{50} &= \hat a_{2}^\dagger\hat a_{4}^\dagger\hat a_{6}^\dagger\hat a_{8}^\dagger\hat a_{9}^\dagger\hat a_{11}^\dagger\hat a_{13}^\dagger\hat a_{15}^\dagger, & w_{50} &= \hat a_{15}\hat a_{13}\hat a_{11}\hat a_{9}\hat a_{8}\hat a_{6}\hat a_{4}\hat a_{2} 
\end{align}
\paragraph{Determinants $D_{51}$--$D_{55}$.}
\begin{align}
f_{51} &= \hat a_{2}^\dagger\hat a_{4}^\dagger\hat a_{7}^\dagger\hat a_{8}^\dagger\hat a_{9}^\dagger\hat a_{11}^\dagger\hat a_{13}^\dagger\hat a_{14}^\dagger, & w_{51} &= \hat a_{14}\hat a_{13}\hat a_{11}\hat a_{9}\hat a_{8}\hat a_{7}\hat a_{4}\hat a_{2} \\
f_{52} &= \hat a_{2}^\dagger\hat a_{5}^\dagger\hat a_{6}^\dagger\hat a_{7}^\dagger\hat a_{9}^\dagger\hat a_{11}^\dagger\hat a_{12}^\dagger\hat a_{16}^\dagger, & w_{52} &= \hat a_{16}\hat a_{12}\hat a_{11}\hat a_{9}\hat a_{7}\hat a_{6}\hat a_{5}\hat a_{2} \\
f_{53} &= \hat a_{2}^\dagger\hat a_{5}^\dagger\hat a_{6}^\dagger\hat a_{8}^\dagger\hat a_{9}^\dagger\hat a_{11}^\dagger\hat a_{12}^\dagger\hat a_{15}^\dagger, & w_{53} &= \hat a_{15}\hat a_{12}\hat a_{11}\hat a_{9}\hat a_{8}\hat a_{6}\hat a_{5}\hat a_{2} \\
f_{54} &= \hat a_{2}^\dagger\hat a_{5}^\dagger\hat a_{7}^\dagger\hat a_{8}^\dagger\hat a_{9}^\dagger\hat a_{11}^\dagger\hat a_{12}^\dagger\hat a_{14}^\dagger, & w_{54} &= \hat a_{14}\hat a_{12}\hat a_{11}\hat a_{9}\hat a_{8}\hat a_{7}\hat a_{5}\hat a_{2} \\
f_{55} &= \hat a_{2}^\dagger\hat a_{6}^\dagger\hat a_{7}^\dagger\hat a_{8}^\dagger\hat a_{9}^\dagger\hat a_{11}^\dagger\hat a_{12}^\dagger\hat a_{13}^\dagger, & w_{55} &= \hat a_{13}\hat a_{12}\hat a_{11}\hat a_{9}\hat a_{8}\hat a_{7}\hat a_{6}\hat a_{2} 
\end{align}
\paragraph{Determinants $D_{56}$--$D_{60}$.}
\begin{align}
f_{56} &= \hat a_{3}^\dagger\hat a_{4}^\dagger\hat a_{5}^\dagger\hat a_{6}^\dagger\hat a_{9}^\dagger\hat a_{10}^\dagger\hat a_{15}^\dagger\hat a_{16}^\dagger, & w_{56} &= \hat a_{16}\hat a_{15}\hat a_{10}\hat a_{9}\hat a_{6}\hat a_{5}\hat a_{4}\hat a_{3} \\
f_{57} &= \hat a_{3}^\dagger\hat a_{4}^\dagger\hat a_{5}^\dagger\hat a_{7}^\dagger\hat a_{9}^\dagger\hat a_{10}^\dagger\hat a_{14}^\dagger\hat a_{16}^\dagger, & w_{57} &= \hat a_{16}\hat a_{14}\hat a_{10}\hat a_{9}\hat a_{7}\hat a_{5}\hat a_{4}\hat a_{3} \\
f_{58} &= \hat a_{3}^\dagger\hat a_{4}^\dagger\hat a_{5}^\dagger\hat a_{8}^\dagger\hat a_{9}^\dagger\hat a_{10}^\dagger\hat a_{14}^\dagger\hat a_{15}^\dagger, & w_{58} &= \hat a_{15}\hat a_{14}\hat a_{10}\hat a_{9}\hat a_{8}\hat a_{5}\hat a_{4}\hat a_{3} \\
f_{59} &= \hat a_{3}^\dagger\hat a_{4}^\dagger\hat a_{6}^\dagger\hat a_{7}^\dagger\hat a_{9}^\dagger\hat a_{10}^\dagger\hat a_{13}^\dagger\hat a_{16}^\dagger, & w_{59} &= \hat a_{16}\hat a_{13}\hat a_{10}\hat a_{9}\hat a_{7}\hat a_{6}\hat a_{4}\hat a_{3} \\
f_{60} &= \hat a_{3}^\dagger\hat a_{4}^\dagger\hat a_{6}^\dagger\hat a_{8}^\dagger\hat a_{9}^\dagger\hat a_{10}^\dagger\hat a_{13}^\dagger\hat a_{15}^\dagger, & w_{60} &= \hat a_{15}\hat a_{13}\hat a_{10}\hat a_{9}\hat a_{8}\hat a_{6}\hat a_{4}\hat a_{3} 
\end{align}
\paragraph{Determinants $D_{61}$--$D_{65}$.}
\begin{align}
f_{61} &= \hat a_{3}^\dagger\hat a_{4}^\dagger\hat a_{7}^\dagger\hat a_{8}^\dagger\hat a_{9}^\dagger\hat a_{10}^\dagger\hat a_{13}^\dagger\hat a_{14}^\dagger, & w_{61} &= \hat a_{14}\hat a_{13}\hat a_{10}\hat a_{9}\hat a_{8}\hat a_{7}\hat a_{4}\hat a_{3} \\
f_{62} &= \hat a_{3}^\dagger\hat a_{5}^\dagger\hat a_{6}^\dagger\hat a_{7}^\dagger\hat a_{9}^\dagger\hat a_{10}^\dagger\hat a_{12}^\dagger\hat a_{16}^\dagger, & w_{62} &= \hat a_{16}\hat a_{12}\hat a_{10}\hat a_{9}\hat a_{7}\hat a_{6}\hat a_{5}\hat a_{3} \\
f_{63} &= \hat a_{3}^\dagger\hat a_{5}^\dagger\hat a_{6}^\dagger\hat a_{8}^\dagger\hat a_{9}^\dagger\hat a_{10}^\dagger\hat a_{12}^\dagger\hat a_{15}^\dagger, & w_{63} &= \hat a_{15}\hat a_{12}\hat a_{10}\hat a_{9}\hat a_{8}\hat a_{6}\hat a_{5}\hat a_{3} \\
f_{64} &= \hat a_{3}^\dagger\hat a_{5}^\dagger\hat a_{7}^\dagger\hat a_{8}^\dagger\hat a_{9}^\dagger\hat a_{10}^\dagger\hat a_{12}^\dagger\hat a_{14}^\dagger, & w_{64} &= \hat a_{14}\hat a_{12}\hat a_{10}\hat a_{9}\hat a_{8}\hat a_{7}\hat a_{5}\hat a_{3} \\
f_{65} &= \hat a_{3}^\dagger\hat a_{6}^\dagger\hat a_{7}^\dagger\hat a_{8}^\dagger\hat a_{9}^\dagger\hat a_{10}^\dagger\hat a_{12}^\dagger\hat a_{13}^\dagger, & w_{65} &= \hat a_{13}\hat a_{12}\hat a_{10}\hat a_{9}\hat a_{8}\hat a_{7}\hat a_{6}\hat a_{3} 
\end{align}
\paragraph{Determinants $D_{66}$--$D_{70}$.}
\begin{align}
f_{66} &= \hat a_{4}^\dagger\hat a_{5}^\dagger\hat a_{6}^\dagger\hat a_{7}^\dagger\hat a_{9}^\dagger\hat a_{10}^\dagger\hat a_{11}^\dagger\hat a_{16}^\dagger, & w_{66} &= \hat a_{16}\hat a_{11}\hat a_{10}\hat a_{9}\hat a_{7}\hat a_{6}\hat a_{5}\hat a_{4} \\
f_{67} &= \hat a_{4}^\dagger\hat a_{5}^\dagger\hat a_{6}^\dagger\hat a_{8}^\dagger\hat a_{9}^\dagger\hat a_{10}^\dagger\hat a_{11}^\dagger\hat a_{15}^\dagger, & w_{67} &= \hat a_{15}\hat a_{11}\hat a_{10}\hat a_{9}\hat a_{8}\hat a_{6}\hat a_{5}\hat a_{4} \\
f_{68} &= \hat a_{4}^\dagger\hat a_{5}^\dagger\hat a_{7}^\dagger\hat a_{8}^\dagger\hat a_{9}^\dagger\hat a_{10}^\dagger\hat a_{11}^\dagger\hat a_{14}^\dagger, & w_{68} &= \hat a_{14}\hat a_{11}\hat a_{10}\hat a_{9}\hat a_{8}\hat a_{7}\hat a_{5}\hat a_{4} \\
f_{69} &= \hat a_{4}^\dagger\hat a_{6}^\dagger\hat a_{7}^\dagger\hat a_{8}^\dagger\hat a_{9}^\dagger\hat a_{10}^\dagger\hat a_{11}^\dagger\hat a_{13}^\dagger, & w_{69} &= \hat a_{13}\hat a_{11}\hat a_{10}\hat a_{9}\hat a_{8}\hat a_{7}\hat a_{6}\hat a_{4} \\
f_{70} &= \hat a_{5}^\dagger\hat a_{6}^\dagger\hat a_{7}^\dagger\hat a_{8}^\dagger\hat a_{9}^\dagger\hat a_{10}^\dagger\hat a_{11}^\dagger\hat a_{12}^\dagger, & w_{70} &= \hat a_{12}\hat a_{11}\hat a_{10}\hat a_{9}\hat a_{8}\hat a_{7}\hat a_{6}\hat a_{5} 
\end{align}

\subsection{Spin-coupled branches in the determinant basis}
\label{app:C2_branch_Dbasis}

Using the fourteen Rumer singlet branches, each spin-coupled structure contains
sixteen determinant terms.  In the determinant labels of
Table~\ref{tab:C2_complete_alpha_beta_bitstrings}, the branches are

\begin{align}
\psi^8_{0,0;1} &= \frac{1}{4\sqrt{8!}}\Bigl(\ket{D_{21}} -\ket{D_{22}} -\ket{D_{23}} +\ket{D_{24}} -\ket{D_{27}} +\ket{D_{28}} +\ket{D_{29}} -\ket{D_{30}}
\nonumber\\ &\hspace{2.6cm} -\ket{D_{41}} +\ket{D_{42}} +\ket{D_{43}} -\ket{D_{44}} +\ket{D_{47}} -\ket{D_{48}} -\ket{D_{49}} +\ket{D_{50}}\Bigr),
\\[1mm]\psi^8_{0,0;2} &= \frac{1}{4\sqrt{8!}}\Bigl(\ket{D_{20}} -\ket{D_{21}} -\ket{D_{24}} +\ket{D_{25}} -\ket{D_{26}} +\ket{D_{27}} +\ket{D_{30}} -\ket{D_{31}}
\nonumber\\ &\hspace{2.6cm} -\ket{D_{40}} +\ket{D_{41}} +\ket{D_{44}} -\ket{D_{45}} +\ket{D_{46}} -\ket{D_{47}} -\ket{D_{50}} +\ket{D_{51}}\Bigr),
\\[1mm]\psi^8_{0,0;3} &= \frac{1}{4\sqrt{8!}}\Bigl(\ket{D_{18}} -\ket{D_{19}} -\ket{D_{21}} +\ket{D_{22}} -\ket{D_{29}} +\ket{D_{30}} +\ket{D_{32}} -\ket{D_{33}}
\nonumber\\ &\hspace{2.6cm} -\ket{D_{38}} +\ket{D_{39}} +\ket{D_{41}} -\ket{D_{42}} +\ket{D_{49}} -\ket{D_{50}} -\ket{D_{52}} +\ket{D_{53}}\Bigr),
\\[1mm]\psi^8_{0,0;4} &= \frac{1}{4\sqrt{8!}}\Bigl(\ket{D_{17}} -\ket{D_{18}} -\ket{D_{20}} +\ket{D_{21}} -\ket{D_{30}} +\ket{D_{31}} +\ket{D_{33}} -\ket{D_{34}}
\nonumber\\ &\hspace{2.6cm} -\ket{D_{37}} +\ket{D_{38}} +\ket{D_{40}} -\ket{D_{41}} +\ket{D_{50}} -\ket{D_{51}} -\ket{D_{53}} +\ket{D_{54}}\Bigr),
\\[1mm]\psi^8_{0,0;5} &= \frac{1}{4\sqrt{8!}}\Bigl(\ket{D_{16}} -\ket{D_{17}} -\ket{D_{21}} +\ket{D_{23}} -\ket{D_{28}} +\ket{D_{30}} +\ket{D_{34}} -\ket{D_{35}}
\nonumber\\ &\hspace{2.6cm} -\ket{D_{36}} +\ket{D_{37}} +\ket{D_{41}} -\ket{D_{43}} +\ket{D_{48}} -\ket{D_{50}} -\ket{D_{54}} +\ket{D_{55}}\Bigr),
\\[1mm]\psi^8_{0,0;6} &= \frac{1}{4\sqrt{8!}}\Bigl(\ket{D_{11}} -\ket{D_{12}} -\ket{D_{13}} +\ket{D_{14}} -\ket{D_{21}} +\ket{D_{22}} +\ket{D_{23}} -\ket{D_{24}}
\nonumber\\ &\hspace{2.6cm} -\ket{D_{47}} +\ket{D_{48}} +\ket{D_{49}} -\ket{D_{50}} +\ket{D_{57}} -\ket{D_{58}} -\ket{D_{59}} +\ket{D_{60}}\Bigr),
\\[1mm]\psi^8_{0,0;7} &= \frac{1}{4\sqrt{8!}}\Bigl(\ket{D_{10}} -\ket{D_{11}} -\ket{D_{14}} +\ket{D_{15}} -\ket{D_{20}} +\ket{D_{21}} +\ket{D_{24}} -\ket{D_{25}}
\nonumber\\ &\hspace{2.6cm} -\ket{D_{46}} +\ket{D_{47}} +\ket{D_{50}} -\ket{D_{51}} +\ket{D_{56}} -\ket{D_{57}} -\ket{D_{60}} +\ket{D_{61}}\Bigr),
\\[1mm]\psi^8_{0,0;8} &= \frac{1}{4\sqrt{8!}}\Bigl(\ket{D_{8}} -\ket{D_{9}} -\ket{D_{11}} +\ket{D_{12}} -\ket{D_{18}} +\ket{D_{19}} +\ket{D_{21}} -\ket{D_{22}}
\nonumber\\ &\hspace{2.6cm} -\ket{D_{49}} +\ket{D_{50}} +\ket{D_{52}} -\ket{D_{53}} +\ket{D_{59}} -\ket{D_{60}} -\ket{D_{62}} +\ket{D_{63}}\Bigr),
\\[1mm]\psi^8_{0,0;9} &= \frac{1}{4\sqrt{8!}}\Bigl(\ket{D_{4}} -\ket{D_{5}} -\ket{D_{8}} +\ket{D_{9}} -\ket{D_{21}} +\ket{D_{22}} +\ket{D_{27}} -\ket{D_{28}}
\nonumber\\ &\hspace{2.6cm} -\ket{D_{43}} +\ket{D_{44}} +\ket{D_{49}} -\ket{D_{50}} +\ket{D_{62}} -\ket{D_{63}} -\ket{D_{66}} +\ket{D_{67}}\Bigr),
\\[1mm]\psi^8_{0,0;10} &= \frac{1}{4\sqrt{8!}}\Bigl(\ket{D_{7}} -\ket{D_{8}} -\ket{D_{10}} +\ket{D_{11}} -\ket{D_{17}} +\ket{D_{18}} +\ket{D_{20}} -\ket{D_{21}}
\nonumber\\ &\hspace{2.6cm} -\ket{D_{50}} +\ket{D_{51}} +\ket{D_{53}} -\ket{D_{54}} +\ket{D_{60}} -\ket{D_{61}} -\ket{D_{63}} +\ket{D_{64}}\Bigr),
\\[1mm]\psi^8_{0,0;11} &= \frac{1}{4\sqrt{8!}}\Bigl(\ket{D_{6}} -\ket{D_{7}} -\ket{D_{11}} +\ket{D_{13}} -\ket{D_{16}} +\ket{D_{17}} +\ket{D_{21}} -\ket{D_{23}}
\nonumber\\ &\hspace{2.6cm} -\ket{D_{48}} +\ket{D_{50}} +\ket{D_{54}} -\ket{D_{55}} +\ket{D_{58}} -\ket{D_{60}} -\ket{D_{64}} +\ket{D_{65}}\Bigr),
\\[1mm]\psi^8_{0,0;12} &= \frac{1}{4\sqrt{8!}}\Bigl(\ket{D_{3}} -\ket{D_{4}} -\ket{D_{7}} +\ket{D_{8}} -\ket{D_{20}} +\ket{D_{21}} +\ket{D_{26}} -\ket{D_{27}}
\nonumber\\ &\hspace{2.6cm} -\ket{D_{44}} +\ket{D_{45}} +\ket{D_{50}} -\ket{D_{51}} +\ket{D_{63}} -\ket{D_{64}} -\ket{D_{67}} +\ket{D_{68}}\Bigr),
\\[1mm]\psi^8_{0,0;13} &= \frac{1}{4\sqrt{8!}}\Bigl(\ket{D_{2}} -\ket{D_{3}} -\ket{D_{6}} +\ket{D_{7}} -\ket{D_{21}} +\ket{D_{23}} +\ket{D_{27}} -\ket{D_{29}}
\nonumber\\ &\hspace{2.6cm} -\ket{D_{42}} +\ket{D_{44}} +\ket{D_{48}} -\ket{D_{50}} +\ket{D_{64}} -\ket{D_{65}} -\ket{D_{68}} +\ket{D_{69}}\Bigr),
\\[1mm]\psi^8_{0,0;14} &= \frac{1}{4\sqrt{8!}}\Bigl(\ket{D_{1}} -\ket{D_{2}} -\ket{D_{7}} +\ket{D_{10}} -\ket{D_{18}} +\ket{D_{21}} +\ket{D_{29}} -\ket{D_{32}}
\nonumber\\ &\hspace{2.6cm} -\ket{D_{39}} +\ket{D_{42}} +\ket{D_{50}} -\ket{D_{53}} +\ket{D_{61}} -\ket{D_{64}} -\ket{D_{69}} +\ket{D_{70}}\Bigr).
\end{align}

\subsection{Reduction of overlaps and Hamiltonian elements to vacuum expectation values}
\label{app:C2_vacuum_reduction}

Using the determinant labels above,
\begin{equation}
  S_{IJ}=\braket{D_I}{D_J}
  =
  \bra{\phi_0}w_I f_J\ket{\phi_0},
  \qquad I,J=1,\ldots,70,
\end{equation}
and
\begin{equation}
  H_{IJ}=\bra{D_I}\hat H\ket{D_J}
  =
  \bra{\phi_0}w_I \hat H f_J\ket{\phi_0},
  \qquad I,J=1,\ldots,70.
\end{equation}

\subsection{NO--JW mapping of annihilation operators for the C$_2$ active space}
\label{app:C2_nojw_annihilation}

For nonorthogonal spin-orbitals, the annihilation operator admits the expansion
\begin{equation}
\hat{a}_{p}
=
\sum_{q=1}^{16}
\mathcal{S}_{pq}
\left(\prod_{k=1}^{q-1} Z_k\right)
\frac{X_q+iY_q}{2}
\left(\prod_{k=q+1}^{16} I_k\right),
\label{eq:C2_nojw_general}
\end{equation}
where $\mathcal{S}_{pq}$ are overlap matrix elements between the 16 active
spin-orbitals.  The spin-orbital ordering is
\begin{equation}
1,\ldots,8\equiv \alpha_1,\ldots,\alpha_8,
\qquad
9,\ldots,16\equiv \beta_1,\ldots,\beta_8 .
\end{equation}

The general mapping in Eq.~\eqref{eq:C2_nojw_general}, together with the spin-orbital ordering above, is sufficient to reconstruct every active-space annihilation operator, so an explicit operator-by-operator listing is unnecessary.

\subsection{Fully explicit usual JW creation operators for the C$_2$ active space}
\label{app:C2_jw_creation_explicit}

For completeness, the creation operators used in the determinant strings $f_I$
are listed below with all $Z$ strings and identities shown explicitly.

\begin{align}
\hat a_{1}^\dagger
={}&
\frac{X_{1}-iY_{1}}{2}\otimes I_{2}\otimes I_{3}\otimes I_{4}\otimes I_{5}\otimes I_{6}\otimes I_{7}\otimes I_{8}\otimes I_{9}\otimes I_{10}\otimes I_{11}\otimes I_{12}\otimes I_{13}\otimes I_{14}\otimes I_{15}\otimes I_{16}.
\end{align}
\begin{align}
\hat a_{2}^\dagger
={}&
Z_{1}\otimes \frac{X_{2}-iY_{2}}{2}\otimes I_{3}\otimes I_{4}\otimes I_{5}\otimes I_{6}\otimes I_{7}\otimes I_{8}\otimes I_{9}\otimes I_{10}\otimes I_{11}\otimes I_{12}\otimes I_{13}\otimes I_{14}\otimes I_{15}\otimes I_{16}.
\end{align}
\begin{align}
\hat a_{3}^\dagger
={}&
Z_{1}\otimes Z_{2}\otimes \frac{X_{3}-iY_{3}}{2}\otimes I_{4}\otimes I_{5}\otimes I_{6}\otimes I_{7}\otimes I_{8}\otimes I_{9}\otimes I_{10}\otimes I_{11}\otimes I_{12}\otimes I_{13}\otimes I_{14}\otimes I_{15}\otimes I_{16}.
\end{align}
\begin{align}
\hat a_{4}^\dagger
={}&
Z_{1}\otimes Z_{2}\otimes Z_{3}\otimes \frac{X_{4}-iY_{4}}{2}\otimes I_{5}\otimes I_{6}\otimes I_{7}\otimes I_{8}\otimes I_{9}\otimes I_{10}\otimes I_{11}\otimes I_{12}\otimes I_{13}\otimes I_{14}\otimes I_{15}\otimes I_{16}.
\end{align}
\begin{align}
\hat a_{5}^\dagger
={}&
Z_{1}\otimes Z_{2}\otimes Z_{3}\otimes Z_{4}\otimes \frac{X_{5}-iY_{5}}{2}\otimes I_{6}\otimes I_{7}\otimes I_{8}\otimes I_{9}\otimes I_{10}\otimes I_{11}\otimes I_{12}\otimes I_{13}\otimes I_{14}\otimes I_{15}\otimes I_{16}.
\end{align}
\begin{align}
\hat a_{6}^\dagger
={}&
Z_{1}\otimes Z_{2}\otimes Z_{3}\otimes Z_{4}\otimes Z_{5}\otimes \frac{X_{6}-iY_{6}}{2}\otimes I_{7}\otimes I_{8}\otimes I_{9}\otimes I_{10}\otimes I_{11}\otimes I_{12}\otimes I_{13}\otimes I_{14}\otimes I_{15}\otimes I_{16}.
\end{align}
\begin{align}
\hat a_{7}^\dagger
={}&
Z_{1}\otimes Z_{2}\otimes Z_{3}\otimes Z_{4}\otimes Z_{5}\otimes Z_{6}\otimes \frac{X_{7}-iY_{7}}{2}\otimes I_{8}\otimes I_{9}\otimes I_{10}\otimes I_{11}\otimes I_{12}\otimes I_{13}\otimes I_{14}\otimes I_{15}\otimes I_{16}.
\end{align}
\begin{align}
\hat a_{8}^\dagger
={}&
Z_{1}\otimes Z_{2}\otimes Z_{3}\otimes Z_{4}\otimes Z_{5}\otimes Z_{6}\otimes Z_{7}\otimes \frac{X_{8}-iY_{8}}{2}\otimes I_{9}\otimes I_{10}\otimes I_{11}\otimes I_{12}\otimes I_{13}\otimes I_{14}\otimes I_{15}\otimes I_{16}.
\end{align}
\begin{align}
\hat a_{9}^\dagger
={}&
Z_{1}\otimes Z_{2}\otimes Z_{3}\otimes Z_{4}\otimes Z_{5}\otimes Z_{6}\otimes Z_{7}\otimes Z_{8}\otimes \frac{X_{9}-iY_{9}}{2}\otimes I_{10}\otimes I_{11}\otimes I_{12}\otimes I_{13}\otimes I_{14}\otimes I_{15}\otimes I_{16}.
\end{align}
\begin{align}
\hat a_{10}^\dagger
={}&
Z_{1}\otimes Z_{2}\otimes Z_{3}\otimes Z_{4}\otimes Z_{5}\otimes Z_{6}\otimes Z_{7}\otimes Z_{8}\otimes Z_{9}\otimes \frac{X_{10}-iY_{10}}{2}\otimes I_{11}\otimes I_{12}\otimes I_{13}\otimes I_{14}\otimes I_{15}\otimes I_{16}.
\end{align}
\begin{align}
\hat a_{11}^\dagger
={}&
Z_{1}\otimes Z_{2}\otimes Z_{3}\otimes Z_{4}\otimes Z_{5}\otimes Z_{6}\otimes Z_{7}\otimes Z_{8}\otimes Z_{9}\otimes Z_{10}\otimes \frac{X_{11}-iY_{11}}{2}\otimes I_{12}\otimes I_{13}\otimes I_{14}\otimes I_{15}\otimes I_{16}.
\end{align}
\begin{align}
\hat a_{12}^\dagger
={}&
Z_{1}\otimes Z_{2}\otimes Z_{3}\otimes Z_{4}\otimes Z_{5}\otimes Z_{6}\otimes Z_{7}\otimes Z_{8}\otimes Z_{9}\otimes Z_{10}\otimes Z_{11}\otimes \frac{X_{12}-iY_{12}}{2}\otimes I_{13}\otimes I_{14}\otimes I_{15}\otimes I_{16}.
\end{align}
\begin{align}
\hat a_{13}^\dagger
={}&
Z_{1}\otimes Z_{2}\otimes Z_{3}\otimes Z_{4}\otimes Z_{5}\otimes Z_{6}\otimes Z_{7}\otimes Z_{8}\otimes Z_{9}\otimes Z_{10}\otimes Z_{11}\otimes Z_{12}\otimes \frac{X_{13}-iY_{13}}{2}\otimes I_{14}\otimes I_{15}\otimes I_{16}.
\end{align}
\begin{align}
\hat a_{14}^\dagger
={}&
Z_{1}\otimes Z_{2}\otimes Z_{3}\otimes Z_{4}\otimes Z_{5}\otimes Z_{6}\otimes Z_{7}\otimes Z_{8}\otimes Z_{9}\otimes Z_{10}\otimes Z_{11}\otimes Z_{12}\otimes Z_{13}\otimes \frac{X_{14}-iY_{14}}{2}\otimes I_{15}\otimes I_{16}.
\end{align}
\begin{align}
\hat a_{15}^\dagger
={}&
Z_{1}\otimes Z_{2}\otimes Z_{3}\otimes Z_{4}\otimes Z_{5}\otimes Z_{6}\otimes Z_{7}\otimes Z_{8}\otimes Z_{9}\otimes Z_{10}\otimes Z_{11}\otimes Z_{12}\otimes Z_{13}\otimes Z_{14}\otimes \frac{X_{15}-iY_{15}}{2}\otimes I_{16}.
\end{align}
\begin{align}
\hat a_{16}^\dagger
={}&
Z_{1}\otimes Z_{2}\otimes Z_{3}\otimes Z_{4}\otimes Z_{5}\otimes Z_{6}\otimes Z_{7}\otimes Z_{8}\otimes Z_{9}\otimes Z_{10}\otimes Z_{11}\otimes Z_{12}\otimes Z_{13}\otimes Z_{14}\otimes Z_{15}\otimes \frac{X_{16}-iY_{16}}{2}.
\end{align}

\section{C$_2$ molecule: extension of the H$_4$ determinant estimator}
\label{subsec:C2_results}

The C$_2$ calculation was obtained by adapting the original H$_4$ code to a larger
nonorthogonal determinant space. The H$_4$ implementation was designed to evaluate
all determinant-overlap and Hamiltonian matrix elements by mapping the relevant
fermionic operator products to Jordan--Wigner Pauli strings and measuring the
resulting vacuum-projected amplitudes. In the present C$_2$ extension, the same
workflow was generalized to 70 SCGVB determinants and 16 qubits. The calculation
therefore provides a direct test of whether the H$_4$ estimator remains operational
for a significantly larger determinant manifold.

The reconstructed values were compared against the reference values obtained
from classical L\"owdin nonorthogonal determinant rules. 
The complete matrices agree to numerical precision. Using the independent L\"owdin/PySCF reference files employed for the main-text comparison, the maximum absolute deviations are $6.66\times10^{-15}$ for the overlap matrix and $4.32\times10^{-11}$~Ha for the total Hamiltonian. Individual entries printed in Tables~\ref{tab:C2_nonzero_overlap_complete} and \ref{tab:C2_nonzero_hamiltonian_complete} may appear identical at the displayed precision. The resource estimation is provided in Table~\ref{tab:C2_resource_summary}. 

The C$_2$ calculations were performed for the $X^1\Sigma_g^+$ state at an internuclear distance of $R_{\mathrm{C-C}}=1.20$~\AA.

\begin{table}[t]
\small
\caption{Resource and matrix summary for the C$_2$ calculation.}
\label{tab:C2_resource_summary}
\centering
\begin{tabular}{|l|c|}
\hline
\textbf{Quantity} & \textbf{Value} \\
\hline
Number of SCGVB determinants & 70 \\
Number of qubits & 16 \\
Pair tasks evaluated & 2485 \\
Shots per circuit & 8192 \\
Total wall time (s) & 50462.43 \\
Maximum RSS (MB) & 3609.00 \\
Tracemalloc peak (MB) & 1015.87 \\
Nonzero unique $S_{ij}$ entries ($i\le j$) & 456 \\
Nonzero unique $H_{ij}$ entries ($i\le j$) & 1864 \\
$S$ circuits & 456 \\
$H_1$ circuits & 488 \\
$H_2$ circuits & 1864 \\
Total Hamiltonian circuits ($H_1+H_2$) & 2352 \\
Maximum circuit depth & 1 \\
Total CX gates & 0 \\
\hline
\end{tabular}
\end{table}
\FloatBarrier
\clearpage

\begin{longtable}{|l|c|c|c|}
\caption{Complete list of nonzero C$_2$ overlap matrix elements. Only unique
upper-triangular entries ($i\le j$) with $|S_{ij}|>10^{-12}$ are reported.}
\label{tab:C2_nonzero_overlap_complete}\\
\hline
$\mathbf{S_{ij}}$ & \textbf{Quantum circuit} & \textbf{L\"owdin reference} & $\mathbf{|\Delta S_{ij}|}$ \\
\hline
\endfirsthead
\hline
$\mathbf{S_{ij}}$ & \textbf{Quantum circuit} & \textbf{L\"owdin reference} & $\mathbf{|\Delta S_{ij}|}$ \\
\hline
\endhead
$\langle \psi_{1}|\psi_{1}\rangle$ & $1.000000000000e+00$ & $1.000000000000e+00$ & $0.000000000000e+00$ \\
\hline
$\langle \psi_{1}|\psi_{2}\rangle$ & $2.072069856713e-01$ & $2.072069856713e-01$ & $0.000000000000e+00$ \\
\hline
$\langle \psi_{1}|\psi_{5}\rangle$ & $-9.592025866574e-02$ & $-9.592025866574e-02$ & $0.000000000000e+00$ \\
\hline
$\langle \psi_{1}|\psi_{8}\rangle$ & $-9.147489225571e-02$ & $-9.147489225571e-02$ & $0.000000000000e+00$ \\
\hline
$\langle \psi_{1}|\psi_{11}\rangle$ & $-1.895423668891e-02$ & $-1.895423668891e-02$ & $0.000000000000e+00$ \\
\hline
$\langle \psi_{1}|\psi_{15}\rangle$ & $8.774295326588e-03$ & $8.774295326588e-03$ & $0.000000000000e+00$ \\
\hline
$\langle \psi_{1}|\psi_{17}\rangle$ & $-9.147489225571e-02$ & $-9.147489225571e-02$ & $0.000000000000e+00$ \\
\hline
$\langle \psi_{1}|\psi_{20}\rangle$ & $-1.895423668891e-02$ & $-1.895423668891e-02$ & $0.000000000000e+00$ \\
\hline
\end{longtable}

\begin{longtable}{|l|c|c|c|}
\caption{Complete list of nonzero C$_2$ total Hamiltonian matrix elements.
Only unique upper-triangular entries ($i\le j$) with $|H_{ij}|>10^{-12}$ are
reported. The total Hamiltonian includes both one-electron and two-electron
contributions.}
\label{tab:C2_nonzero_hamiltonian_complete}\\
\hline
$\mathbf{H_{ij}}$ & \textbf{Quantum circuit} & \textbf{L\"owdin reference} & $\mathbf{|\Delta H_{ij}|}$ \\
\hline
\endfirsthead
\hline
$\mathbf{H_{ij}}$ & \textbf{Quantum circuit} & \textbf{L\"owdin reference} & $\mathbf{|\Delta H_{ij}|}$ \\
\hline
\endhead
$\langle \psi_{1}|\hat{H}|\psi_{1}\rangle$ & $-2.667919484820e+01$ & $-2.667919484820e+01$ & $0.000000000000e+00$ \\
\hline
$\langle \psi_{1}|\hat{H}|\psi_{2}\rangle$ & $-5.618433603783e+00$ & $-5.618433603783e+00$ & $0.000000000000e+00$ \\
\hline
$\langle \psi_{1}|\hat{H}|\psi_{3}\rangle$ & $4.742744597738e-03$ & $4.742744597738e-03$ & $0.000000000000e+00$ \\
\hline
$\langle \psi_{1}|\hat{H}|\psi_{4}\rangle$ & $-4.742744597738e-03$ & $-4.742744597738e-03$ & $0.000000000000e+00$ \\
\hline
$\langle \psi_{1}|\hat{H}|\psi_{5}\rangle$ & $2.538651243349e+00$ & $2.538651243349e+00$ & $0.000000000000e+00$ \\
\hline
$\langle \psi_{1}|\hat{H}|\psi_{6}\rangle$ & $1.430160545135e-02$ & $1.430160545135e-02$ & $0.000000000000e+00$ \\
\hline
$\langle \psi_{1}|\hat{H}|\psi_{7}\rangle$ & $-1.418661321859e-03$ & $-1.418661321859e-03$ & $0.000000000000e+00$ \\
\hline
$\langle \psi_{1}|\hat{H}|\psi_{8}\rangle$ & $2.467037238828e+00$ & $2.467037238828e+00$ & $0.000000000000e+00$ \\
\hline
\end{longtable}

\clearpage


\section*{References}
\begin{thebibliography}{99}

\bibitem{chirgwin1950}
B.~Chirgwin and C.~A.~Coulson,
``The electronic structure of conjugated systems. VI. The evaluation of valence bond structures and resonance energies,''
\textit{Proc. R. Soc. A} \textbf{201}, 196 (1950).
\newblock DOI: \href{https://doi.org/10.1098/rspa.1950.0053}{10.1098/rspa.1950.0053}

\bibitem{lowdin1950}
P.~O.~L\"owdin, ``On the non-orthogonality problem connected with the use of atomic wave functions in the theory of molecules and crystals,''
\textit{J. Chem. Phys.} \textbf{18}, 365 (1950).
\newblock DOI: \href{https://doi.org/10.1063/1.1747632}{10.1063/1.1747632}

\bibitem{thorsteinsson1995}
T.~Thorsteinsson,
\emph{Development of Methods in Spin-Coupled Theory},
Ph.D. thesis, University of Liverpool, 1995.
\newblock URL: \url{https://livrepository.liverpool.ac.uk/3176065/}

\bibitem{shaik2012}
S.~Shaik and P.~C.~Hiberty,
``Valence Bond Theory, Its History, Fundamentals, and Applications: A Primer,''
in \textit{Reviews in Computational Chemistry},
Vol.~20, edited by K.~B.~Lipkowitz and T.~R.~Cundari
(John Wiley \& Sons, Ltd, 2004), Chap.~1, pp.~1--100.
\newblock DOI: \href{https://doi.org/10.1002/0471678856.ch1}{10.1002/0471678856.ch1}

\bibitem{penotti2019}
F.~E.~Penotti, D.~L.~Cooper, and P.~B.~Karadakov,
``Is the S$_2$N$_2$ ring a singlet diradical? Critical analysis of alternative valence bond descriptions,''
\textit{Int. J. Quantum Chem.} \textbf{119}, e25845 (2019).
\newblock DOI: \href{https://doi.org/10.1002/qua.25845}{10.1002/qua.25845}

\bibitem{araujo2025compactifyingelectronicwavefunctionsi}
B.~G.~M.~Ara\'ujo and A.~M.~S.~Macedo,
``Compactifying Electronic Wavefunctions I: Error-Mitigated Transcorrelated DMRG,''
arXiv:2503.00627 [quant-ph] (2025).
\newblock DOI: \href{https://doi.org/10.48550/arXiv.2503.00627}{10.48550/arXiv.2503.00627}
\newblock URL: \url{https://arxiv.org/abs/2503.00627}

\bibitem{wiki_cc}
\textit{Chirgwin--Coulson weights}, Wikipedia, accessed November 2025.
\newblock URL: \url{https://en.wikipedia.org/wiki/Chirgwin%E2%80%93Coulson_weights}

\bibitem{McWeeny1992}
R.~McWeeny,
\textit{Methods of Molecular Quantum Mechanics}, 2nd ed.
(Academic Press, London, 1992).
\newblock ISBN: 9780124865525
\newblock URL: \url{https://openlibrary.org/books/OL21124027M/Methods_of_molecular_quantum_mechanics}

\bibitem{Pauncz1979}
R.~Pauncz,
\textit{Spin Eigenfunctions: Construction and Use}
(Plenum Press, New York, 1979).
\newblock ISBN: 9780306401411
\newblock URL: \url{https://archive.org/details/spineigenfunctio0000paun}

\bibitem{araujo2022lqot}
B.~G.~M. Ara\'ujo, M.~M. Taddei, D.~Cavalcanti, and A.~Ac\'{\i}n,
\newblock ``Local quantum overlapping tomography,''
\newblock \textit{Phys. Rev. A} \textbf{106}, 062441 (2022).
\newblock DOI: \href{https://doi.org/10.1103/PhysRevA.106.062441}{10.1103/PhysRevA.106.062441}.

\bibitem{perezobiol2022}
A.~P\'erez-Obiol, A.~P\'erez-Salinas, S.~S\'anchez-Ram\'{\i}rez, B.~G.~M. Ara\'ujo, and A.~Garcia-Saez,
\newblock ``Adiabatic quantum algorithm for artificial graphene,''
\newblock \textit{Phys. Rev. A} \textbf{106}, 052408 (2022).
\newblock DOI: \href{https://doi.org/10.1103/PhysRevA.106.052408}{10.1103/PhysRevA.106.052408}.

\bibitem{marruzzo2025nojw}
A.~Marruzzo, M.~Casalegno, P.~Macchi, F.~Mascherpa,
B.~Tirri, G.~Raos, and A.~Genoni,
\newblock ``Extension of the Jordan--Wigner mapping to nonorthogonal spin orbitals for quantum computing application to valence bond approaches,''
\newblock in \textit{A Snapshot of Molecular Electronic Structure Theory and its Applications},
edited by C.~Coletti and P.~E.~Hoggan,
\textit{Adv. Quantum Chem.} \textbf{92}, 245--266 (2025).
\newblock DOI: \href{https://doi.org/10.1016/bs.aiq.2025.07.007}{10.1016/bs.aiq.2025.07.007}.


\bibitem{lowdin1955}
P.~O.~L\"owdin,
``Quantum theory of many-particle systems. I. Physical interpretations by means of density matrices, natural spin-orbitals, and convergence problems in the method of configurational interaction,''
\textit{Phys. Rev.} \textbf{97}, 1474--1489 (1955).
\newblock DOI: \href{https://doi.org/10.1103/PhysRev.97.1474}{10.1103/PhysRev.97.1474}

\bibitem{lowdin1962}
P.~O.~L\"owdin,
``Studies in perturbation theory. IV. Solution of eigenvalue problem by projection operator formalism,''
\textit{J. Math. Phys.} \textbf{3}, 969--982 (1962).
\newblock DOI: \href{https://doi.org/10.1063/1.1724312}{10.1063/1.1724312}

\bibitem{goddard1973}
W.~A.~Goddard~III, T.~H.~Dunning~Jr., W.~J.~Hunt, and P.~J.~Hay,
``Generalized valence bond description of bonding in low-lying states of molecules,''
\textit{Acc. Chem. Res.} \textbf{6}, 368--376 (1973).
\newblock DOI: \href{https://doi.org/10.1021/ar50071a002}{10.1021/ar50071a002}
\bibitem{hiberty2002}
P.~C.~Hiberty and S.~Shaik, 
``Comment on the relationship between valence bond and molecular orbital methods,'' 
\textit{J. Comput. Chem.} \textbf{23}, 973 (2002).

\bibitem{kutzelnigg1997}
W.~Kutzelnigg and D.~Mukherjee,
``Normal order and extended Wick theorem for a multiconfiguration reference wave function,''
\textit{J. Chem. Phys.} \textbf{107}, 432--449 (1997).
\newblock DOI: \href{https://doi.org/10.1063/1.474405}{10.1063/1.474405}

\bibitem{ShaikHiberty2007}
S.~Shaik and P.~C.~Hiberty,
\textit{A Chemist's Guide to Valence Bond Theory}
(John Wiley \& Sons, Inc., Hoboken, NJ, 2007).
\newblock DOI: \href{https://doi.org/10.1002/9780470192597}{10.1002/9780470192597}

\bibitem{szabo1996}
A.~Szabo and N.~S.~Ostlund,
\textit{Modern Quantum Chemistry: Introduction to Advanced Electronic Structure Theory}
(Dover Publications, New York, 1996).
\newblock ISBN: 9780486691862
\newblock URL: \url{https://store.doverpublications.com/products/9780486691862}

\bibitem{helgaker2000}
T.~Helgaker, P.~J{\o}rgensen, and J.~Olsen,
\textit{Molecular Electronic-Structure Theory}
(Wiley, Chichester, 2000).
\newblock DOI: \href{https://doi.org/10.1002/9781119019572}{10.1002/9781119019572}

\bibitem{Cleve1998}
R.~Cleve, A.~Ekert, C.~Macchiavello, and M.~Mosca,
``Quantum algorithms revisited,''
\textit{Proc. R. Soc. A} \textbf{454}, 339--354 (1998).
\newblock DOI: \href{https://doi.org/10.1098/rspa.1998.0164}{10.1098/rspa.1998.0164}

\bibitem{Kitaev1997}
A.~Yu.~Kitaev,
``Quantum measurements and the Abelian stabilizer problem,''
arXiv:quant-ph/9511026 (1995).
\newblock DOI: \href{https://doi.org/10.48550/arXiv.quant-ph/9511026}{10.48550/arXiv.quant-ph/9511026}
\newblock URL: \url{https://arxiv.org/abs/quant-ph/9511026}

\bibitem{NielsenChuang}
M.~A.~Nielsen and I.~L.~Chuang,
\textit{Quantum Computation and Quantum Information}, 10th anniversary ed.
(Cambridge University Press, Cambridge, 2010).
\newblock ISBN: 9781107002173
\newblock URL: \url{https://www.cambridge.org/highereducation/books/quantum-computation-and-quantum-information/01E10196D0A682A6AEFFEA52D53BE9AE}

\bibitem{Baek2023NO}
U.~Baek, D.~Hait, J.~Shee, O.~Leimkuhler, W.~J.~Huggins, T.~F.~Stetina,
M.~Head-Gordon, and K.~B.~Whaley,
``Say NO to Optimization: A Nonorthogonal Quantum Eigensolver,''
\textit{PRX Quantum} \textbf{4}, 030307 (2023).
\newblock DOI: \href{https://doi.org/10.1103/PRXQuantum.4.030307}{10.1103/PRXQuantum.4.030307}

\bibitem{pyscf}
Q.~Sun, T.~C.~Berkelbach, N.~S.~Blunt, \textit{et al.},
``PySCF: the Python-based simulations of chemistry framework,''
\textit{WIREs Comput. Mol. Sci.} \textbf{8}, e1340 (2018).
\newblock DOI: \href{https://doi.org/10.1002/wcms.1340}{10.1002/wcms.1340}

\bibitem{qiskit}
H.~Abraham \textit{et al.},
``Qiskit: An Open-source Framework for Quantum Computing,''
Zenodo (2019).
\newblock DOI: \href{https://doi.org/10.5281/zenodo.2562111}{10.5281/zenodo.2562111}


\bibitem{dunning2021_scgvb}
T.~H.~Dunning, Jr., L.~T.~Xu, D.~L.~Cooper, and P.~B.~Karadakov,
``Spin-Coupled Generalized Valence Bond Theory: New Perspectives on the Electronic Structure of Molecules and Chemical Bonds,''
\textit{J. Phys. Chem. A} \textbf{125}, 2021--2050 (2021).
\newblock DOI: \href{https://doi.org/10.1021/acs.jpca.0c10472}{10.1021/acs.jpca.0c10472}

\bibitem{dunning2023_scgvb}
T.~H.~Dunning, Jr., D.~L.~Cooper, L.~T.~Xu, and P.~B.~Karadakov,
``Spin-Coupled Generalized Valence Bond Theory: An Appealing Orbital Theory of the Electronic Structure of Atoms and Molecules,''
in \textit{Comprehensive Computational Chemistry (First Edition)},
edited by M.~Yáñez and R.~J.~Boyd
(Elsevier, Oxford, 2024), pp.~354--402.
\newblock DOI: \href{https://doi.org/10.1016/B978-0-12-821978-2.00017-9}{10.1016/B978-0-12-821978-2.00017-9}
\newblock URL: \url{https://www.sciencedirect.com/science/article/pii/B9780128219782000179}

\bibitem{cooper_introVB}
D.~L.~Cooper,
``Valence Bond Theory: Introduction to \emph{Ab Initio} Modern VB Methods,''
lecture notes, Sorbonne/LCT Workshop (2014).
\newblock URL: \url{https://wiki.lct.jussieu.fr/workshop/images/8/85/Cooper_Intro.pdf}

\bibitem{cooper_rumer}
J.~Gerratt, D.~L.~Cooper, and M.~Raimondi,
``The spin-coupled valence bond theory of molecular electronic structure,''
in \textit{Valence Bond Theory and Chemical Structure},
edited by D.~J.~Klein and N.~Trinajsti\'c
(Elsevier, Amsterdam, 1990), pp.~287--329.
\newblock URL: \url{https://books.google.com/books/about/Valence_Bond_Theory_and_Chemical_Structu.html?id=ZttkAAAAIAAJ}

\bibitem{valence2019}
G.~D.~Fletcher, C.~Bertoni, M.~Ke\c{c}eli, and M.~D'Mello,
``VALENCE: A Massively Parallel Implementation of the Variational Subspace Valence Bond Method,''
\textit{J. Comput. Chem.} \textbf{40}, 1664--1673 (2019).
\newblock URL: \url{https://www.osti.gov/servlets/purl/1510723}

\bibitem{rumer_spin_functions}
G.~Rumer,
``Zur Theorie der Spinvalenz,''
\textit{Nachrichten von der Gesellschaft der Wissenschaften zu G\"ottingen, Mathematisch-Physikalische Klasse} \textbf{1932}, 337--341 (1932).
\newblock URL: \url{http://eudml.org/doc/59385}

\bibitem{XuDunning_C2_GVB}
L.-T. Xu and T. H. Dunning, Jr.,
``Insights into the perplexing nature of the bonding in C$_2$ from generalized valence bond calculations,''
\textit{J. Chem. Theory Comput.} \textbf{10}, 195--201 (2014).
DOI:~\href{https://doi.org/10.1021/ct400867h}{10.1021/ct400867h}.

\bibitem{cooper2015C2}
D. L. Cooper, F. E. Penotti, and R. Ponec,
``Why is the bond multiplicity in C$_2$ so elusive?''
\textit{Comput. Theor. Chem.} \textbf{1053}, 189--194 (2015).
DOI:~\href{https://doi.org/10.1016/j.comptc.2014.08.024}{10.1016/j.comptc.2014.08.024}.

\bibitem{RaosGerrattCooperRaimondi1993_spinbases}
G. Raos, J. Gerratt, D. L. Cooper, and M. Raimondi,
``Spin correlation in $\pi$-electron systems from spin-coupled wavefunctions. II. Further applications,''
\textit{Chem. Phys.} \textbf{186}, 251--273 (1994).
DOI:~\href{https://doi.org/10.1016/0301-0104(94)00178-2}{10.1016/0301-0104(94)00178-2}.

\end{thebibliography}

\begin{thebibliography}{99}


\bibitem{marruzzo2025nojw_SI}
A.~Marruzzo, M.~Casalegno, P.~Macchi, F.~Mascherpa,
B.~Tirri, G.~Raos, and A.~Genoni,
\newblock ``Extension of the Jordan--Wigner mapping to nonorthogonal spin orbitals for quantum computing application to valence bond approaches,''
\newblock in \textit{A Snapshot of Molecular Electronic Structure Theory and its Applications},
edited by C.~Coletti and P.~E.~Hoggan,
\textit{Adv. Quantum Chem.} \textbf{92}, 245--266 (2025).
\newblock DOI: \href{https://doi.org/10.1016/bs.aiq.2025.07.007}{10.1016/bs.aiq.2025.07.007}.

\bibitem{XuDunning_C2_GVB_SI}
L.-T. Xu and T. H. Dunning, Jr.,
``Insights into the perplexing nature of the bonding in C$_2$ from generalized valence bond calculations,''
\textit{J. Chem. Theory Comput.} \textbf{10}, 195--201 (2014).
DOI:~\href{https://doi.org/10.1021/ct400867h}{10.1021/ct400867h}.


\bibitem{cooper2015C2_SI}
D. L. Cooper, F. E. Penotti, and R. Ponec,
``Why is the bond multiplicity in C$_2$ so elusive?''
\textit{Comput. Theor. Chem.} \textbf{1053}, 189--194 (2015).
DOI:~\href{https://doi.org/10.1016/j.comptc.2014.08.024}{10.1016/j.comptc.2014.08.024}.


\bibitem{RaosGerrattCooperRaimondi1993_spinbases_SI}
G. Raos, J. Gerratt, D. L. Cooper, and M. Raimondi,
``Spin correlation in $\pi$-electron systems from spin-coupled wavefunctions. II. Further applications,''
\textit{Chem. Phys.} \textbf{186}, 251--273 (1994).
DOI:~\href{https://doi.org/10.1016/0301-0104(94)00178-2}{10.1016/0301-0104(94)00178-2}.

\end{thebibliography}
\end{document}